\documentclass{elsarticle}

\usepackage{mathrsfs}
\usepackage{hyperref}
\usepackage{tikz}
\usepackage[export]{adjustbox}
\usepackage{amsmath}
\usepackage{geometry,tikz}
\usetikzlibrary{arrows,matrix,positioning,fit}
\usepackage{amssymb}
\usepackage{physics}
\usepackage{float}
\usepackage{subcaption}
\usepackage{array}
\usepackage{multirow}
\usepackage{relsize}
\usepackage{ulem}
\journal{Ocean Engineering (accepted version)}

\newcommand{\fig}[1]{Fig.~\ref{#1}}
\newcommand{\eq}[1]{Eq.~(\ref{#1})}
\newcommand{\eqp}[1]{Eq.~\ref{#1}}
\newcommand{\tab}[1]{Tab.~\ref{#1}}
\newcommand{\mypar}[1]{\vspace{0.3cm}\noindent\textbf{#1}}
\newcommand{\rev}[1]{#1}

\DeclareCaptionFormat{cont}{#1 #2#3\par}

\newcommand{\vx}{u} 
\newcommand{\vy}{v} 
\newcommand{\vz}{w} 
\newcommand{\ax}{u_{,t}} 
\newcommand{\ay}{v_{,t}} 
\newcommand{\az}{w_{,t}} 
\newcommand{\sx}{\eta_{,x}} 
\newcommand{\sy}{\eta_{,y}} 
\newcommand{\de}{\eta_{,t}} 

\newcommand{\etaN}{\eta_0} 
\newcommand{\dEtaN}{\dot{\eta}_0} 

\newcommand{\Tz}{T_z}
\newcommand{\cZ}{\check{Z}}
\newcommand{\cZs}{\check{Z}_s}
\newcommand{\cZb}{\check{Z}_b}
\newcommand{\cX}{\check{X}}
\newcommand{\vxs}{u} 
\newcommand{\vys}{v} 
\newcommand{\axs}{u_{,t}} 
\newcommand{\ays}{v_{,t}}  
\newcommand{\sxs}{\eta_{,x}} 
\newcommand{\sys}{\eta_{,y}} 
\newcommand{\nuz}{\mu_{_0}^{_\uparrow}}
\newcommand{\nus}{\mu_{s}^{_\uparrow}}
\newcommand{\nub}{\mu_{b}^{_\uparrow}}
\newcommand{\mmt}{m_{\tilde{2}}}
\newcommand{\etaS}{\eta_{s}} 
\newcommand{\dEtaS}{\dot{\eta}_{s}} 
\newcommand{\etaB}{\eta_b} 
\newcommand{\dEtaB}{\dot{\eta}_b}

\begin{document}

\begin{frontmatter}

\title{Effect of forward speed on the level-crossing distribution of 
kinematic variables in multidirectional ocean waves}

% %%%%%%%%%%%%%%%%%%%%%%%%%%%%%%%

\author[enstaAddress]{Romain Hasco\"{e}t}
\cortext[mycorrespondingauthor]{Corresponding author}
\ead{romain.hascoet@ensta-bretagne.fr}

\author[ifremerAddress]{Nicolas Raillard}
\author[enstaAddress]{Nicolas Jacques}

\address[enstaAddress]{ENSTA Bretagne, CNRS UMR 6027, IRDL, 2 rue Fran\c{c}ois Verny, 29806 Brest Cedex 9, France}
\address[ifremerAddress]{IFREMER -- LCSM, ZI Pointe du Diable, 29280 Plouzan\'{e} CS 10070, France}

\begin{abstract}

The influence of forward speed on 
stochastic free-surface crossing, 
in a Gaussian wave field, is investigated.
The case of a material point moving with a constant forward speed is considered;
the wave field is assumed stationary in time, and homogeneous in space.
The focus is on up-crossing events, which are defined as the material point crossing the free surface, into the water domain.
The effect of the Doppler shift (induced by the forward speed)
on the up-crossing frequency, and the related conditional joint distribution 
of wave kinematic variables is analytically investigated.
Some general trends are illustrated through different examples, 
where three kinds of wave direction distribution are considered:  
unidirectional, short-crested anisotropic, and isotropic.
The way the developed approach may be used in the context of 
slamming on marine structures is briefly discussed.

\end{abstract}

\begin{keyword}
water wave \sep
Gaussian \sep 
level crossing  \sep 
Doppler effect \sep
forward speed \sep
slamming
\end{keyword}

\end{frontmatter}

%\linenumbers

\section{Introduction}

From an engineering standpoint, 
two quantities related to the chance of free-surface crossing
for an object traveling in an ocean wave field
may be of interest: 
(i) the average frequency of crossing events; 
(ii) the related joint probability distribution of wave kinematic variables, given crossing. 
\rev{
The present study focuses on up-crossing events, 
which are defined as 
follows: the free surface up-crosses the object,
or equivalently the object down-crosses the free surface (into the water domain).
}
%the crossing of the free surface by the object, into the water domain.
An up-crossing event will lead to a water entry phenomenon, 
which may induce significant hydrodynamic loads on the structure of the object 
(see e.g. \cite{korobkin_1988, faltinsen_2006,kapsenberg_2011,wang_2017}).
Therefore, the knowledge of (i) and (ii) may be valuable 
for the design of a marine structure that will be exposed to water wave impacts. 
Down-crossing events (i.e. water-exit events) may also be of practical interest in ship design;
for example to assess the risk of efficiency loss due to the (partial) emersion 
of an appendage or a propeller.
Moreover, water exit events can also generate high-intensity transient hydrodynamic loads 
(see e.g. \cite{baarholm_2004,korobkin_2017,breton_2020}).
The theoretical approach, 
developed in the present paper for up-crossing events, 
may be readily transposed to down-crossing events.

If the considered object (for instance, a ship appendage or a hull section)
is sufficiently small compared to water wave wavelengths,
the body geometry may be reduced to a single material point regarding the risk of free-surface crossing. 
Then, the problem becomes more tractable and
may be addressed by using the level-crossing theory of stochastic processes,
based on the pioneering work of Rice (1944,1945) \cite{rice_1944, rice_1945}
and subsequent works (see for example \cite{lindgren_2012}, Chapter 8, and references therein).
When the motions of the water waves are modelled at the first order (Airy wave theory), 
the randomness of the related kinematic variables can be modelled through Gaussian processes.
Then, the average up-crossing frequency and the related 
joint probability distribution of kinematic variables are both analytically tractable.

When the material point moves with a given forward velocity, 
the motion-induced Doppler shift has a non-trivial effect 
on the encounter wave spectrum.
Lindgren et al. (1999) \cite{lindgren_1999} investigated the effect of Doppler shift 
on the distribution of the zero crossing wave period 
(the time between successive mean level down- and up-crossings) 
measured in the frame of a moving body,
but they did not consider the related conditional distribution of kinematic variables.
\rev{ 
More recently, Aberg et al. (2008) \cite{aberg_2008} also investigated 
the effect of Doppler shift on the distribution of some wave characteristics 
measured along the direction of body motion
(namely the wave slope, waveheight and wavelength).
}
In the context of slamming on ships, 
following the pioneering work of Ochi \cite{ochi_1964a, ochi_1964b} and Ochi and Motter \cite{ochi_1971,ochi_1973},
different authors investigated the stochastic properties of slamming-induced 
loads and structural stresses, 
including the effect of forward speed in the analysis.
In most studies, however, 
the vertical component of the relative fluid velocity is considered
as the only kinematic variable relevant to the estimate of slamming loads
(see for example \cite{ochi_1973,rassinot_1995,wang_2002,hermundstad_2007,dessi_2013,wang_2016}).
This assumption greatly simplifies the problem
by reducing the conditional distribution of kinematic 
variables (used as an input for the impact model), 
given up-crossing, 
to a univariate distribution, which is of Rayleigh type in the framework of linear wave theory.
However, this univariate approach may provide unreliable predictions,
since slamming loads may be 
sensitive to other kinematic variables (e.g. \cite{scolan_2015}),
such as the acceleration (added-mass effect), 
the tangential velocity,
or the angular position of the object relative to the local free surface (e.g. \cite{hascoet_2019}).
Going beyond a univariate approach,
the early study of Belik and Price (1982) \cite{belik_1982} investigated the effect of
accounting for the tangential component of the water entry velocity, 
in the context of slamming on high-speed vessels.
They considered only long-crested (unidirectional) sea states and carried out their investigation through numerical experiments, 
where the random sampling of impact kinematic variables was obtained from 
the numerical realisation of Gaussian waves and ship responses.
More recently, Helmers et al. (2012) \cite{helmers_2012}
 analytically considered the joint distribution of several kinematic variables 
(namely the fluid vertical velocity, the fluid vertical acceleration, the wave slope, and the seakeeping roll angle) 
to estimate the probability distribution of impact loads on a wedge-shaped body exposed to irregular waves;
however, these authors did not consider scenarii where the body has a forward speed and focused on unidirectional sea states.

The present study is devoted to
the analytical investigation of the effect of forward speed
on the conditional joint distribution of wave kinematic variables, given up-crossing.
The effect of  forward motion on the related up-crossing frequency is also analysed.
The case of a material point moving through a Gaussian wave field, 
at a constant velocity, in a horizontal plane (i.e. at a given altitude), is investigated.
Both long-crested (unidirectional) and short-crested (multidirectional) wave fields are considered.
Section \ref{sec_framework} sets the framework of the present study. 
In Section \ref{sect_body_rest}, the case of a body at rest is first considered as a preamble.
Then, in Section \ref{sec_forward}, the effect of forward speed on the up-crossing frequency 
and the related conditional joint distribution of kinematic variables is analytically investigated;
it is illustrated through a few examples. 
Section \ref{sect_discussion} briefly discusses how the 
developed framework may be used in the context of slamming. 
The paper ends with a concluding summary in Section \ref{sec_conclusion}.

\section{Framework and assumptions}
\label{sec_framework}

The present section sets the framework of the analysis
to be developed in Sections \ref{sect_body_rest} and \ref{sec_forward}.

\subsection{Water waves modelled as a Gaussian field}
\label{subsec_seastate_gauss_process}

% %%%%%%%%%%%%%%%%%%%%%%%%
% %%%%%%%%%%%%%%%%%%%%%%%%
% %%%%%        FIGURE    BEGIN         %%%%%
% %%%%%%%%%%%%%%%%%%%%%%%%
% %%%%%%%%%%%%%%%%%%%%%%%%

\begin{figure}[t]
\begin{center}
\begin{tabular}{c}
        \includegraphics[width=0.45\textwidth]{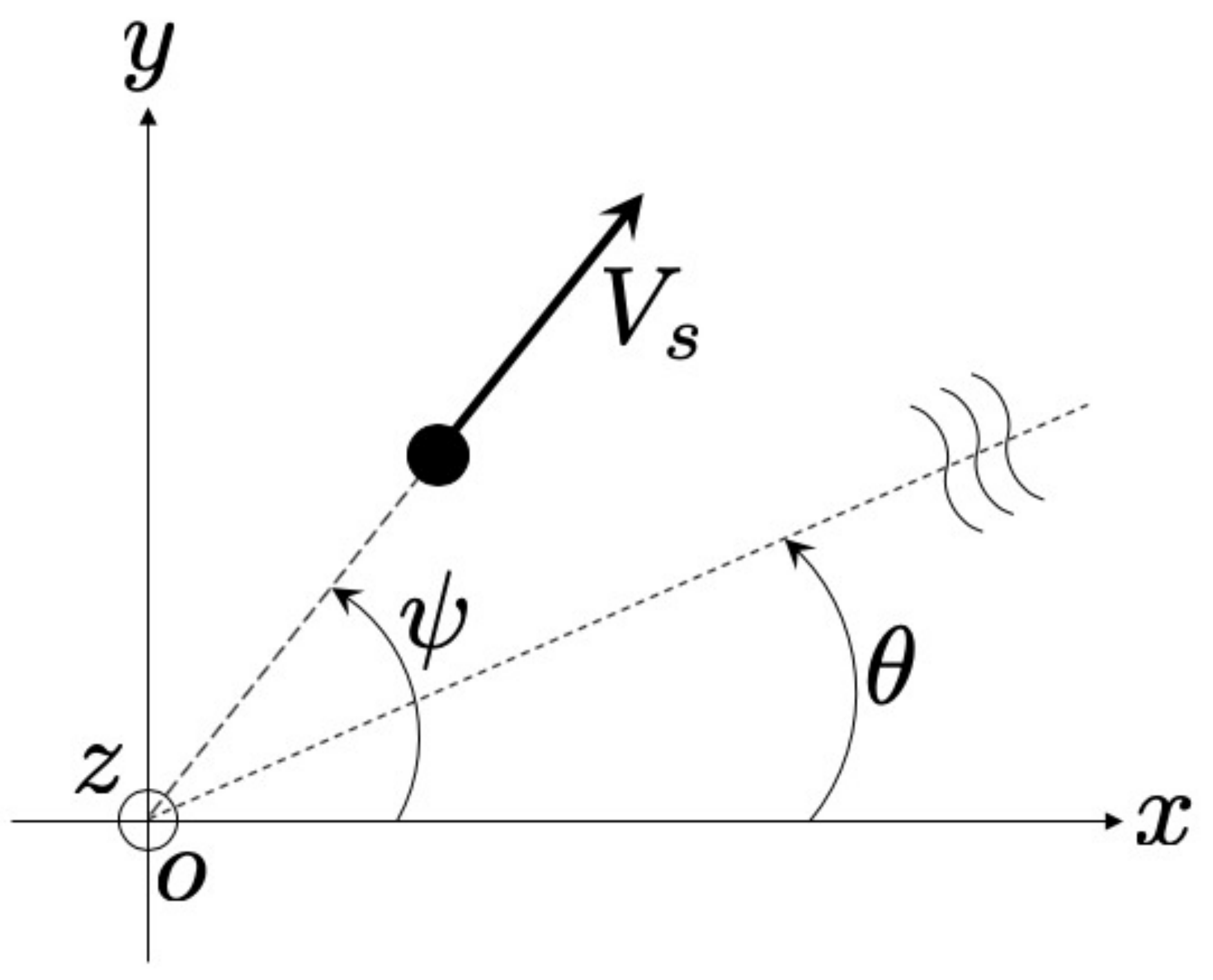}  
\end{tabular}
\end{center}
\caption{
Sketch of the problem. 
A material point (black dot) moves uniformly, at an altitude $z=a$, in a stationary and homogeneous Gaussian wave field.
The problem is formulated in a space coordinate system $Oxyz$ attached to the reference frame of the mean flow 
(i.e. the reference frame where there is no mean current).
The origin of the space coordinate system, $O$, and the time origin are chosen so that
the material point coordinates are $(x_0=0,y_0=0,z=a)$, at $t=0$.
Unless otherwise specified, the flow velocity and acceleration are measured in the reference frame of the mean flow.  
The multidirectional sea state is modelled as the superposition of Airy waves,
whose direction of propagation is noted $\theta$.
The material point moves at a speed $V_s$ (relative to the mean flow) and heads along the direction $\psi$.
}
\label{fig_frame}
\end{figure}

% %%%%%%%%%%%%%%%%%%%%%%%%
% %%%%%%%%%%%%%%%%%%%%%%%%
% %%%%%        FIGURE    END         %%%%%
% %%%%%%%%%%%%%%%%%%%%%%%%
% %%%%%%%%%%%%%%%%%%%%%%%%

In the present paper both long-crested (unidirectional) and short-crested (multidirectional) seas are considered.
Let the wave motions be described in a spacetime coordinate system $(t,x,y,z)$,
where the origin of the space coordinates, $O$,
is a fixed point in the reference frame of the mean flow (i.e. the reference frame where there is no mean current),
located on the plane of the mean free surface, and $z$ is directed along the ascending vertical 
(see \fig{fig_frame} for a sketch of the problem formulation).
In the most general case, the free surface elevation $\eta$ (measured along the vertical direction $z$) can be modelled as a stochastic process 
which depends on the time $t$ and on the two horizontal space coordinates, $x$ and $y$.
If a linear wave model is assumed, 
this stochastic process can be modelled as Gaussian.
Besides, 
in the present study, the considered sea states are assumed to be stationary in time and 
homogeneous in space. 
Then the Gaussian process $\eta$ is fully characterised -- in a probabilistic sense -- 
by its mean (which is zero in the present case) and
its two-dimensional one-sided variance density spectrum $G(\omega,\theta)$ 
(defined for $\omega>0$ and $\theta \in ]-\pi,\pi]$),
where $\omega$ is the intrinsic wave angular frequency 
(measured in the reference frame of the mean flow), 
and $\theta$ the direction of wave propagation in the plane $Oxy$.
In this framework, a realisation of the random wave field  
may be numerically approximated as the sum of independent Airy waves 
(see e.g. \cite{ochi_2005, holthuijsen_2007})
\begin{equation}
\eta(x,y,t) \simeq \sum\limits_{n=1}^{N} \sum\limits_{q=1}^{Q} 
a_{nq} \cos \left[ \omega_n t - (k_n \cos \theta_q) x - (k_n \sin \theta_q) y + \phi_{nq} \right] \, ,
\end{equation}
where the frequencies and directions, $\omega_n$, $\theta_q$, 
account for the discretisation of the two-dimensional wave spectrum.
The wave amplitudes and phases, $a_{nq}$ and $\phi_{nq}$, are independent random variables. 
The wave amplitudes, $a_{nq}$, follow individual Rayleigh distributions of modes 
\begin{equation}
\sigma_{nq} = \sqrt{G(\omega_n,\theta_q) \Delta \omega_n \Delta \theta_q } \, ,
\end{equation}
where $\Delta \omega_n$ and  $\Delta \theta_q$ 
are the sizes of frequency and direction discretisation intervals.
The wave phases, $\phi_{nq}$, are uniformly distributed over $]-\pi,\pi]$.
The wave numbers, $k_n$, are related to the wave frequencies, $\omega_n$, through the dispersion relation
\begin{equation}
\label{eq_disp_rel}
\omega_n^2 = g k_n \tanh k_n h \, ,
\end{equation}
where $g$ is the acceleration due to gravity
($g=9.81 \ {\rm m\cdot s^{-2}}$ is assumed in the present paper), 
and $h$ is the water depth. 
In the following sections, the transfer functions of different kinematic variables will be introduced.
To that end, it is convenient to introduce the complex notation and express the surface elevation as
\begin{equation}
\label{eq_airy_cpx}
\eta(x,y,t) \simeq \sum\limits_{n=1}^{N} \sum\limits_{q=1}^{Q} 
{\rm Re} \left\{ A_{nq}   \exp \left( i \left[ \omega_n t - (k_n \cos \theta_q) x - (k_n \sin \theta_q) y\right] \right) \right\} \, ,
\end{equation}
where $i$ is the imaginary unit, and
\begin{equation}
\label{eq_cpx_amp}
A_{nq} = a_{nq} \exp \left\{ i \phi_{nq} \right\} \, ,
\end{equation}
are the wave complex amplitudes. 

\rev{
In the present approach, the space coordinate system $(x,y,z)$ is Eulerian.
A Lagrangian-type linear wave model could have been an attractive alternative, 
as this kind of model has been shown to account for interesting wave features
(see e.g. \cite{gjosund_2003, fouques_2006, lindgren_2011}), 
which are missed by the Eulerian linear model (e.g. the steepening of crests and flattening of troughs).
However the Lagrangian approach leads to a model which is nonlinear 
(with respect to wave amplitudes) when expressed in an Eulerian coordinate system,
even when it is restricted to the first order.
This nonlinear feature would have hindered the analytical developments 
to be set out in Sections \ref{sect_body_rest}-\ref{sec_forward}.
}

\subsection{Wave direction distribution}
\label{subsec_Dspread}

Most equations developed in Sections \ref{sect_body_rest}-\ref{sec_forward}
assume 
that the frequency and direction dependencies in the two-dimensional spectrum can be separated as follows:
\begin{equation}
\label{eq_G_dir_spectrum}
G(\omega,\theta) = D(\theta) S(\omega).
\end{equation}
In the above equation, $S$ is the 
frequency spectrum,
and $D$ is a normalised function, defined for $\theta \in ]-\pi,\pi]$, 
and satisfying,
\begin{equation}
\int_{-\pi}^{\pi}
{\rm d} \theta \ D(\theta) = 1 \, .
\end{equation}
This frequency-direction separation is assumed, in order to further analytical developments.
However, it is not theoretically restrictive and formulae valid in the general case
(where this assumption would not hold) are always introduced beforehand.

In the different illustrative examples given below, three different types 
of wave direction distribution are considered:
\begin{enumerate}
\item Unidirectional sea state,
with a spreading function given by
\begin{equation}
\label{eq_D1}
D_1(\theta) = \delta(\theta) \, ,
\end{equation}
where $\delta$ denotes the Dirac delta function.
Following this distribution, all waves propagate in the direction of increasing $x$-coordinate.
\item Multidirectional anisotropic sea state, with a spreading function given by
\begin{equation}
\label{eq_D2}
\begin{array}{cc|ll}
D_2(\theta) & = & (2/\pi) \cos^2 \theta & , \ {\rm for} \ \abs{\theta} < \pi/2 \\
  &  &  0 \,  & , \ {\rm for} \ \abs{\theta} > \pi/2 \, .
\end{array}
\end{equation}
\item Isotropic sea state, with a spreading function
\begin{equation}
\label{eq_D3}
D_3(\theta) = 1/2\pi  \, , \ \theta \in ]-\pi,\pi] \, .
\end{equation}
\end{enumerate}

\subsection{Spectral shape}
\label{subsec_specShape}

In the illustrative examples reported below, 
the considered sea states are assumed to have 
a JONSWAP frequency spectrum \cite{hasselmann_1973}:
\begin{equation}
\label{eq_jonswap}
\begin{array}{c}
S(\omega) = N_0 H_s^2 \omega_p^4 \omega^{-5} \exp \left[ -\frac{5}{4} \left( \frac{\omega}{\omega_p} \right)^{-4} \right]
\gamma^{ \exp\left[-(\omega-\omega_p)^2/2\sigma^2\omega_p^2\right]    } 
\end{array}
\end{equation}
with
\begin{equation}
\begin{array}{cc|ll}
 \sigma & = & 0.07 & \, , {\rm for} \ \omega \le \omega_p  \\
             &    & 0.09 & \, , {\rm for} \ \omega > \omega_p \, .
\end{array} 
\end{equation}
$H_s$ is the significant wave height and $\omega_p$ is the peak angular frequency of the spectrum. 
$\gamma$ is a parameter which controls the magnitude of the peak of the spectrum 
relative to its high-frequency tail;
in all examples it is set to its ``typical'' value $\gamma=3.3$.
The normalisation factor, $N_0$, is such that
\begin{equation}
\label{eq_jonswap_norm}
H_s^2 = 16 \int_{0}^{+\infty} S(\omega) \ {\rm d} \omega \, .
\end{equation}
Besides, the wave spectrum is truncated at low and high frequencies 
in order to discard excessively long and short waves: 1\% of wave variance (which is proportional to wave energy) 
is truncated at the low-frequency and high-frequency ends (in total 2\% of wave energy is disregarded).\footnote{
This assumption is not critical for the theoretical developments which follow; 
another truncation level may have been adopted.
However, note that the truncation of the high-frequency end will matter when coming to practical applications, 
such as the study of wave-induced slamming loads.
In such applications, the exposed part of the solid body should be sufficiently small compared to water wave wavelengths, 
so that it can be modelled as a single material point regarding the risk of free surface crossing.
}
The normalisation of the spectrum following \eq{eq_jonswap_norm} is performed before the truncation.
For $\gamma = 3.3$, this leads to a low-frequency cutoff $\omega_{\rm min}  \simeq 0.74 \omega_p$
and a high-frequency cutoff $\omega_{\rm max}  \simeq 3.0 \omega_p$.
The average zero-crossing wave period is given by 
\begin{equation}
\Tz = 2 \pi \sqrt{\frac{m_0}{m_2}} \, ,
\end{equation}
where $m_p$ denotes the $p$-th moment of the wave spectra,
\begin{equation}
\label{eq_spectrum_moment}
m_p = \int_0^{+\infty} \omega^p S(\omega) {\rm d} \omega \, .
\end{equation}
Following these assumptions about the spectral shape, 
the numerical values of the first five 
moments are
\begin{equation}
\label{eq_num_moments}
\begin{array}{l}
m_0 \simeq 0.0613 \ {H_s}^2 \, ,  \\
m_1 \simeq  0.0720 \ {H_s}^2 \omega_p \, ,  \\
 m_2  \simeq 0.0918 \ {H_s}^2 {\omega_p}^2 \, ,  \\ 
 m_3  \simeq 0.130 \ {H_s}^2 {\omega_p}^3 \, ,  \\  
 m_4  \simeq 0.208 \ {H_s}^2 {\omega_p}^4 \, .  \\   
\end{array} 
\end{equation}

\subsection{Considered kinematic variables}
\label{subsec_kin_vars}

In Sections \ref{sect_body_rest} and \ref{sec_forward}, analytical formulae 
will be provided for 
the joint probability distribution of wave kinematic variables, given up-crossing.
The variables considered in the present study are the following:
\begin{itemize}
\item $\eta(x,y,t)$: the free surface elevation
\item $\vx(x,y,t)$, $\vy(x,y,t)$: the horizontal components of the fluid velocity in the plane $z=0$, 
along the x-axis and y-axis, respectively.
\item $\ax(x,y,t)$, $\ay(x,y,t)$: the components of the fluid acceleration in the plane $z=0$, along x-axis and y-axis respectively.\footnote{
In the present paper, $Q_{,v}$ denotes the derivative of the function $Q$ with respect to the variable $v$.
}
\item $\vz(x,y,t)$: the vertical component of the fluid velocity in the plane $z=0$. In the linear wave model, $\vz=\de$.
\item $\az(x,y,t)$: the vertical component of the fluid acceleration in the plane $z=0$. 
\item $\sx(x,y,t)$, $\sy(x,y,t)$: the slope components of the free surface along x-axis and y-axis respectively.
\end{itemize}
Including other first-order kinematic variables (such as the free surface curvature) 
in the analysis to be developed in Sections \ref{sect_body_rest}-\ref{sec_forward},
would be straightforward.
In the present study, the first-order flow velocity and acceleration components at $z = 0$, 
are considered as a direct proxy for the kinematics at the free surface.
This assumption is in line with several ``stretching'' schemes used to alleviate the deficiencies 
of the linear wave theory in describing the near-surface fluid kinematics for irregular sea states 
(e.g. \cite{wheeler_1970,chakrabarti_1971,horng_1991}).
Conversely, other techniques such as linear extrapolation above $z=0$ (e.g. \cite{xu_1995}) or ``Delta stretching'' \cite{rodenbusch_1986} 
would yield a different proxy 
for the fluid kinematics at the free surface.

The kinematic variables listed above were selected as potentially relevant 
for the computation of hydrodynamic loads on a marine structure 
\rev{(see section \ref{sect_discussion} for a discussion in the context of slamming).
Hence, the considered horizontal velocity and acceleration components are those of the fluid.
When the interest would be on the horizontal motions of the free surface itself
(e.g. for remote sensing applications), 
the approach developed below may still be used to compute 
the probability distribution of the relevant variables.
Note however that different definitions are admissible to describe the horizontal motions of the free surface.
%In terms of horizontal velocity components, 
These different definitions translate into nonlinear relations involving partial derivatives of $\eta$, 
which would complicate the computation of the resulting distributions 
(for more details, see Baxevani et al. 2003 \cite{baxevani_2003}).}

\section{Free surface up-crossing at a fixed material point}
\label{sect_body_rest}

\subsection{Non-conditional distribution of kinematic variables}
\label{subsec_body_rest}

When the body is assumed to be at rest, located at $(x_0,y_0,z=a)$,
the wetting of the material point stands for the up-crossing of the level $a$ by the stochastic process
\begin{equation}
\label{eq_eta}
\etaN(t) = \eta(x_0,y_0,t) \, ,
\end{equation}
whose time derivative is given by
\begin{equation}
\label{eq_eta}
\dEtaN(t) = \de(x_0,y_0,t) \, .
\end{equation}
The kinematic variables introduced in \S\ref{subsec_kin_vars} 
-- measured at the location of the material point \rev{at a given time} -- may be gathered  in a random vector:\rev{\footnote{
\rev{
In the present paper, when considering a time stochastic process, 
$\mathcal{P}(t)$ and $\mathcal{P}$ are respectively used 
to denote the stochastic process itself and its value at a given time (which is a random variable).
Similarly, in the case  a random field $\mathcal{F}(x,y,t)$, $\mathcal{F}$ denotes its value at a given location and at a given time.
}
} 
}
\begin{equation}
\label{eq_Ztot}
Z_A = \left[ 
\begin{array}{l}
\eta = \etaN \\ 
\vx  \\
\vy  \\ 
 \az \\ 
  \vz  = \dEtaN  \\
 \sx \\
 \sy \\  
 \ax \\
 \ay \\ 
\end{array} 
\right] \, .
\end{equation}
As the different kinematic variables are obtained from linear transformations of $\eta$,
the random vector, $Z_A$, is Gaussian. 
The mean vector of $Z_A$ is zero.
The coefficients of its covariance matrix, $\Sigma_{Z_A}$, 
may be computed as 
\begin{equation}
\label{eq_covmat_expr}
\displaystyle
\left[ \Sigma_{Z_A} \right]_{k,l} 
= \int_{-\pi}^{\pi} {\rm d} \theta 
\int_{0}^{+\infty}
{\rm d} \omega \ 
{\rm Re}\left\{ \mathcal{H}_k(\omega, \theta) \bar{\mathcal{H}}_l(\omega, \theta)  \right\} G(\omega,\theta) \, ,
\end{equation}
where $\mathcal{H}_k$ (resp. $\mathcal{H}_l$) is the complex transfer function whose input and output are respectively 
the sea surface elevation $\eta$ and the $k$-th (resp. $l$-th) variable of the random vector $Z_A$; 
$\bar{\mathcal{H}}_l$ denotes the complex conjugate of $\mathcal{H}_l$.
The linear wave theory yields the following transfer functions 
(following the complex notation adopted in \eqp{eq_airy_cpx}):
\begin{equation}
\label{eq_transFunc_0}
\begin{array}{l}
\mathcal{H}_{\eta} (\omega, \theta) = 1\\
\mathcal{H}_{\vx} (\omega, \theta) = \cos \theta \cdot g k (\omega) / \omega \\
\mathcal{H}_{\vy} (\omega, \theta) = \sin \theta \cdot g k (\omega) / \omega \\
\mathcal{H}_{\az} (\omega, \theta) = - \omega^2 \\
\mathcal{H}_{\vz} (\omega, \theta) = i \omega \\
\mathcal{H}_{\sx} (\omega, \theta) = - i \cos \theta \cdot k(\omega) \\
\mathcal{H}_{\sy} (\omega, \theta) = - i \sin \theta \cdot k(\omega) \\
\mathcal{H}_{\ax} (\omega, \theta) = i \cos \theta \cdot g k (\omega) \\
\mathcal{H}_{\ay} (\omega, \theta) = i \sin \theta \cdot g k (\omega) \, .
\end{array}
\end{equation}
In the linear wave model, the horizontal acceleration components
and slope components are linearly related through
\begin{equation}
\label{eq_linrel_sa}
\left[
\begin{array}{l}
\ax \\ 
\ay
\end{array} 
\right]
= -g \left[
\begin{array}{l}
\sx \\ 
\sy
\end{array} 
\right] \, .
\end{equation}
Hence, two of these four variables should be discarded when considering 
the joint normal distribution of 
kinematic variables (otherwise the covariance matrix would be singular).
In the subsequent development,  $\ax$ and $\ay$ are discarded,
and the remaining variables are collected in a reduced vector
\begin{equation}
\label{eq_Zred}
Z = \left[ 
\begin{array}{l}
\eta = \etaN \\ 
\vx  \\
\vy  \\ 
 \az  \\ 
  \vz  = \dEtaN  \\
 \sx \\
 \sy  
\end{array} 
\right] \, .
\end{equation}
Besides, the transfer functions of \eq{eq_transFunc_0} are either real or imaginary, 
which implies that $Z$ can be split in two independent Gaussian vectors, 
\begin{equation}
\label{eq_vectX}
X = \left[ 
\begin{array}{c}
\eta = \etaN \\
\vx \\
\vy \\
\az 
\end{array} 
\right] \, ,
\end{equation}
and 
\begin{equation}
\label{eq_vectY_0}
Y = \left[ 
\begin{array}{c}
 \vz = \dEtaN \\
  \sx \\
  \sy 
\end{array} 
\right] \, ,
\end{equation}
whose covariance matrices are given by expressions of the form of \eq{eq_covmat_expr}.
In terms of probability density function, the above considerations translate into
\begin{equation}
f_Z(\eta, \vx, \vy, \az, \vz, \sx, \sy) = f_X(\eta, \vx,\vy, \az) \times f_{Y}( \vz, \sx, \sy) \, ,
\end{equation}
where $f_Z$, $f_X$, $f_{Y}$, 
are the respective multivariate normal density functions of the Gaussian vectors $Z$, $X$, and $Y$.

\mypar{Case of infinite water depth with frequency-direction separation.}
\noindent 
In the case of infinite water depth, the dispersion relation simplifies into 
\begin{equation}
k = \frac{\omega^2}{g} \, .
\end{equation}
Then, by further assuming the independence of wave direction and frequency distributions (\eqp{eq_G_dir_spectrum}), 
the covariance matrix of the vectors $X$ and $Y$ 
can be expressed in terms of wave spectrum moments:
\begin{equation}
\label{eq_sigmaX}
\Sigma_X = 
\left[
\begin{array}{cccc}
m_0 & \alpha_{10} m_1 & \alpha_{01} m_1  & -m_2 \\
 & \alpha_{20} m_2 & \alpha_{11} m_2 & - \alpha_{10} m_3  \\
 &  & \alpha_{02} m_2 & - \alpha_{01} m_3 \\
\multicolumn{2}{c}{\text{\smash{\raisebox{1.5ex}{(Sym.)}}}} &  & m_4 \\
\end{array}
\right]
\end{equation}
\begin{equation}
\label{eq_sigmaY}
\Sigma_{Y} = 
\left[
\begin{array}{ccc}
m_2  & - \alpha_{10}m_3/g & - \alpha_{01}m_3/g  \\
 &  \alpha_{20}m_4 / g^2 & \alpha_{11} m_4 / g^2 \\
 \multicolumn{1}{c}{\text{\smash{\raisebox{1.5ex}{(Sym.)}}}}  & & \alpha_{02}m_4 / g^2 \\
\end{array}
\right] \, .
\end{equation}
In Eqs.~(\ref{eq_sigmaX}-\ref{eq_sigmaY}), $m_p$ denotes the $p$-th moment 
of the wave frequency spectrum (see \eqp{eq_spectrum_moment}),
and $\alpha_{pq}$ are numerical factors accounting for the directional spreading of waves:
\begin{equation}
\label{eq_alpha_pq}
\alpha_{pq} = \int_{-\pi}^{\pi} {\rm d} \theta \ D(\theta) \cos^p\theta \sin^q\theta \,.
\end{equation}

\subsection{Conditional distribution given up-crossing}
\label{subsec_cond_dist}

Let $\cZ$ and  $\cX$ denote respectively the random vectors containing the variables of $Z$ and $X$, 
except for $\eta = \etaN$.
The conditional density function of $\cZ$,  given that 
$\etaN\rev{(t)}$ is up-crossing the level $a$, can be written as
(see for example \cite{lindgren_2012}):
\begin{equation} 
\label{eq_fZ_rest}
f_{\cZ|\etaN\rev{(t)} \uparrow a} = \frac{\vz f_{\cZ|\eta= a}}
{\displaystyle \int_0^{+\infty} \! \! \! \!  \ \xi  f_{\vz|\eta= a}(\xi) \ {\rm d} \xi } , \ \vz > 0 \, ,
\end{equation} 
where $f_{\cZ|\eta= a}$ and $f_{\vz|\eta= a}$ are the conditional density functions of $\cZ$ and $\vz = \dEtaN$, given $\eta=a$.
Taking advantage of the independence of the vectors $\cX$ and $Y$, \eq{eq_fZ_rest} may also be written :
\begin{equation} 
\label{eq_fXY_rest}
f_{\cZ|\etaN\rev{(t)} \uparrow a}  = 
f_{\cX|\eta=a} \times
\frac{\displaystyle \vz  f_{Y} }
{\displaystyle 
\int_0^{+\infty} \! \! \! \!  \ \xi  f_{\vz}(\xi) \ {\rm d} \xi  } 
\, , \ \vz > 0 \, ,
\end{equation} 
where $f_{\cX|\eta=a}$ is the conditional density function of $\cX$, given $\eta=a$, (which is Gaussian) 
and $f_w$ is the non-conditional density function of $w$. 
The normalisation factor appearing in \eq{eq_fXY_rest} can be readily calculated, yielding:
\begin{equation} 
\label{eq_fZ_2}
f_{\cZ|\etaN\rev{(t)} \uparrow a} = \sqrt{\frac{2\pi}{m_2}} f_{\check{X}|\eta=a}  \times
  \vz  f_{Y}  
\, , \ \vz > 0 \, .
\end{equation} 

\subsection{Up-crossing frequency}

The average up-crossing frequency of the level $z=a$, 
by the sea surface elevation, 
is given by Rice's formula \cite{rice_1945}:
\begin{equation}
\nuz(a) 
= \int_0^{+\infty}  \xi f_{\eta, \vz} (a, \xi) \ {\rm d} \xi \, ,
\end{equation}
where $f_{\eta, \vz}$ is the non-conditional bivariate density function of $\eta$ and $\vz$.
In the present case, where the considered stochastic process $\etaN\rev{(t)}$ is Gaussian,
the up-crossing frequency can be further expressed as:
\begin{equation}
\label{eq_fip_rest}
\nuz(a) 
= \frac{1}{2\pi} \sqrt{\frac{m_2}{m_0}} \exp \left( - \frac{a^2}{2 m_0} \right) \, .
\end{equation}

\subsection{Illustrative examples}
\label{subsec_illust_rest}

This subsection illustrates the effect of the up-crossing conditioning 
on the distribution of the wave kinematic variables.
The water depth is assumed to be infinite.
Three different sea states are considered: 
these sea states have the same wave frequency distribution (see \S\ref{subsec_specShape}) 
but have a different wave direction distribution, following  Eqs.~(\ref{eq_D1}-\ref{eq_D2}-\ref{eq_D3}).
As a result, from one case to another, the non-conditional covariance matrices of the kinematic variables, 
$\Sigma_X$ and $\Sigma_Y$, 
differ only through the 
coefficients $\alpha_{pq}$ (see Eqs.~\ref{eq_sigmaX}-\ref{eq_sigmaY}).
The numerical values of these 
coefficients, obtained for each considered sea state, 
are reported in \tab{table_alpha_value}.
The numerical values of the first five wave spectrum moments, 
also necessary to compute $\Sigma_X$ and $\Sigma_Y$,
have been reported in \eq{eq_num_moments}.

% %%%%%%%%%%%%%%%%%%%%%%%%
% %%%%%        TABLE    BEGIN         %%%%%
% %%%%%%%%%%%%%%%%%%%%%%%%

\begin{table}[h]
\centering
\begin{tabular}{|c|c|c|c|c|c|} 
 \cline{2-6}
\multicolumn{1}{c|}{}   & $\alpha_{10}$ & $\alpha_{01}$ & $\alpha_{11}$ & $\alpha_{20}$ & $\alpha_{02}$ \\ 
   \hline
$D1$ & $1$ & $0$ & $0$ & $1$ & $0$    \\ 
 \hline
$D2$ & $8/3\pi$ & $0$ & $0$ & $3/4$ & $1/4$     \\
 \hline
$D3$ & $0$ & $0$ & $0$ & $1/2$ & $1/2$ \\
\hline
\end{tabular}
\caption{
Values of the coefficients $\alpha_{pq}$ (\eqp{eq_alpha_pq}), 
for the three wave direction distributions which are considered as illustrative examples 
in \S\ref{subsec_illust_rest}. 
These coefficients account for the effect of the wave direction distribution 
on the covariance matrices of the kinematic variables (see Eqs.~\ref{eq_sigmaX}-\ref{eq_sigmaY}). 
Each column corresponds to a coefficient, and each line to a wave direction distribution 
(see Eqs.~\ref{eq_D1}-\ref{eq_D2}-\ref{eq_D3}). 
}
\label{table_alpha_value}
\end{table}

% %%%%%%%%%%%%%%%%%%%%%%%%
% %%%%%        TABLE    END             %%%%%
% %%%%%%%%%%%%%%%%%%%%%%%%

% %%%%%%%%%%%%%%%%%%%%%%%%
% %%%%%%%%%%%%%%%%%%%%%%%%
% %%%%%        FIGURE    BEGIN         %%%%%
% %%%%%%%%%%%%%%%%%%%%%%%%
% %%%%%%%%%%%%%%%%%%%%%%%%

% scaling of the figure
\def\scaleF{0.32}

\begin{figure}[h!]
\begin{center}
\begin{tabular}{ccc}
        \includegraphics[width=\scaleF\textwidth]{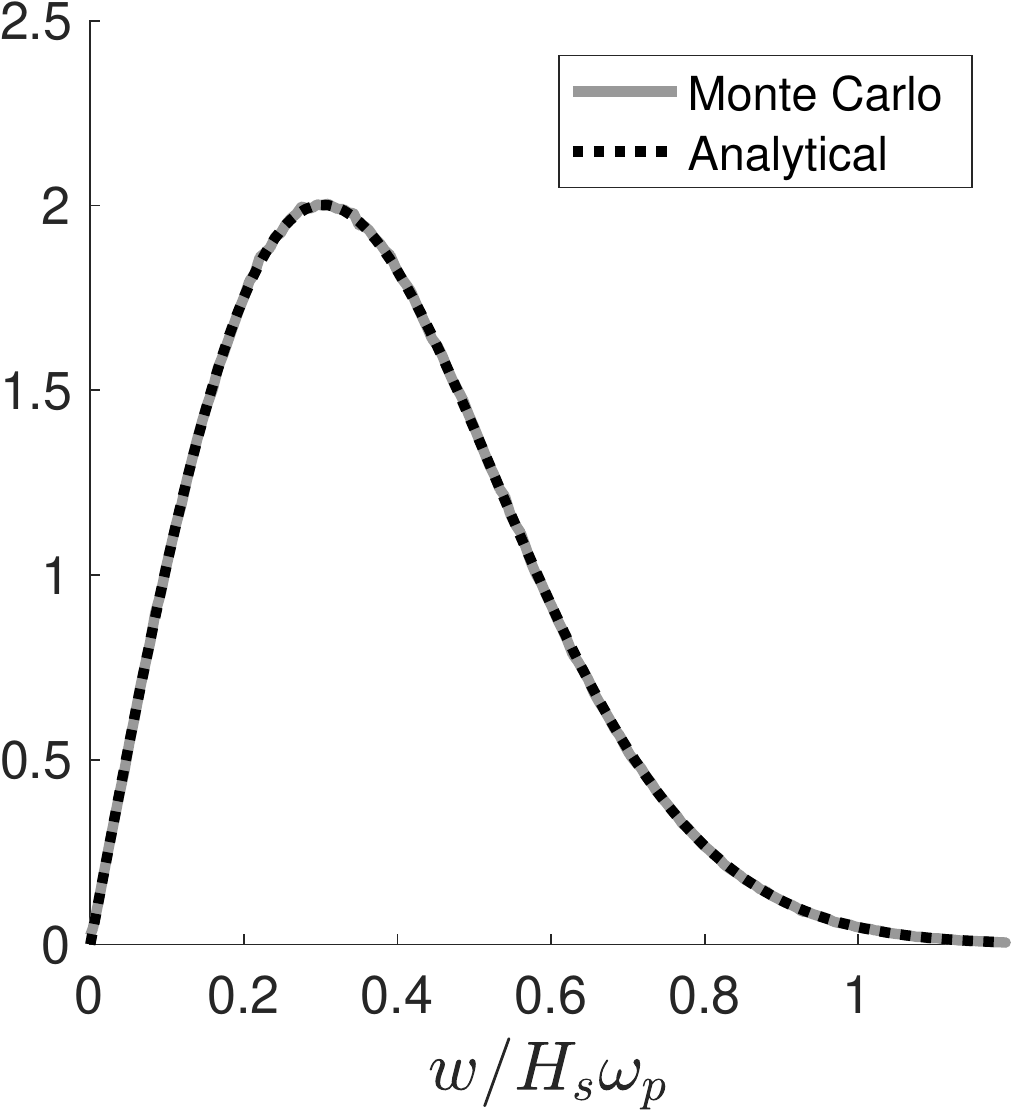} & 
        \includegraphics[width=\scaleF\textwidth]{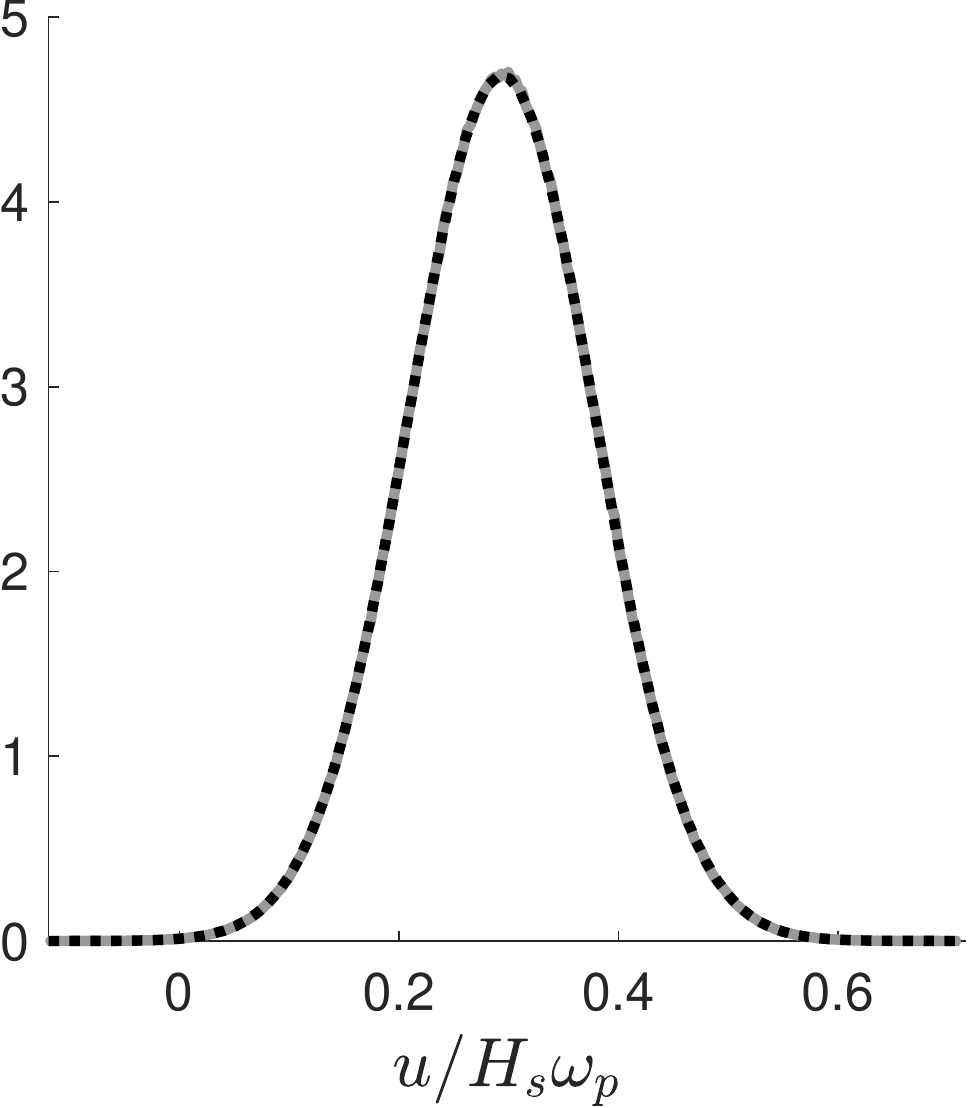} &
        \includegraphics[width=\scaleF\textwidth]{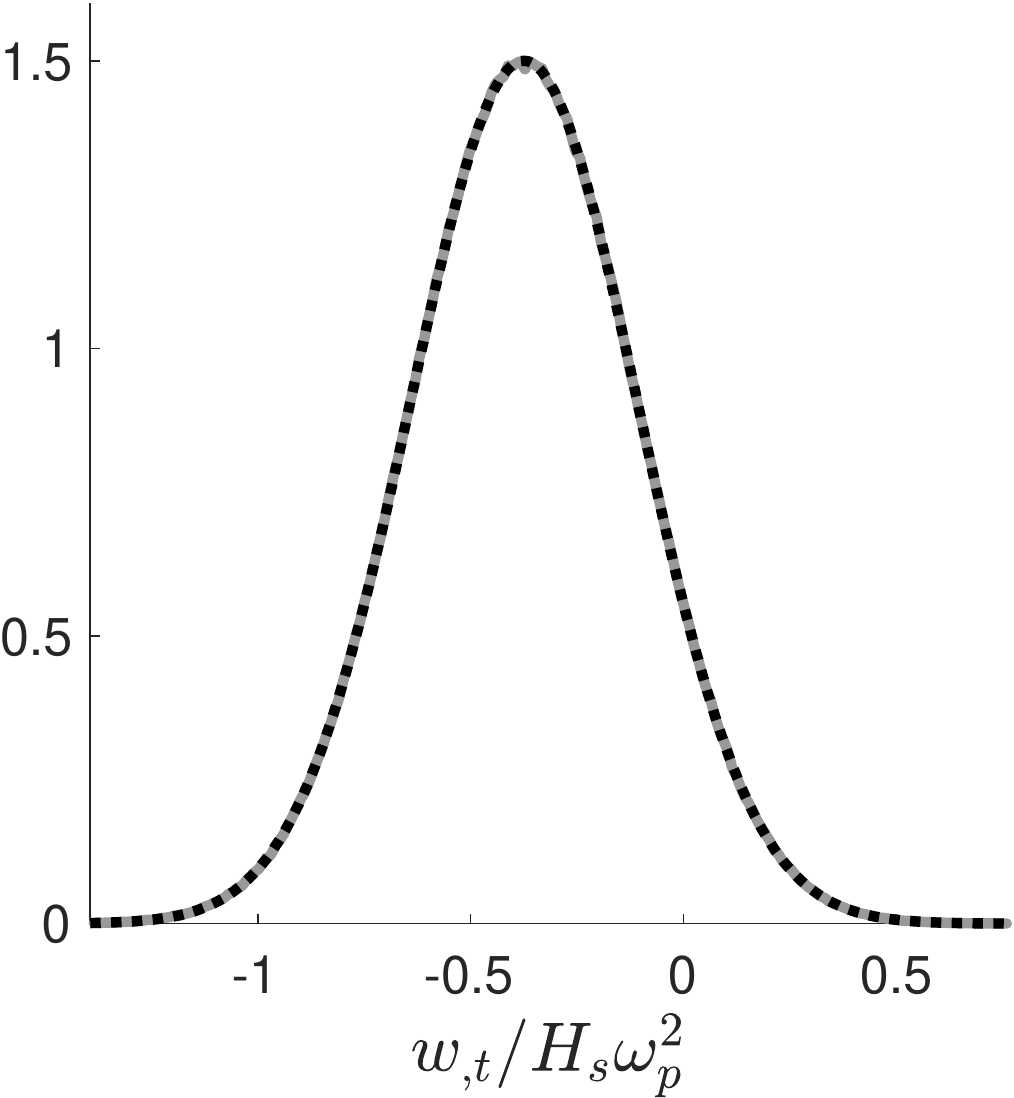} \\
        \includegraphics[width=\scaleF\textwidth]{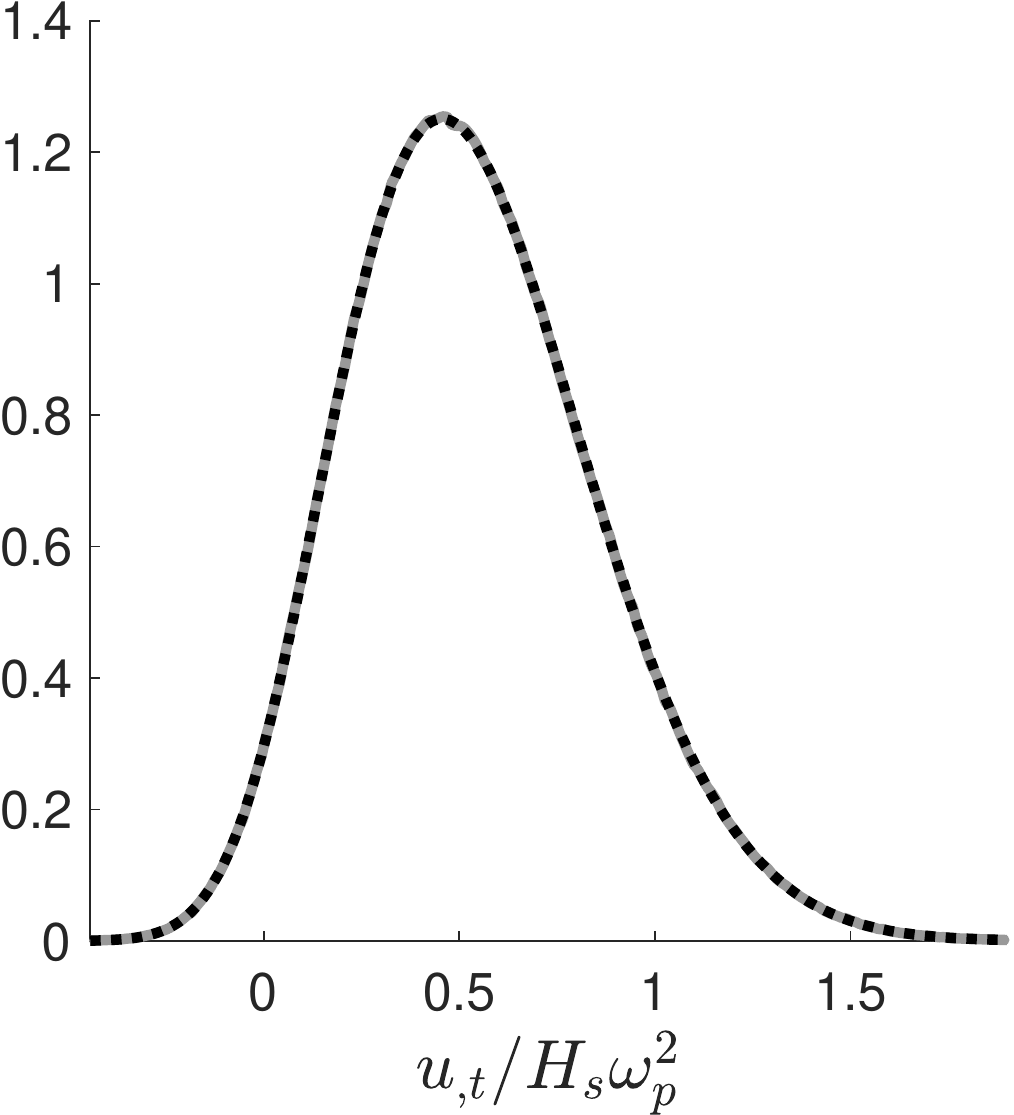} &        
        \includegraphics[width=\scaleF\textwidth]{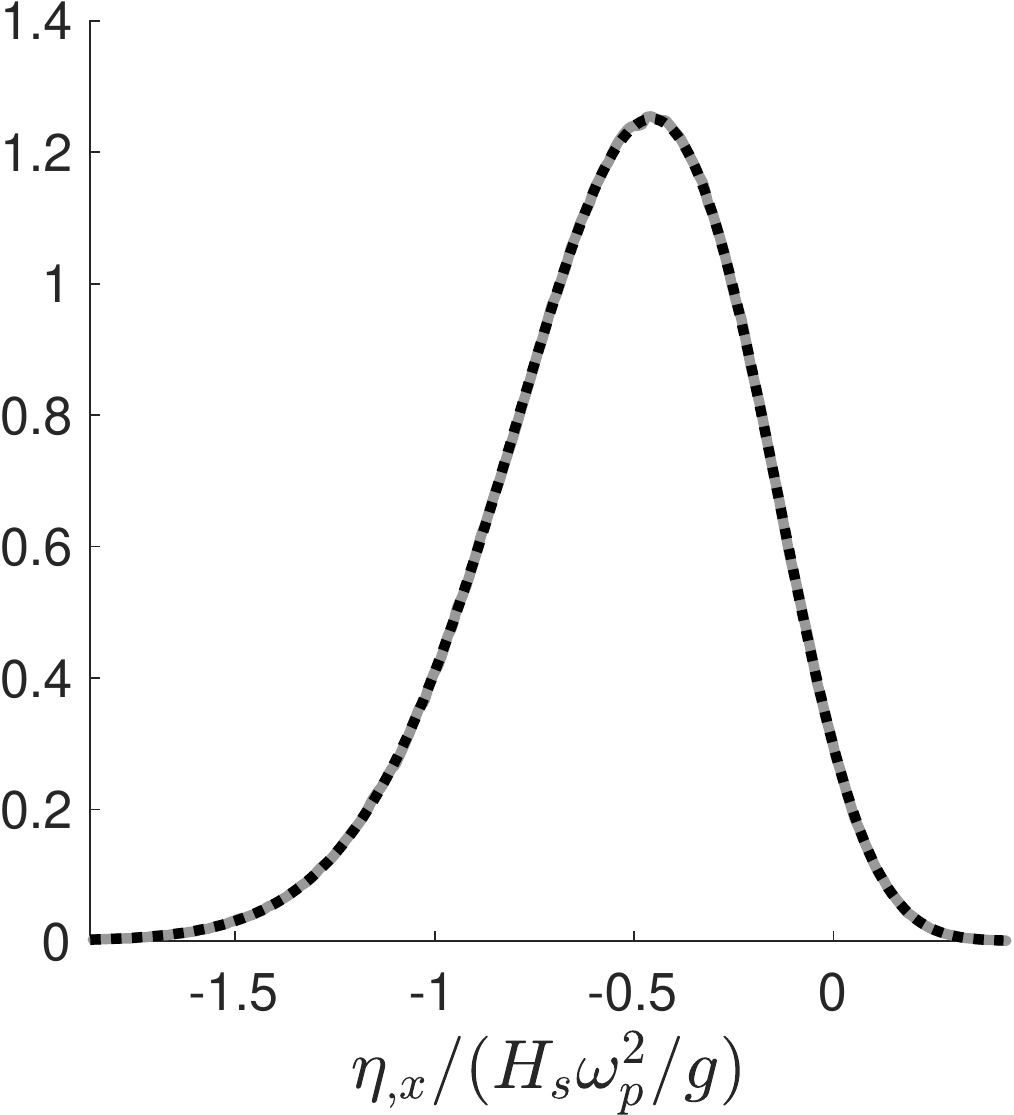} &                 
\end{tabular}
\end{center}
\caption{
Univariate density functions of random kinematic variables, given up-crossing -- body at rest 
in a unidirectional sea state. 
The material point is fixed at an altitude $a=H_s/4$.
The dashed black lines show the analytical density functions 
obtained from successive integrations of \eq{eq_fZ_2}.
The solid grey lines show the results obtained from the Monte Carlo simulation of 
sea state realisations,
based on the random phase/amplitude model (\S\ref{subsec_seastate_gauss_process}).
As the sea is assumed to be unidirectional along the $x$-axis, 
the variables $\ay$ and $\sy$ are identically zero.
}
\label{fig_example_rest_unidir}
\end{figure}

% %%%%%%%%%%%%%%%%%%%%%%%%
% %%%%%%%%%%%%%%%%%%%%%%%%
% %%%%%        FIGURE    END         %%%%%
% %%%%%%%%%%%%%%%%%%%%%%%%
% %%%%%%%%%%%%%%%%%%%%%%%%

\subsubsection{Unidirectional sea state}
\label{subsubsec_unidir}

As a first illustrative example, \fig{fig_example_rest_unidir} 
shows the univariate density functions of the different variables, given up-crossing,
for a material point standing at an altitude $a=H_s/4$, in a unidirectional sea state
(see \eqp{eq_D1}).
The kinematic variables have been nondimensionalised, 
so that the density functions reported in \fig{fig_example_rest_unidir} 
do not depend on the actual values of $H_s$ and $T_p$.
For testing purpose, these conditional univariate distributions have been computed by using two different methods:
\begin{enumerate} 
\item Analytical calculation by successive integration of the multivariate density function given in \eq{eq_fZ_2}.
The detailed expressions of the the resulting univariate density functions are reported in appendix \ref{app_marg_dist_rest}.
\item Level-crossing detection in Monte-Carlo realisations of the sea state,
following the random phase/amplitude model briefly described in \S\ref{subsec_seastate_gauss_process}.
\rev{
For each realisation, complex wave amplitudes (see Eq. \ref{eq_cpx_amp}) are drawn randomly 
and the corresponding free surface elevation is obtained from \eq{eq_airy_cpx}.
The other kinematic variables ($\sx$, $\vz$, $\az$, $\vx$, $\ax$) are computed
from expressions similar to \eq{eq_airy_cpx}, 
where the appropriate complex amplitudes are obtained 
by using the transfer functions given in \eq{eq_transFunc_0}.
To compute the time sequence of the different variables, 
the fast Fourier transform has been advantageously used 
(see for example \S5.6 in \cite{lindgren_2013}).
}
For the results reported in \fig{fig_example_rest_unidir},
$1000$ sea state realisations were simulated, each of them over a physical duration of $10^4 T_p$,
leading to the detection of $7348281$ up-crossings in total.
\end{enumerate}
Both methods show an excellent agreement on the distributions.
Besides, the Monte Carlo ``empirical'' up-crossing frequency is also in good agreement 
with the analytical expression (\eqp{eq_fip_rest}), 
which yields, using the frequency moments given in \eq{eq_num_moments}, 
the numerical value $\nuz(a=H_s /4) \simeq 0.735 / Tp$.
Note that the Monte Carlo approach is numerically demanding,
since the total number of detected up-crossings should be quite large to 
reach a reasonable statistical precision (especially in the tails of the distributions).
The present calculation required a total CPU time of about one hour (on a personal computer).

In order to check the analytical developments provided throughout the present paper, 
Monte-Carlo simulations were also used to estimate the results reported in 
\fig{fig_example_rest_multidir} 
(conditional distributions, given up-crossing, for a material point at rest in a multidirectional sea state),
\fig{fig_example_fImp} (up-crossing frequency as a function of forward speed), 
and \fig{fig_example_vs_dists} 
(conditional distributions, given up-crossing, for a material point with forward motion).
The agreement between analytical and numerical results was good; 
however, for reasons of clarity, Monte Carlo results are not shown in 
Figs.~\ref{fig_example_rest_multidir}-\ref{fig_example_fImp}-\ref{fig_example_vs_dists}

The nature of the different conditional univariate distributions, shown in \fig{fig_example_rest_unidir}, is as follows:
\begin{itemize}
\item The vertical component of the fluid velocity, $\vz$, given up-crossing, follows a Rayleigh distribution of mode $\sqrt{m_2}$.
\item The univariate distributions of $\vx$ and $\az$, given up-crossing, are normal 
\rev{(see appendix \ref{subsubsec_doeta})}.
\item The univariate distributions of $\ax$ and $\sx$ 
\rev{(which are linearly related through Eq. \ref{eq_linrel_sa})}, 
given up-crossing, 
result from the \rev{convolution} of a normal distribution with a Rayleigh distribution 
\rev{(see appendix \ref{subsubsec_xi0})}.
\end{itemize}
The analytical expressions of these distributions 
(see appendix \ref{app_marg_dist_rest})
show that a change in the assumed crossing level (here $a=H_s/4$ has been assumed),
would affect the conditional distributions of $\vx$ and $\az$,
only through their mean value (see \S\ref{subsubsec_doeta}).
The conditional distributions of $\vz$, $\ax$ and $\sx$, do not depend on the value of the crossing level $a$.

% %%%%%%%%%%%%%%%%%%%%%%%%
% %%%%%%%%%%%%%%%%%%%%%%%%
% %%%%%        FIGURE    BEGIN         %%%%%
% %%%%%%%%%%%%%%%%%%%%%%%%
% %%%%%%%%%%%%%%%%%%%%%%%%

% scaling of the figure
\def\scaleF{0.32}

\begin{figure}[t]
\begin{center}
\begin{tabular}{ccc}
        \includegraphics[width=\scaleF\textwidth]{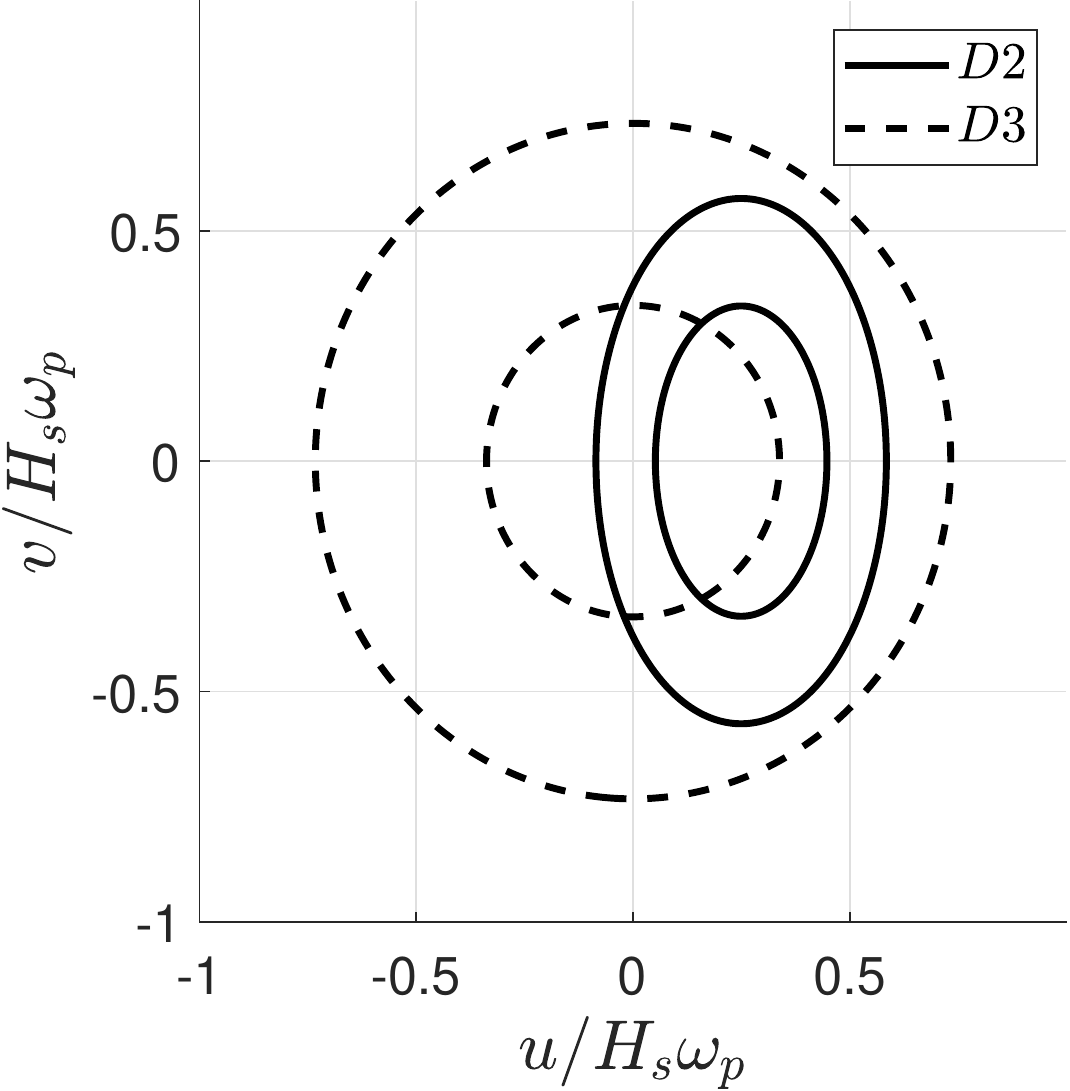} & 
        \includegraphics[width=\scaleF\textwidth]{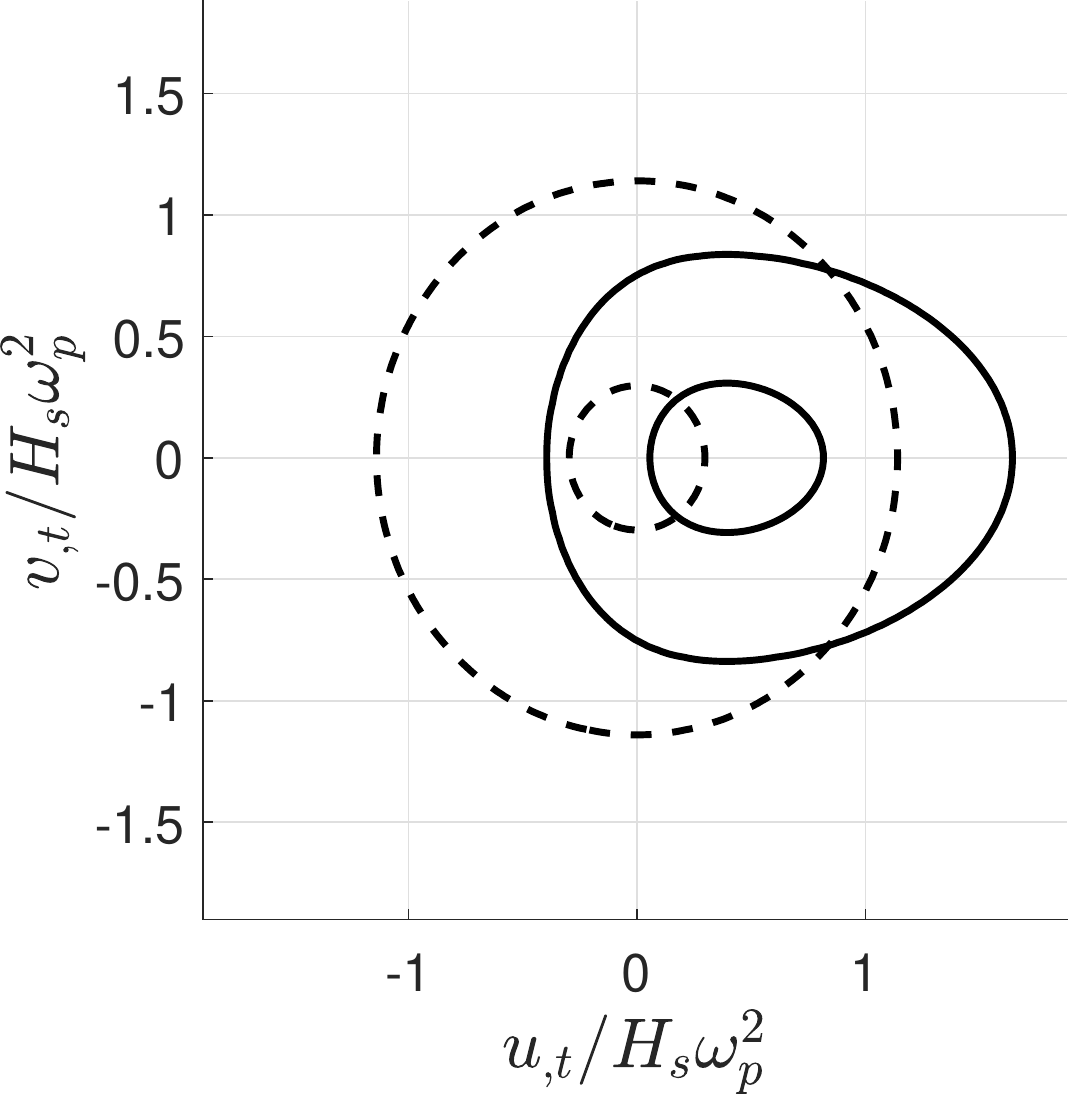} &
        \includegraphics[width=\scaleF\textwidth]{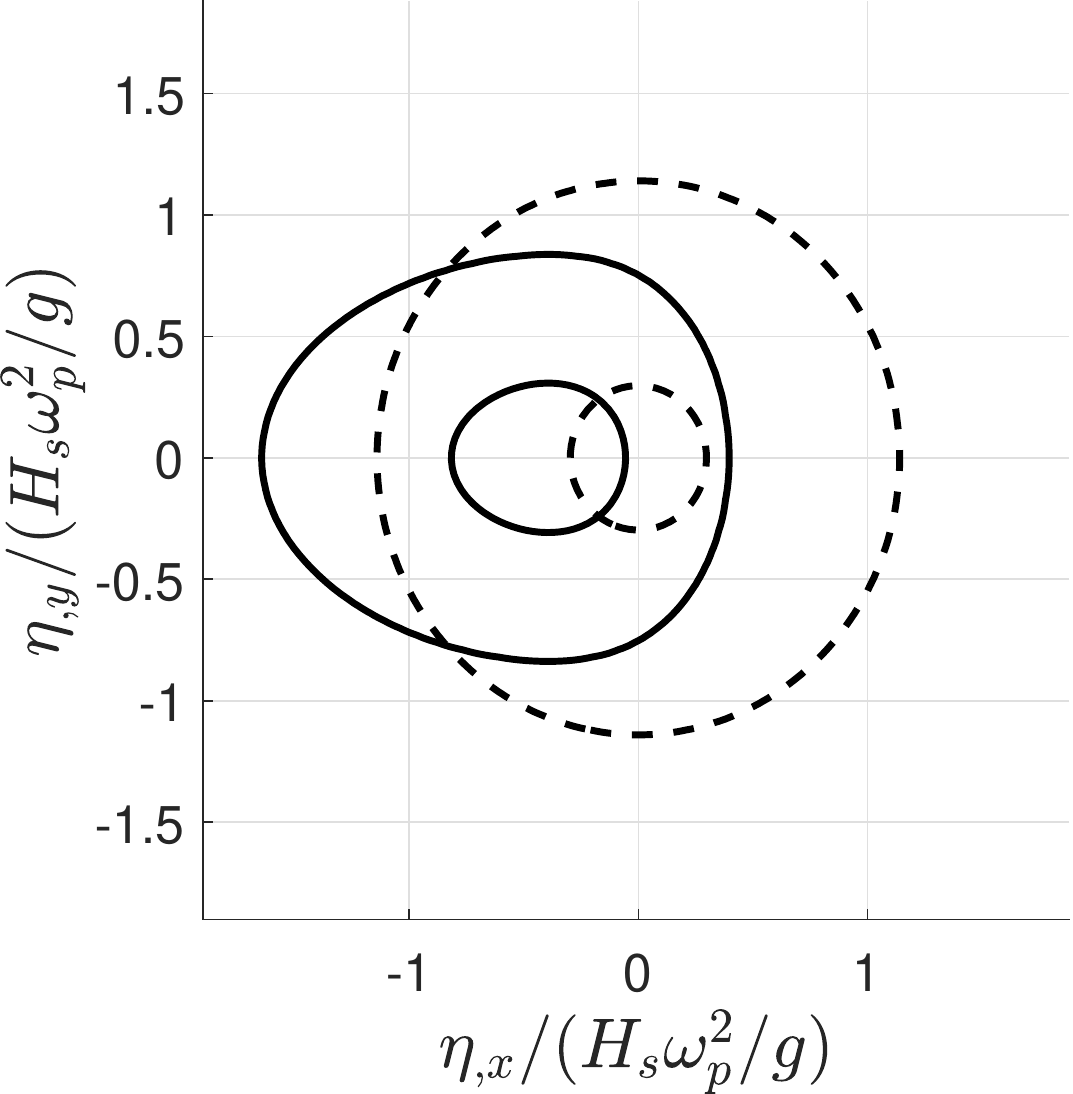}               
\end{tabular}
\end{center}
\caption{
Density functions of random kinematic variables, given up-crossing --
multidirectional sea states. 
The material point is at rest, at an altitude $a=H_s/4$.
The three figures show isodensity lines of the bivariate distributions 
of the pairs $(\vx,\vy)$,  $(\ax,\ay)$,  $(\sx,\sy)$, given up-crossing, in nondimensional form.
Results are shown for two different wave direction distributions:
anisotropic (\eqp{eq_D2}), shown as solid lines, 
and isotropic (\eqp{eq_D3}), shown as dashed lines.
The levels of the isodensity lines are (0.01;1) for the velocity distribution, and (0.003;1) 
for the acceleration and slope distributions.
}
\label{fig_example_rest_multidir}
\end{figure}

% %%%%%%%%%%%%%%%%%%%%%%%%
% %%%%%%%%%%%%%%%%%%%%%%%%
% %%%%%        FIGURE    END         %%%%%
% %%%%%%%%%%%%%%%%%%%%%%%%
% %%%%%%%%%%%%%%%%%%%%%%%%

\subsubsection{Multidirectional sea states}
\label{subsubsec_multidir}

To illustrate the effect of wave directional spreading, 
the bivariate density functions of 
the horizontal velocity components $(\vx,\vy)$, 
the horizontal acceleration components $(\ax,\ay)$, 
and the slope components $(\sx,\sy)$, 
given up-crossing, are shown in \fig{fig_example_rest_multidir}. 
Two different direction distributions are considered: 
anisotropic and isotropic ($D2$ and $D3$ defined in Eqs.~\ref{eq_D2}-\ref{eq_D3}).
The univariate distributions of the vertical components of the fluid velocity and acceleration 
($\vz$ and  $\az$), given up-crossing, 
are not sensitive to the direction distribution of random waves 
(see appendix \ref{app_marg_dist_rest} along with \eqp{eq_sigmaX}): 
therefore, they are identical to the ones shown in \fig{fig_example_rest_unidir} 
and are not reproduced in \fig{fig_example_rest_multidir}.

For both considered sea states,
the symmetry of the direction distribution about the $y=0$ plane 
translates into $\alpha_{11} = \alpha_{01} = 0$ (see \tab{table_alpha_value}). 
The nullity of $\alpha_{11}$ implies that the variables $\vx,\ax, \sx$ 
respectively do not depend on $\vy, \ay, \sy$, when non-conditioned
(see Eqs.~\ref{eq_sigmaX}-\ref{eq_sigmaY}).
Besides, the nullity of $\alpha_{01}$ implies that $\vy, \ay, \sy$ do not depend neither on $\eta$, nor on $\vz$. 
Therefore, their univariate distributions are not affected by the level-crossing conditioning.
It also implies that the level-crossing conditioning does not alter the independence of the pairs 
$(\vx,\vy)$, $(\ax, \ay)$, $(\sx,\sy)$: 
then, their conditional bivariate density functions are equal 
to the product of the respective conditional univariate density functions.

In the case of the isotropic sea state ($D_3$), 
two additional remarkable properties can be noted: 
$\alpha_{10} = 0$ and $\alpha_{20} = \alpha_{02}$.
Then, the random variables $\vx, \ax, \sx, \vy, \ay, \sy$ are not affected by the level-crossing conditioning 
(due to $\alpha_{10} = \alpha_{01} = 0$). 
Moreover, $\vx, \ax, \sx$ have the same variance as $\vy, \ay, \sy$ (due to $\alpha_{20} = \alpha_{02}$).
This explains why the corresponding isodensity contours, 
shown as dashed lines in \fig{fig_example_rest_multidir}, are centred circles.

In the case of the anisotropic sea state, $\vx$ non-conditionally depends on $\eta$ (correlation coefficient $\simeq 0.94$), 
and $\ax, \sx$ non-conditionally depend on $\vz$ (correlation coefficients $\simeq +0.92$ and $\simeq -0.92$, respectively).
The random variable $\vx$, given up-crossing, follows a normal distribution with a non-zero mean and a reduced variance 
(see Eqs. \ref{eq_mean_cn}-\ref{eq_sigma_cn} in appendix);
the conditional variance of $\vx$ is smaller than the one of $\vy$, although it is the opposite for non-conditional variances.
The ``egglike'' shape of the isodensity contours 
of the conditional bivariate distributions of $(\ax,\ay)$ and $(\sx,\sy)$,
is due to the conditional univariate distributions of $\ax$ and $\sx$ 
being the \rev{convolution} of a normal distribution with a Rayleigh distribution
(see \S\ref{subsubsec_xi0} in appendix).

\section{Body with horizontal motion}
\label{sec_forward}

The present section analytically investigates the effect of Doppler shift on stochastic up-crossing,
when the considered material point has a forward speed in the reference frame of the mean flow.
The paragraph \S\ref{subsec_encounter} is a preamble, where the concepts of encounter wave frequency 
and encounter wave spectrum are briefly reminded.
The new up-crossing condition
is defined in \S\ref{subsec_upc_cond}. 
The following paragraph (\S\ref{subsec_non_cond_fs}) 
introduces an extra kinematic variable that need to be considered in the problem,
along with the related non-conditional distribution.
Then, the general expression of the conditional distribution of kinematic variables, given up-crossing, 
is specified in \S\ref{subsec_cond_dist_fs};
the related up-crossing frequency is considered in \S\ref{subsec_upfreq_fs}.
The section ends with illustrative examples (\S\ref{subsec_illust_fs}).

\subsection{Encounter wave frequency and encounter wave spectrum}
\label{subsec_encounter}

\mypar{Encounter wave frequency.} 
The material point is now assumed to move at a constant speed, $V_s$ (which may be positive or negative),  
in a plane parallel to the mean free surface,
at a given altitude $z=a$.
The heading of the material point, $\psi$, 
is measured relative to the $x$-axis of the frame in which the problem is formulated (see \fig{fig_frame}).
The encounter frequency between the moving body and a water wave of angular frequency $\omega$, 
propagating along a direction $\theta$, 
can be expressed as
\begin{equation}
\label{eq_omt_fct_theta_om}
\tilde{\omega}(\omega,\theta) = \omega - V_s \cos(\psi-\theta) k(\omega) \, ,
\end{equation}
where $k(\omega)$ is the dispersion relation.
For $V_s \cos(\psi-\theta)~>~0$, waves approach from behind, 
and two different intrinsic wave frequencies may give the same encounter frequency.
When $V_s \cos(\psi-\theta) > \omega / k$, 
the material point catches up with the water waves and the encounter frequency becomes negative.

\mypar{Encounter wave spectrum.} The two-dimensional encounter wave spectrum 
is then given by
\begin{equation}
\label{eq_spec_2d_ap}
\tilde{G}(\tilde{\omega},\theta) = 
\abs{\frac{\partial \omega}{\partial \tilde{\omega}}}(\tilde{\omega},\theta) \times 
\sum\limits_{r=1}^{R(\tilde{\omega},\theta)} 
G(\omega_r(\tilde{\omega},\theta),\theta)  \, ,
\end{equation}
where $\abs{\partial \omega/\partial \tilde{\omega}}$ 
is the Jacobian related to the variable substitution.
$R(\tilde{\omega},\theta)$ is the number of distinct intrinsic frequencies 
corresponding to the encounter frequency, $\tilde{\omega}$,
for a given direction, $\theta$.
Or, stated differently, $R$ is the number of real solutions of \eq{eq_omt_fct_theta_om}, 
when $\omega$ is the unknown.
For a given water depth, $h$, a given heading, $\psi$, and a given forward speed, $V_s$, 
the number of solutions may be $0$, $1$ or $2$, 
depending on the values of $\tilde{\omega}$ and $\theta$. 
When two solutions exist, 
the two corresponding contributions of the intrinsic wave spectrum have to be summed,
which explains the summation operator in \eq{eq_spec_2d_ap}.

If the water depth is finite, the Jacobian, $\abs{\partial \omega/\partial \tilde{\omega}}$, 
has no simple closed-form expression; 
when infinite water depth can be assumed, the Jacobian reads
\begin{equation}
\abs{\frac{\partial \omega}{\partial \tilde{\omega}}}(\tilde{\omega}, \theta) = 
\left[ 1 - 4 \cos (\psi - \theta) \tilde{\omega} / \omega_s \right]^{-1/2} \, ,
\end{equation}
with
\begin{equation}
\label{eq_oms}
\omega_s = \frac{g}{V_s} \, .
\end{equation}
\eq{eq_spec_2d_ap} shows that the motion-induced Doppler shift has a non-trivial effect 
on the encounter wave spectrum.
However, as it is developed below, 
the computation of the conditional distribution of kinematic variables, given up-crossing,
does not require the explicit use of the encounter spectrum $\tilde{G}$.
Hence, the properties of the encounter spectrum are not further investigated here;
more details on this subject can be found in Lindgren et al. (1999) \cite{lindgren_1999}.

\subsection{Up-crossing condition}
\label{subsec_upc_cond}

When the material point moves at a constant velocity, in a horizontal plane, 
a water entry event stands for the up-crossing of the level $a$
by the stochastic process
\begin{equation}
\label{eq_etas}
\etaS(t) = \eta\left(x_0+ [V_s \cos \psi] t, \, y_0+ [V_s \sin \psi] t, \, t\right) \, ,
\end{equation}
where $(x_0,y_0)$ is the position of the body at $t=0$.

\subsection{Non-conditional distribution of kinematic variables}
\label{subsec_non_cond_fs}

As $\eta(x,y,t)$ is a Gaussian field, stationary in time and homogeneous in space, 
$\etaS(t)$ is a stationary Gaussian process; its variance density spectrum 
corresponds to the encounter wave frequency spectrum 
\begin{equation}
\label{eq_spec_1d_ap}
\tilde{S}(\tilde{\omega}) = \int_{-\pi}^{\pi} {\rm d} \theta \tilde{G}(\tilde{\omega},\theta) \, .
\end{equation}
The time derivative of this stochastic process, $\dEtaS\rev{(t)}$, 
may be physically interpreted as the velocity of the free surface elevation measured
in the reference frame of the moving material point. 
Besides, as the material point moves in a horizontal plane, 
the vertical velocity (and acceleration) of the fluid particles 
is the same when measured in the reference frame of the mean flow 
and the reference frame of the moving material point.
As a consequence,
for a moving material point, $\dEtaS$ is not equal to the vertical component of the fluid velocity, 
to the leading order (this point is further discussed in \S\ref{subsec_vars_we}).
Then, compared to the case of a material point at rest
(see \S\ref{subsec_body_rest}, \eqp{eq_Zred}), 
it is necessary to introduce $\dEtaS$
as an extra variable in the considered random vector of kinematic variables:
\begin{equation}
\label{eq_Zs}
Z_s = \left[ 
\begin{array}{l}
\eta = \etaS \\ 
\vx  \\
\vy  \\ 
\az \\  
\vz  \\
 \sx \\
 \sy \\
 \dEtaS 
\end{array}
\right] \, .
\end{equation}
The different kinematic variables collected in $Z_s$ 
are measured at the instant location of the moving material point\rev{, at a given time}.
The components of the fluid velocity and acceleration, $\vx$, $\vy$, $\vz$, $\az$, $\ax$, $\ay$ 
(the last two variables are absent from $Z_s$, due to their linear relationship with $\sx$ and $\sy$; see \eqp{eq_linrel_sa})
are still measured in the reference frame of the mean flow.
As the extra component, $\dEtaS$, results from a linear transformation of $\eta$, 
the random vector $Z_s$ is Gaussian. Its mean vector is zero.
The transfer functions of the first 7 components are still given by \eq{eq_transFunc_0}, 
while the transfer function of $\dEtaS$ can be expressed as
\begin{equation}
\mathcal{H}_{\dEtaS} (\omega,\theta) = i \tilde{\omega}(\omega,\theta) \, .
\end{equation}
As $\mathcal{H}_{\dEtaS}$ is imaginary, 
it is again possible to split the random vector $Z_s$ into two independent Gaussian random vectors.
A first vector, $X$, whose components and covariance matrix  
remain the same (see Eqs.~\ref{eq_vectX}-\ref{eq_sigmaX}). 
A second vector, $Y_s$, which includes the additional random variable $\dEtaS$, compared to the vector $Y$ 
defined in \eq{eq_vectY_0},
\begin{equation}
\label{eq_vectY_s}
Y_s = \left[ 
\begin{array}{c}
  \vz \\
  \sx \\
  \sy \\
  \dEtaS
\end{array} 
\right] \, .
\end{equation}
$Y_s$ is a zero-mean Gaussian vector, whose covariance matrix, $\Sigma_{Y_s}$,
has an expression of the form of \eq{eq_covmat_expr}.
The non-conditional probability density function of kinematic variables can be expressed as
\begin{equation}
f_{Z_s}(\eta, \vx, \vy, \az, \vz, \sx, \sy, \dEtaS) = f_X(\eta, \vx,\vy, \az) \times f_{Y_s}( \vz, \sx, \sy, \dEtaS) \, ,
\end{equation}
where $f_{Y_s}$ is the probability density function of $Y_s$.
The covariance matrix of $Y_s$ may be further expressed as 
\begin{equation}
\label{eq_sigmaY_s}
\Sigma_{Y_s} = 
\left[
\begin{array}{cccc}
  \multicolumn{3}{c}{\multirow{3}{*}{\text{\smash{{\huge \ $\left(\Sigma_{Y}\right)$}}}}}  & m_{1\tilde{1}} \\
  & & &  - \tau_{10} / g \\
  & &  & - \tau_{01} / g \\
  m_{1\tilde{1}} & - \tau_{10} / g & - \tau_{01} / g & \mmt 
\end{array}
\right] \, ,
\end{equation}
where $\mmt$ denotes the 2nd order moment of the encounter wave spectrum,
\begin{equation}
\label{eq_m2t}
\begin{split}
\mmt & =  \int_{-\pi}^{\pi} {\rm d} \theta  \int_{-\infty}^{+\infty} {\rm d} \tilde{\omega} \ \tilde{\omega}^2 \tilde{G}(\tilde{\omega}, \theta)  \\
         & =  \int_{-\pi}^{\pi} {\rm d} \theta \int_{0}^{+\infty}  {\rm d} \omega \  \tilde{\omega}(\omega,\theta)^2 G(\omega,\theta)  \, ,
\end{split}
\end{equation}
$m_{1\tilde{1}}$ is defined as 
\begin{equation}
\label{eq_mpqt}
m_{1\tilde{1}} 
= \int_{-\pi}^{\pi} {\rm d} \theta  \int_{0}^{+\infty}  {\rm d} \omega \ \omega \cdot \tilde{\omega}(\omega,\theta) G(\omega,\theta)  \, ,
\end{equation}
and $\tau_{pq}$ is defined as
\begin{equation}
\label{eq_taupq}
\tau_{pq} = \int_{-\pi}^{\pi} {\rm d} \theta \ \cos^p(\theta) \sin^q(\theta) 
\int_{0}^{+\infty} {\rm d} \omega \ gk(\omega) \tilde{\omega}(\omega,\theta) G(\omega,\theta) \, .
\end{equation}
\eq{eq_m2t} gives two alternative expressions,
depending on which wave spectrum ($\tilde{G}$ or $G$) is used for the integration.
Conversely,  $m_{1\tilde{1}}$ and $\tau_{pq}$ can not be readily expressed in terms of encounter spectrum,
because $\tilde{\omega}$ may not be an injective function of $\omega$ 
(for a given angle $\theta$, see \S\ref{subsec_encounter}).

From Eqs.~(\ref{eq_sigmaY_s}-\ref{eq_m2t}-\ref{eq_mpqt}-\ref{eq_taupq}), 
it appears that the computation of $\Sigma_{Y_s}$ 
does not require the use of the encounter spectrum $\tilde{G}$.
The knowledge of $\Sigma_{Y_s}$ is sufficient to compute 
the conditional distribution of kinematic variables, given up-crossing,
and the related up-crossing frequency.
Hence, as anticipated in \S\ref{subsec_encounter}, 
the explicit computation of the encounter spectrum $\tilde{G}$ is not required 
for the present matter.

\mypar{Case of infinite water depth with frequency-direction separation.}
If the water depth is infinite and the wave frequency/direction distributions are independent, 
Eqs.~(\ref{eq_m2t}-\ref{eq_mpqt}-\ref{eq_taupq}) can be further expressed as 
\begin{equation}
\label{eq_m11t_pec}
m_{1\tilde{1}} = m_2 - \frac{\beta_{1}}{\omega_s}  m_3
\, ,
\end{equation}
\begin{equation}
\label{eq_tau10_pec}
\tau_{10} = \alpha_{10} m_3 - \left[ \cos(\psi) \alpha_{20} + \sin(\psi) \alpha_{11} \right] \frac{m_4}{\omega_s}
\, ,
\end{equation}
\begin{equation}
\label{eq_tau01_pec}
\tau_{01} = \alpha_{01} m_3 - \left[ \cos(\psi) \alpha_{11} + \sin(\psi) \alpha_{02} \right] \frac{m_4}{\omega_s}
\, ,
\end{equation}
\begin{equation}
\label{eq_m2t_pec}
\mmt = 
m_2
- \frac{2\beta_{1}}{\omega_s}  m_3
+ \frac{\beta_{2}}{{\omega_s}^2}   m_4
\, ,
\end{equation}
where the coefficients $\beta_1$ and $\beta_2$ are given by
\begin{equation}
\label{eq_beta1}
\begin{split}
\beta_1  & = \int_{-\pi}^{\pi} {\rm d} \theta  \ \cos(\theta-\psi) D(\theta) \\
              & = \cos(\psi) \alpha_{10} + \sin(\psi) \alpha_{01}  \, ,
\end{split}
\end{equation}
and
\begin{equation}
\label{eq_beta2}
\begin{split}
\beta_2 & = \int_{-\pi}^{\pi} {\rm d} \theta  \ \cos^2(\theta-\psi) D(\theta) \\
             & = \frac{1}{2} \left[ 
1 
+ \cos(2\psi) (\alpha_{20}-\alpha_{02})  
+ 2 \sin(2\psi) \alpha_{11}    
\right] \, .
\end{split}
\end{equation}

\subsection{Conditional distribution given up-crossing} 
\label{subsec_cond_dist_fs}

Adopting similar notations as in \S\ref{subsec_cond_dist},
let $\cZs$ 
(resp. $\cX$) 
denote the random vector containing the variables of $Z_s$
(resp. $X$), 
except for $\etaS = \eta$. 
The conditional density function of $\cZs$,  given that $\etaS\rev{(t)}$ is up-crossing the level $a$, reads
\begin{equation} 
\label{eq_fZ_fs}
f_{\cZs|\etaS\rev{(t)} \uparrow a} = \frac{\dEtaS f_{\cZs|\eta= a}}
{\displaystyle \int_0^{+\infty} \! \! \! \! {\rm d} \xi \ \xi  f_{\dEtaS|\eta= a}(\xi)} , \ \dEtaS > 0 \, ,
\end{equation} 
which may also be written as
\begin{equation} 
\label{eq_fZ_2_fs}
f_{\cZs | \etaS\rev{(t)}  \uparrow a} = 
\sqrt{\frac{2\pi}{m_{\tilde2}}}  f_{\cX|\eta=a}  \times
  \dEtaS  f_{Y_s}  
\, , \ \dEtaS > 0 \, .
\end{equation} 

\rev{
When comparing the case of a fixed material point (see Section \ref{sect_body_rest})
and the case of a moving material point (present section),
up-crossings are checked for two different stochastic processes,
$\etaN(t)$ and $\etaS(t)$ respectively.
Hence the population sampled from up-crossing conditioning is also different \textit{a priori}.
However, the random vector $\cX$ depends solely on the variable $\etaS$, 
but not on the variable $\dEtaS$ -- 
here it is important to differentiate the stochastic process $\etaS(t)$ 
from the random variable $\etaS$, which is the value of $\etaS(t)$ at a given time.
As the probabilistic properties of $\etaS$ are the same as those of $\etaN$ 
(including its dependency relation with $\cX$),
the conditional distribution of $\cX$, given up-crossing,
turns out to be unaffected by the horizontal motion of the material point 
(it depends solely on the crossing level, $a$).
Conversely, the random vector $Y$ (defined in Eq. \ref{eq_vectY_0})
does not depend on $\etaS$, but depends on $\dEtaS$,
whose probabilistic properties are affected by the velocity ($V_s$) and heading ($\psi$) of the material point.
Hence, the conditional distribution of $Y$, given up-crossing, 
is found to be affected by the horizontal motion of the material point,
which reflects the fact that the underlying sampled population 
is indeed statistically different when the material is given a horizontal motion
(see \S\ref{subsec_illust_fs} for illustrative examples).
%as it depends on the up-crossing velocity $\dEtaS$.
}

\subsection{Up-crossing frequency}
\label{subsec_upfreq_fs}

Similarly to \eq{eq_fip_rest}, the up-crossing frequency for a material point moving at a constant speed is given by:
\begin{equation}
\label{eq_fip_fs}
\nus = \frac{1}{2\pi} \sqrt{\frac{\mmt}{m_0}} \exp \left( - \frac{a^2}{2 m_0} \right) \, .
\end{equation}

\mypar{Case of infinite water depth with frequency-direction separation.}
If the water depth is infinite and the wave frequency/direction distributions are independent,
Eqs.~(\ref{eq_oms}-\ref{eq_m2t_pec}) may be combined to express $\mmt$ as follows:
\begin{equation}
\label{eq_m2t_quadratic}
\mmt = 
m_2
- 2\frac{m_3}{g} \beta_{1} V_s
+ \frac{m_4}{g^2} \beta_{2} V_s^2
\, .
\end{equation}
From \eq{eq_m2t_quadratic} it appears that, 
for a given sea state and given heading, 
$\tilde{m}_2$ is a quadratic function of $V_s$, 
which reaches a minimum value 
\begin{equation}
\label{eq_m2t_min}
{\tilde{m}_2}^{\rm min}  = m_2  - \frac{{m_3}^2}{m_4} \frac{{\beta_{1}}^2}{\beta_{2}}
\end{equation}
for a velocity  
\begin{equation}
\label{eq_vs_min}
{V_s}^{\rm min} = g \frac{m_3}{m_4} \frac{\beta_{1}}{\beta_{2}} \, . 
\end{equation}
The corresponding minimum up-crossing frequency can be readily obtained 
by substituting \eq{eq_m2t_min} into \eq{eq_fip_fs}.
When the ship velocity becomes much larger than the phase velocity of waves, 
the up-crossing frequency tends to the asymptote
\begin{equation}
\label{eq_nu_asympt}
\nus 
\ \  \underset{\text{\scriptsize $V_s \rightarrow \pm \infty$}}{\text{\LARGE $\sim$}} \
\frac{1}{2\pi} \frac{\abs{V_s}}{g} \sqrt{\frac{m_4}{m_0}} \sqrt{\beta_{2}} 
\exp \left( - \frac{a^2}{2 m_0} \right) 
\, ,
\end{equation}
where $\nus$ becomes linearly dependent on $\abs{V_s}$.
Physically, it corresponds to a situation where the wave field
can be considered as ``frozen'' at a given time.

\subsection{Illustrative examples}
\label{subsec_illust_fs}

Figs.~\ref{fig_example_fImp}-\ref{fig_example_vs_dists} 
illustrate how the up-crossing frequency 
and the related conditional distribution of kinematic variables 
are affected by the forward speed of the material point.
Five different configurations are considered; they are listed in \tab{table_cases}.
The considered sea states are the same as in \S\ref{subsec_illust_rest}.
Compared to the case of a material point at rest, 
the consideration of forward motion required to introduce the additional kinematic variable, $\dEtaS$,
along with four additional covariance coefficients (see \S\ref{subsec_non_cond_fs}).
For infinite water depth, as assumed in the present examples, 
Eqs.~(\ref{eq_m11t_pec} to \ref{eq_beta2}) 
provide closed-form expressions for these four additional covariance coefficients, 
as functions of frequency moments, $m_p$, 
coefficients $\alpha_{pq}$, and material point heading, $\psi$.
For the assumed spectrum shape, the numerical values of the first five 
frequency moments are given in \eq{eq_num_moments}.
The numerical values of the coefficients $\alpha_{pq}$, for the different direction distributions, 
have been reported in \tab{table_alpha_value}.

% %%%%%%%%%%%%%%%%%%%%%%%%
% %%%%%        TABLE    BEGIN         %%%%%
% %%%%%%%%%%%%%%%%%%%%%%%%

\begin{table}[h]
\centering
\begin{tabular}{||l|l|l||} 
 \hline
Name   & Direction distribution & heading \\ 
   \hline
$C1$ & $D_1$ (\eqp{eq_D1}) & $\psi = 0$    \\ 
$C2$ & $D_2$ (\eqp{eq_D2}) & $\psi = 0$     \\
$C3$ & $D_2$ (\eqp{eq_D2}) & $\psi = \pi/4$ \\
$C4$ & $D_2$ (\eqp{eq_D2}) & $\psi = \pi/2$ \\
$C5$ & $D_3$ (\eqp{eq_D3}) & $\psi = 0 $ \\  
 \hline
\end{tabular}
\caption{List of the different configurations considered for illustrative purpose in \S\ref{subsec_illust_fs}. 
Each line of the table corresponds to a configuration.
The first column specifies a name which is used to identify the configuration. 
The second column specifies the assumed direction distribution of waves (see \S\ref{subsec_Dspread}).
The third column specifies the heading of the material point.
For these five configurations,
Figs.~\ref{fig_example_fImp}-\ref{fig_example_vs_dists} respectively
show how the up-crossing frequency and the related conditional distribution of 
kinematic variables are affected by the forward speed of the material point.
}
\label{table_cases}
\end{table}

% %%%%%%%%%%%%%%%%%%%%%%%%
% %%%%%        TABLE    END             %%%%%
% %%%%%%%%%%%%%%%%%%%%%%%%

% %%%%%%%%%%%%%%%%%%%%%%%%
% %%%%%%%%%%%%%%%%%%%%%%%%
% %%%%%        FIGURE    BEGIN         %%%%%
% %%%%%%%%%%%%%%%%%%%%%%%%
% %%%%%%%%%%%%%%%%%%%%%%%%

\begin{figure}[t]
\begin{center}
\begin{tabular}{c}
        \includegraphics[width=0.4\textwidth]{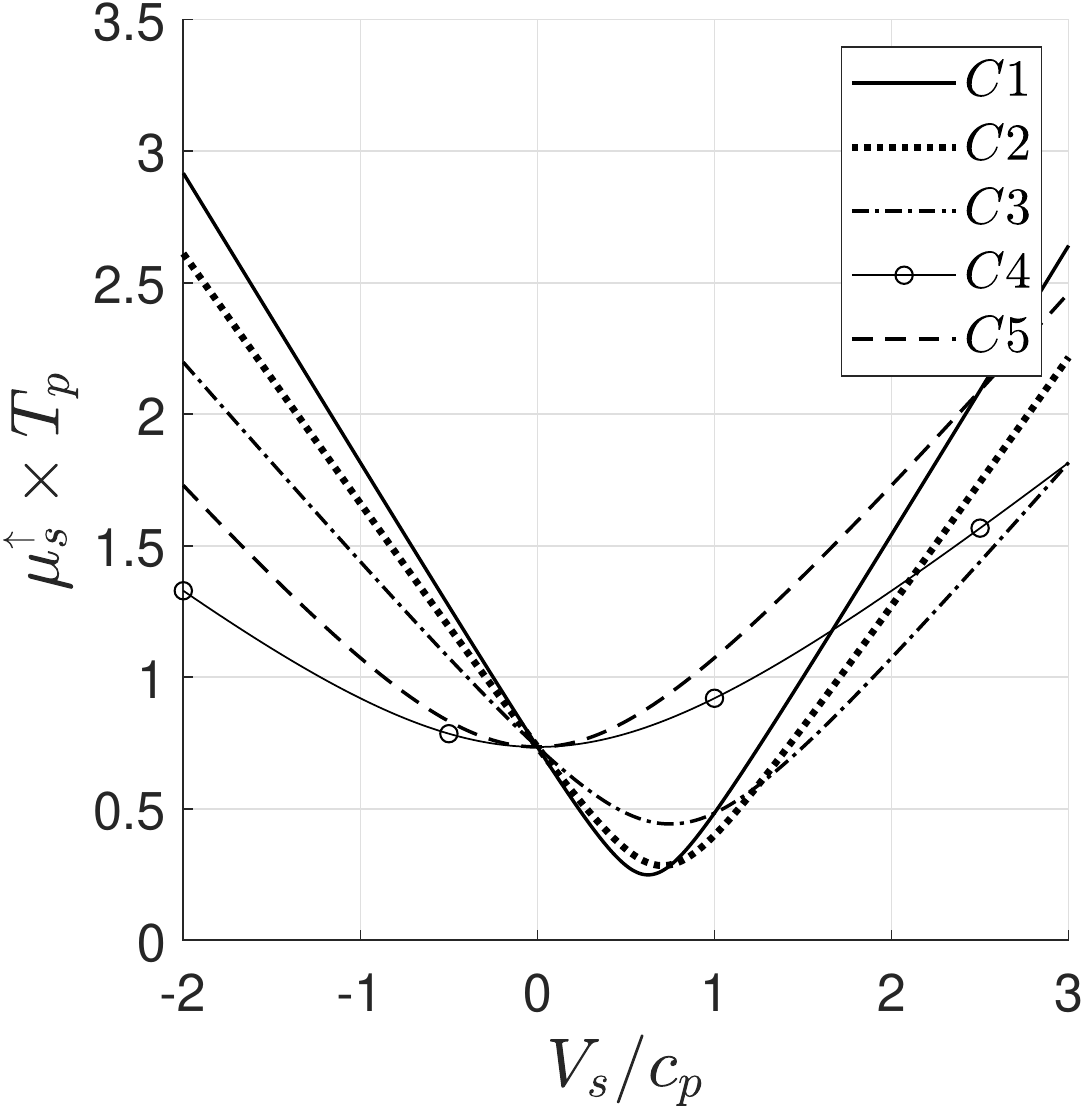}          
\end{tabular}
\end{center}
\caption{
Up-crossing frequency, $\nus$, as a function of the forward speed, $V_s$. 
Both quantities are nondimensionalised by using 
the peak wave period, $T_p = 2\pi / \omega_p$, and the related phase velocity, $c_p = g / \omega_p$.
The material point is assumed to move at an altitude $a=H_s/4$.
Five different configurations are considered, as listed in \tab{table_cases}. 
}
\label{fig_example_fImp}
\end{figure}

% %%%%%%%%%%%%%%%%%%%%%%%%
% %%%%%%%%%%%%%%%%%%%%%%%%
% %%%%%        FIGURE    END         %%%%%
% %%%%%%%%%%%%%%%%%%%%%%%%
% %%%%%%%%%%%%%%%%%%%%%%%%

\subsubsection{Up-crossing frequency}

\fig{fig_example_fImp} shows the evolution of the up-crossing frequency, $\nus$,
as a function of the velocity of the material point, $V_s$, for the different configurations listed in \tab{table_cases}.
As the wave frequency spectrum, $S(\omega)$, 
is the same for the five configurations, all the curves intersect at $V_s = 0$. 
Indeed, when the material point is at rest, 
the up-crossing frequency is not sensitive to the directional spreading of the sea state.
In all cases, the function $\nus(V_s)$ shows a minimum, 
whose coordinates are given by Eqs.~(\ref{eq_fip_fs}-\ref{eq_m2t_min}-\ref{eq_vs_min}).
All curves are symmetrical about the vertical axis passing through their minimum, 
which reflects the fact that $\nus(V_s)$ is the square root of a quadratic function (see Eqs.~\ref{eq_fip_fs}-\ref{eq_m2t_min}).
For the configurations $C4$-$C5$, the minimum up-crossing frequency is obtained for $V_s=~0$:
this is due to the symmetry of the wave direction distribution 
about the direction perpendicular to the material point heading,
which implies $\beta_{1} = 0$, resulting in ${V_s}^{\rm min}=0$ (see \eqp{eq_vs_min}).

The minimum up-crossing frequency is strictly greater than zero for all considered configurations, 
which is related to two different effects.
First, the dispersion relation of water waves implies that individual waves of different frequencies propagate with different phase speeds;
then a material moving at a constant velocity cannot maintain a constant position relative to all individual waves.
In the present illustrative examples, an infinite water depth has been assumed;
the effect of wave dispersion would be much reduced in shallow water.
Second, the projections of wave propagation speeds, along the material point heading,
are further scattered by the directional spreading of waves. 
This second effect is accounted for, in \eq{eq_m2t_min}, by the factor ${\beta_1}^2 / {\beta_2}$.
This factor is equal to $1$ for a unidirectional sea 
(except for a material moving exactly abeam a unidirectional sea, 
degenerate case where the up-crossing frequency does not depend on the forward speed)
and less than $1$ otherwise.
Then, for a given frequency spectrum, $S(\omega)$, 
the smallest possible minimum up-crossing frequency is obtained for a unidirectional sea,
which is consistent with the results reported in \fig{fig_example_fImp}.
In the case of the unidirectional sea ($C1$), 
it is also interesting to note that the minimum up-crossing frequency is reached for ${V_s}^{\rm min} < c_p$,
the former being the phase velocity corresponding to the peak of the wave frequency spectrum.
This velocity ordering is due to the contribution of short waves in the tail of the assumed JONSWAP spectrum.

Finally, when $\abs{V_s} \gg c_p$, 
the ordering of the asymptotic slopes obtained for the different configurations can be understood from \eq{eq_nu_asympt}.
As the wave frequency spectrum, $S(\omega)$, is the same in all considered cases, 
the difference in the asymptotic slopes is governed by the geometrical factor $\sqrt{\beta_{2}}$. 
For the five configurations considered here, $C1$ to $C5$,
the respective geometrical factors can be analytically computed 
(by using \eqp{eq_beta2} and \tab{table_alpha_value}), 
which gives $\sqrt{\beta_{2}} = 1; \sqrt{3}/2; 1/\sqrt{2}; 1/2; 1/\sqrt{2}$.

\subsubsection{Kinematic variables which are not affected by the speed of advance}

The conditional distribution of the kinematic variables ($\vxs$, $\vys$, $\az$), given up-crossing,
is not affected by the horizontal motion 
of the material point, but depends solely on its altitude, $a$ 
\rev{(see \S\ref{subsec_cond_dist_fs} for an explanation)}.
Therefore, the effect of level-crossing conditioning on these variables, 
is identical to the one illustrated in \S\ref{subsec_illust_rest}, 
for the three considered wave direction distributions, $D1$, $D2$, $D3$.

\subsubsection{Kinematic variables which are affected by the speed of advance}

Among the kinematic variables considered in the present paper,
the conditional distributions of $\vz$, $\sx$, $\sy$, $\ax$ and $\ay$, 
given up-crossing, 
are affected by the speed of advance of the material point, 
through their dependence on $\dEtaS$ (see \S\ref{subsec_non_cond_fs}-\ref{subsec_cond_dist_fs}).
Conversely, they do not depend on the assumed altitude of the material point, $a$.
\fig{fig_example_vs_dists} illustrates the effect of the forward speed 
on the conditional distributions of these five kinematic variables, given up-crossing,
for the five configurations listed in \tab{table_cases}.

The random variable $\dEtaS$, given up-crossing, 
follows a Rayleigh distribution with a mode equal to $\sqrt{\mmt}$ (see \eqp{eq_m2t_pec}).
This distribution is not represented in \fig{fig_example_vs_dists}, 
but the value of its mode, computed for the different considered configurations, 
is reported in \tab{table_mode_dEtaS}.
As the up-crossing frequency depends on $V_s$ solely through the term $\sqrt{\mmt}$ 
(see \eqp{eq_fip_fs}),
the curves shown in \fig{fig_example_fImp} can be used as a direct proxy 
(up to a numerical factor) for the evolution of the mode $\sqrt{\mmt}$ as a function of $V_s$. 
Conversely, when $V_s \ne 0$, 
the velocity of the free surface elevation measured in the reference frame of the mean flow, 
$\de = \vz$, given up-crossing, is not Rayleigh-distributed anymore.
Indeed, the moving material point may 
cross the sloped free surface even though the sea elevation, $\eta$, is locally decreasing.
Instead, the conditional distribution of $\vz$, given up-crossing,
results from the \rev{convolution} of a Rayleigh distribution with a normal distribution 
(see \S\ref{subsubsec_marg_does}, in appendix).

The evolution of the conditional bivariate density function of $(\axs, \ays)$, 
as a function of $V_s$, 
follows the one of $(\sx, \sy)$ in a symmetric way, 
since these variables are linearly related through \eq{eq_linrel_sa}.
The effect of the forward speed on the bivariate distribution of 
$(\sx, \sy)$, given up-crossing, is discussed below,
individually for the different considered configurations.

% %%%%%%%%%%%%%%%%%%%%%%%%
% %%%%%        TABLE    BEGIN         %%%%%
% %%%%%%%%%%%%%%%%%%%%%%%%

\begin{table}[h]
\centering
\begin{tabular}{|c|c|c|c|c|} 
 \cline{2-5}
\multicolumn{1}{c|}{}   & $V_s / c_p = -3$ & $V_s / c_p = 0$ & $V_s / c_p = 0.7$ & $V_s / c_p = 4$ \\ 
   \hline
$C1$ & $1.66$ & $0.303$ & $0.108$ & $1.54$    \\ 
 \hline
$C2$ & $1.47$ & $0.303$ & $0.117$ & $1.31$     \\
 \hline
$C3$ & $1.22$ & $0.303$ & $0.183$ & $1.06$ \\
\hline
\multicolumn{5}{c}{}  \\ [-2ex] 
 \cline{2-5}
\multicolumn{1}{c|}{}   & $V_s / c_p = 0$ & $V_s / c_p = 0.6$ & $V_s / c_p = 1.3$ & $V_s / c_p = 4$ \\ 
   \hline
$C4$ & $0.303$ & $0.332$ & $0.424$ & $0.962$ \\
 \hline
$C5$ & $0.303$ & $0.360$ & $0.517$ & $1.33$ \\
 \hline
\end{tabular}
\caption{
Mode of the conditional distribution of $\dEtaS$, given up-crossing.
The kinematic variable $\dEtaS$, given up-crossing, follows a Rayleigh distribution, 
whose mode is equal to $\sqrt{ \mmt }$ (see \eqp{eq_m2t_pec}).
The present table reports the values taken by this mode, in nondimensional form ($\sqrt{\mmt} / H_s \omega_p$), 
for the different cases illustrated in \fig{fig_example_vs_dists}. 
In \fig{fig_example_vs_dists}, for each of the five configurations listed in \tab{table_cases}, 
four different forward speeds are considered.
Each line of the present table corresponds to a configuration, and each column corresponds to a forward speed.
As the chosen forward speeds are different for the configurations $C1$-$C2$-$C3$ and $C4$-$C5$, 
the reported values are divided into two different subtables. 
When the material point is at rest, $\dEtaS$ coincides with $\vz$, 
variable which does not depend on the directional spreading of the sea state; 
this explains why the values reported for $V_s / c_p = 0$ 
are identical (equal to $\simeq 0.303$) for all configurations.
}
\label{table_mode_dEtaS}
\end{table}

% %%%%%%%%%%%%%%%%%%%%%%%%
% %%%%%        TABLE    END             %%%%%
% %%%%%%%%%%%%%%%%%%%%%%%%

% %%%%%%%%%%%%%%%%%%%%%%%%
% %%%%%%%%%%%%%%%%%%%%%%%%
% %%%%%        FIGURE    BEGIN         %%%%%
% %%%%%%%%%%%%%%%%%%%%%%%%
% %%%%%%%%%%%%%%%%%%%%%%%%

% scaling of the figure
\def\scaleF{0.315}
\def\scaleSpeLine{0.75}
\def\widthSpeLine{0.1}

\def\caseNameOne{Hs_04_Tp_10_ga_33EM1_tr_010EM3_hT_InfEM3_psiOpi_0000EM3_DspType_01_CalcType_01}
\def\caseNameTwo{Hs_04_Tp_10_ga_33EM1_tr_010EM3_hT_InfEM3_psiOpi_0000EM3_DspType_02_CalcType_01}
\def\caseNameThree{Hs_04_Tp_10_ga_33EM1_tr_010EM3_hT_InfEM3_psiOpi_0250EM3_DspType_02_CalcType_01}
\def\caseNameFour{Hs_04_Tp_10_ga_33EM1_tr_010EM3_hT_InfEM3_psiOpi_0500EM3_DspType_02_CalcType_01}
\def\caseNameFive{Hs_04_Tp_10_ga_33EM1_tr_010EM3_hT_InfEM3_psiOpi_0000EM3_DspType_03_CalcType_01}

\begin{figure}[h!]
\begin{center}
\begin{tabular}{ccc}
	 \multicolumn{3}{c}{\textbf{Configuration $C1$}}  \\
        \includegraphics[width=\scaleF\textwidth]{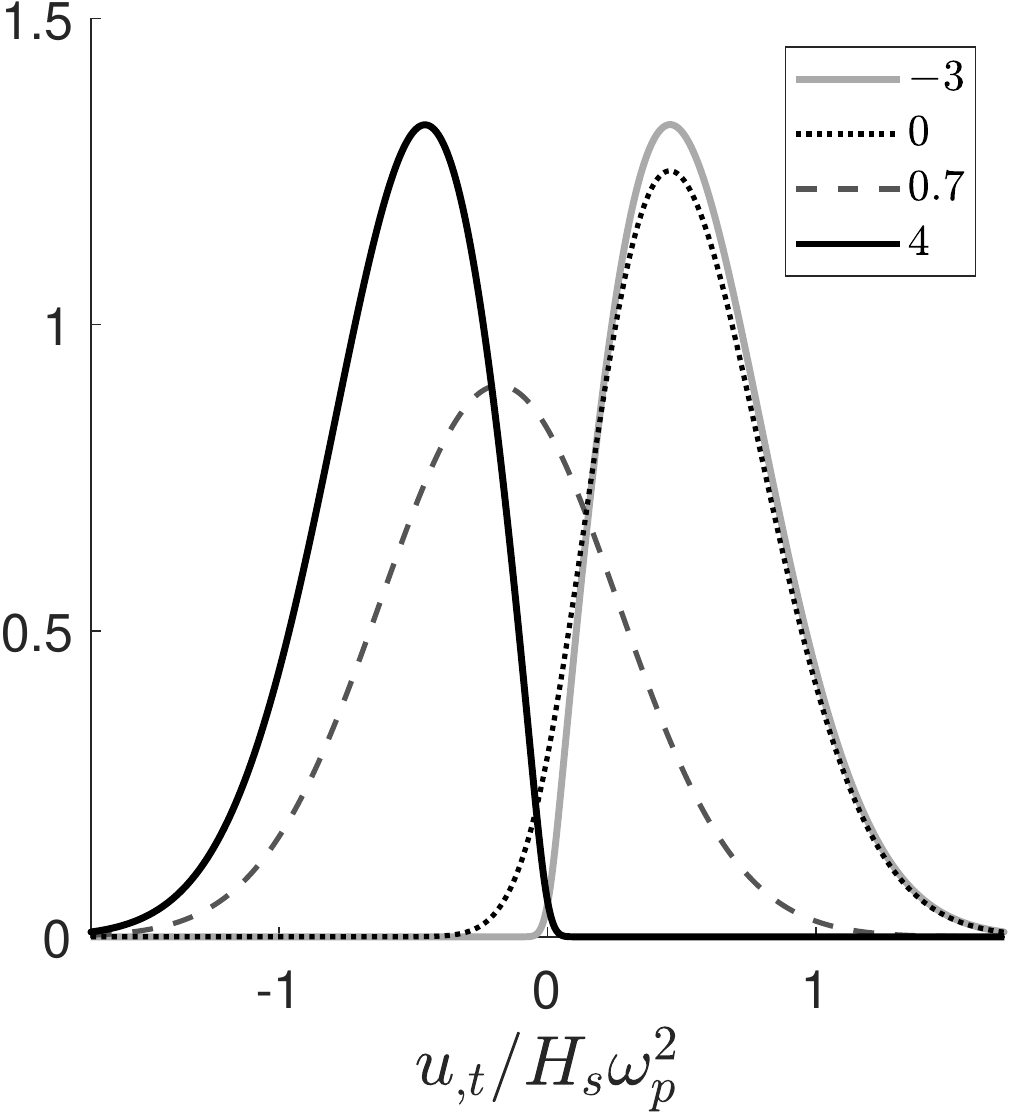} & 
        \includegraphics[width=\scaleF\textwidth]{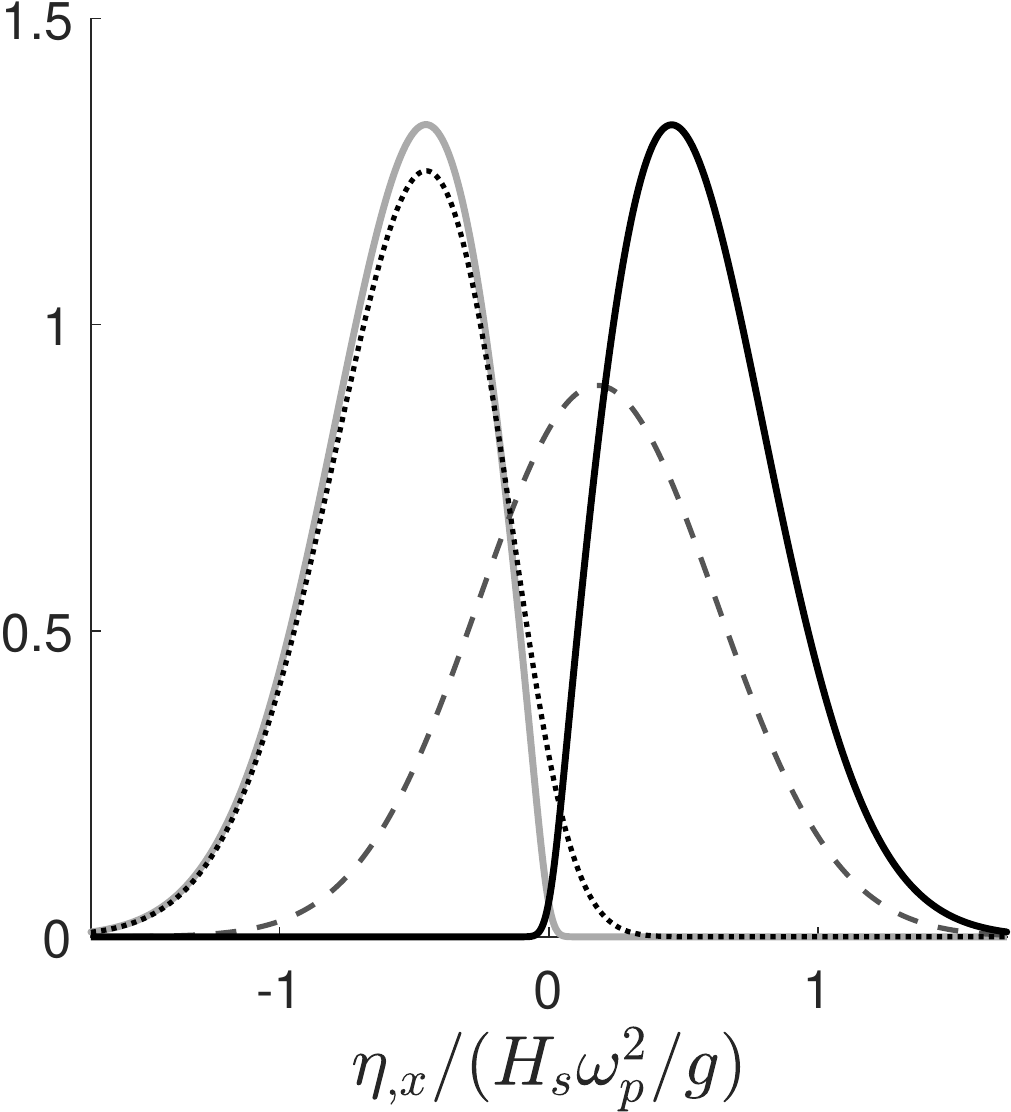} &
        \includegraphics[width=\scaleF\textwidth]{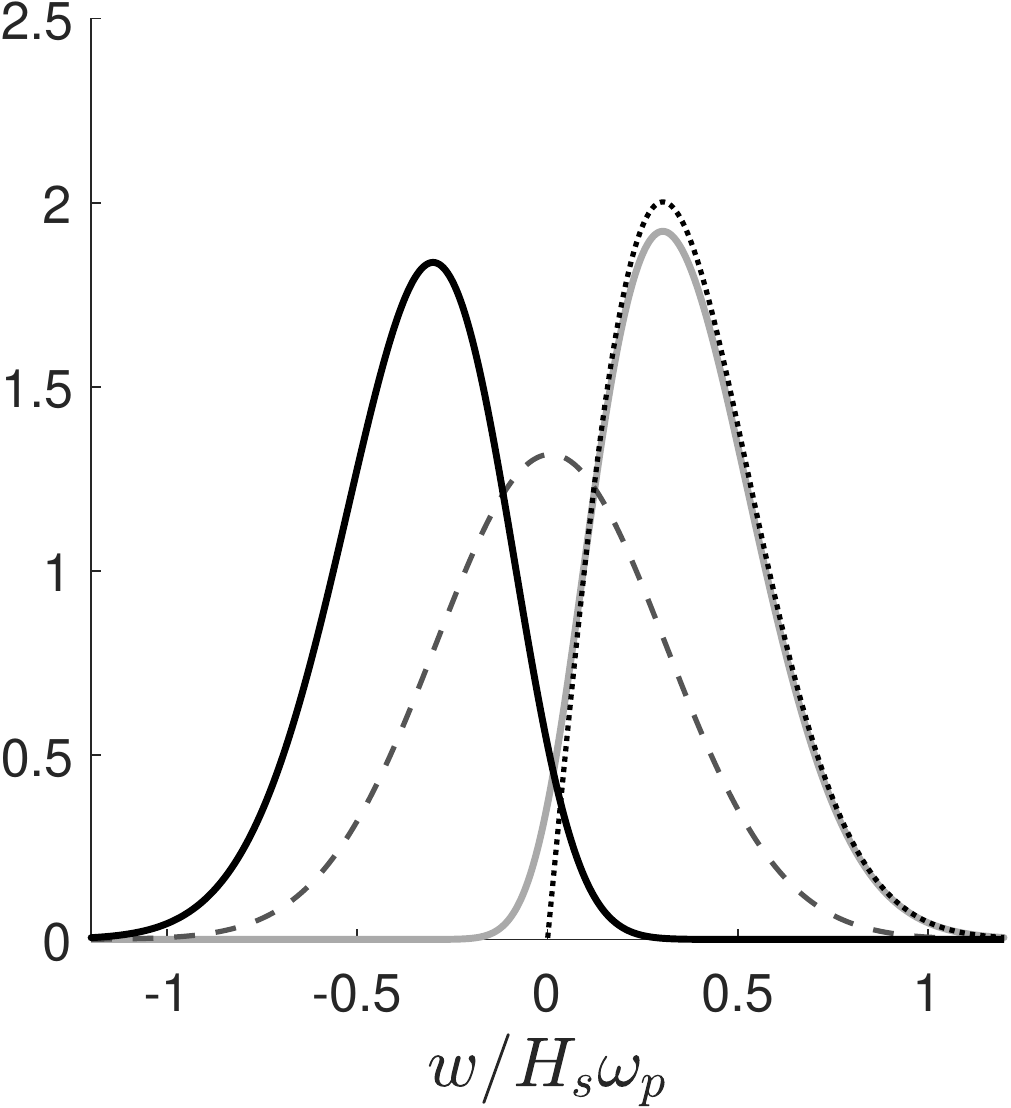}  \\        
        	 \multicolumn{3}{c}{\textbf{Configuration $C2$}}  \\
        \includegraphics[width=\scaleF\textwidth]{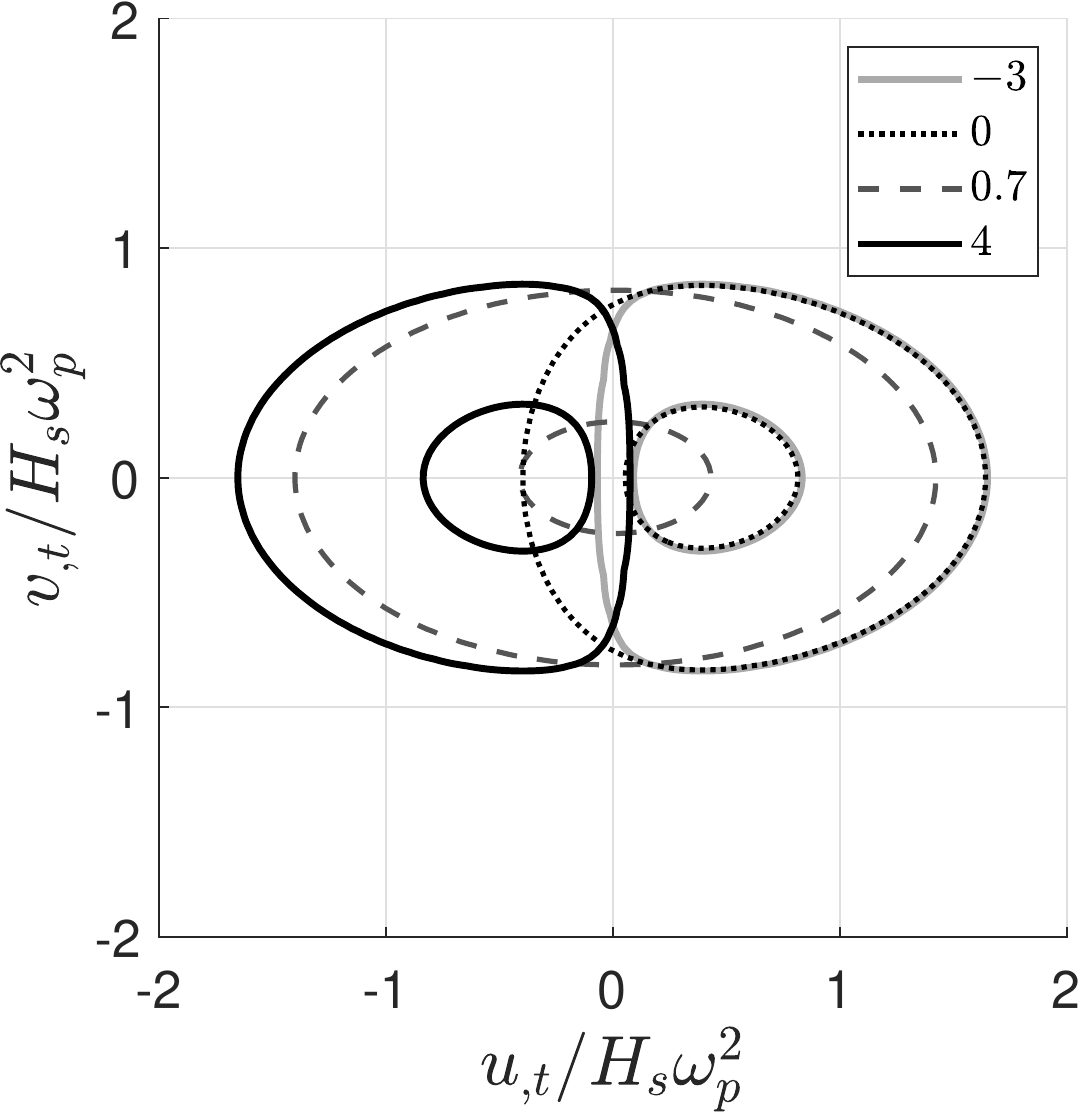} & 
        \includegraphics[width=\scaleF\textwidth]{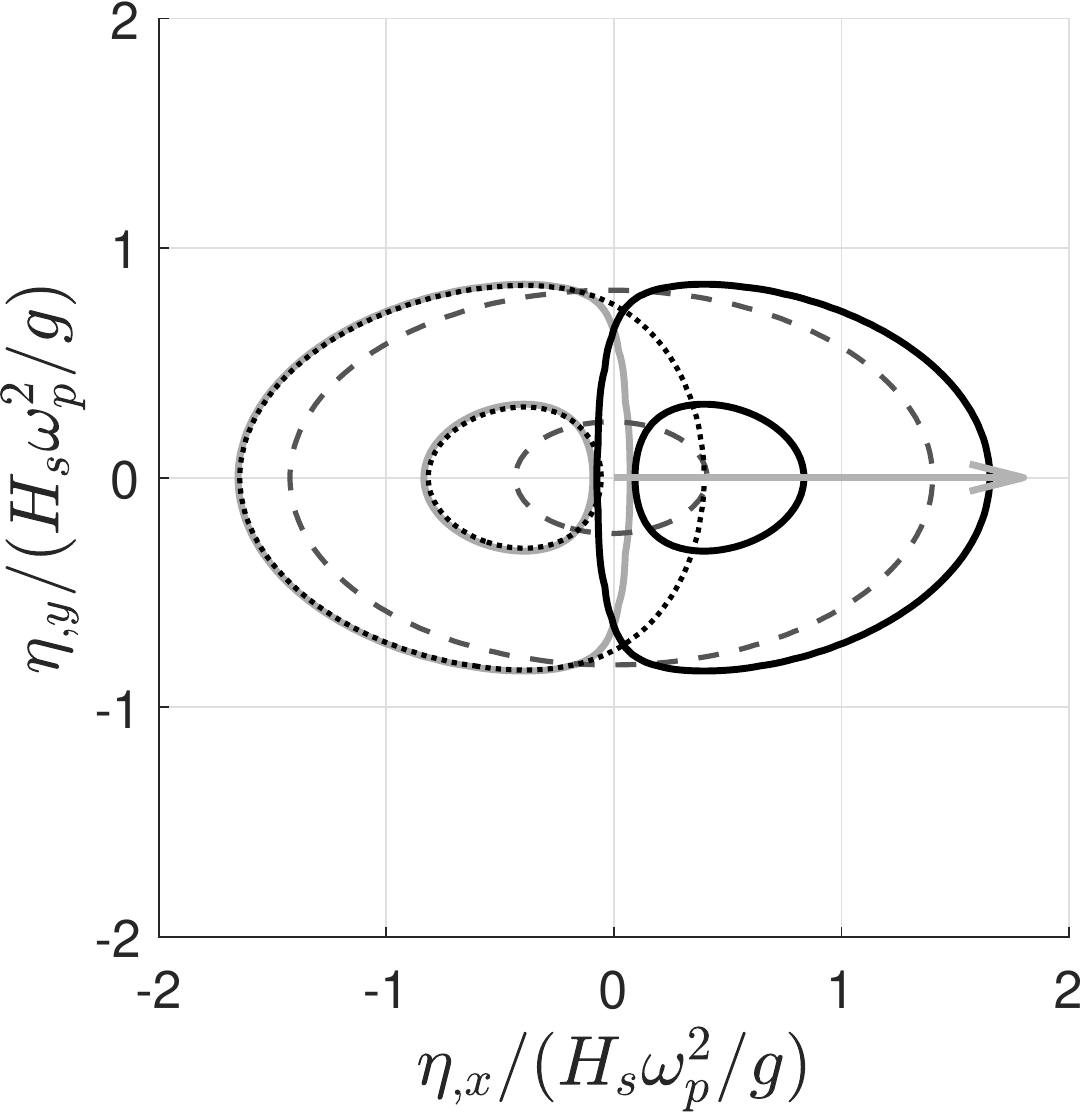} &
        \includegraphics[width=\scaleF\textwidth]{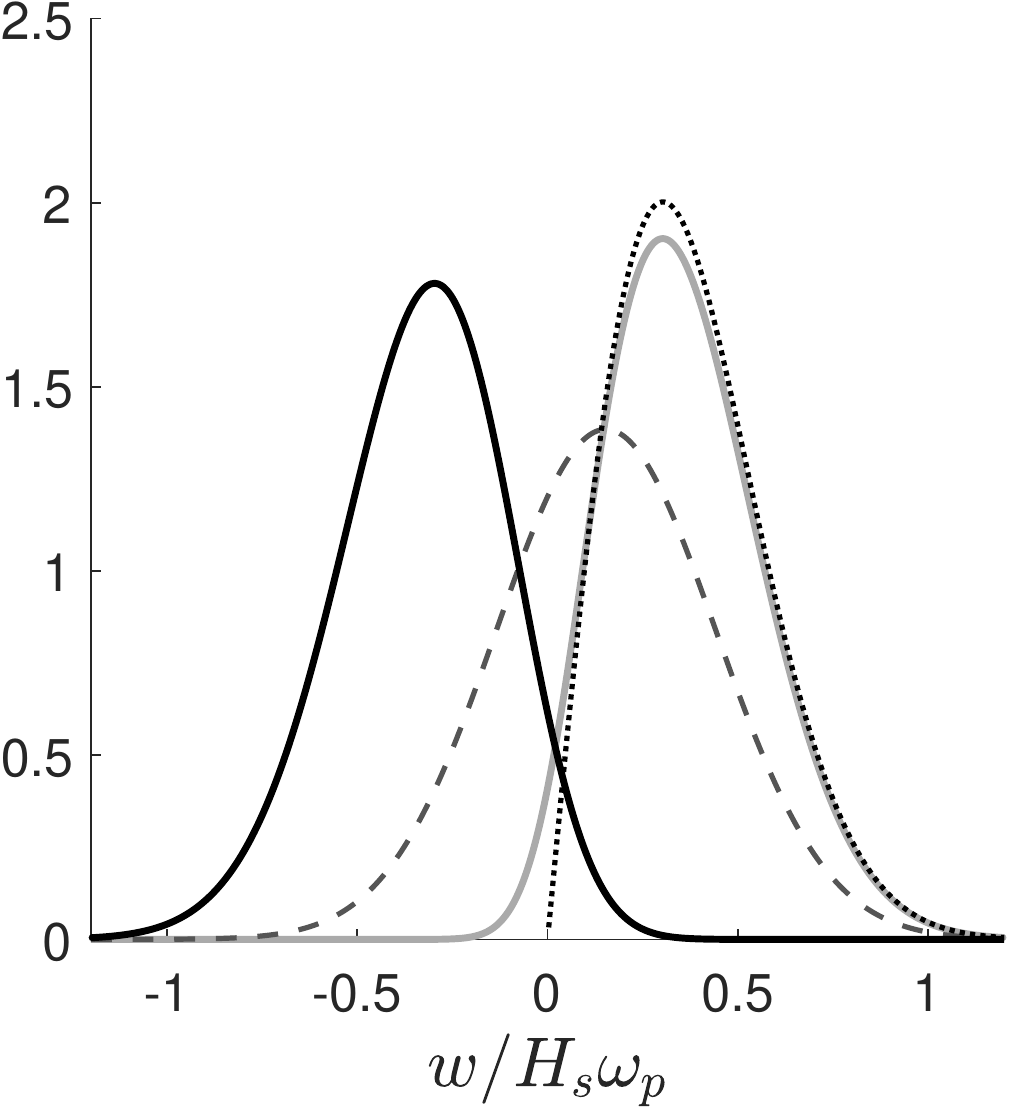}  \\        
        	 \multicolumn{3}{c}{\textbf{Configuration $C3$}}  \\
        \includegraphics[width=\scaleF\textwidth]{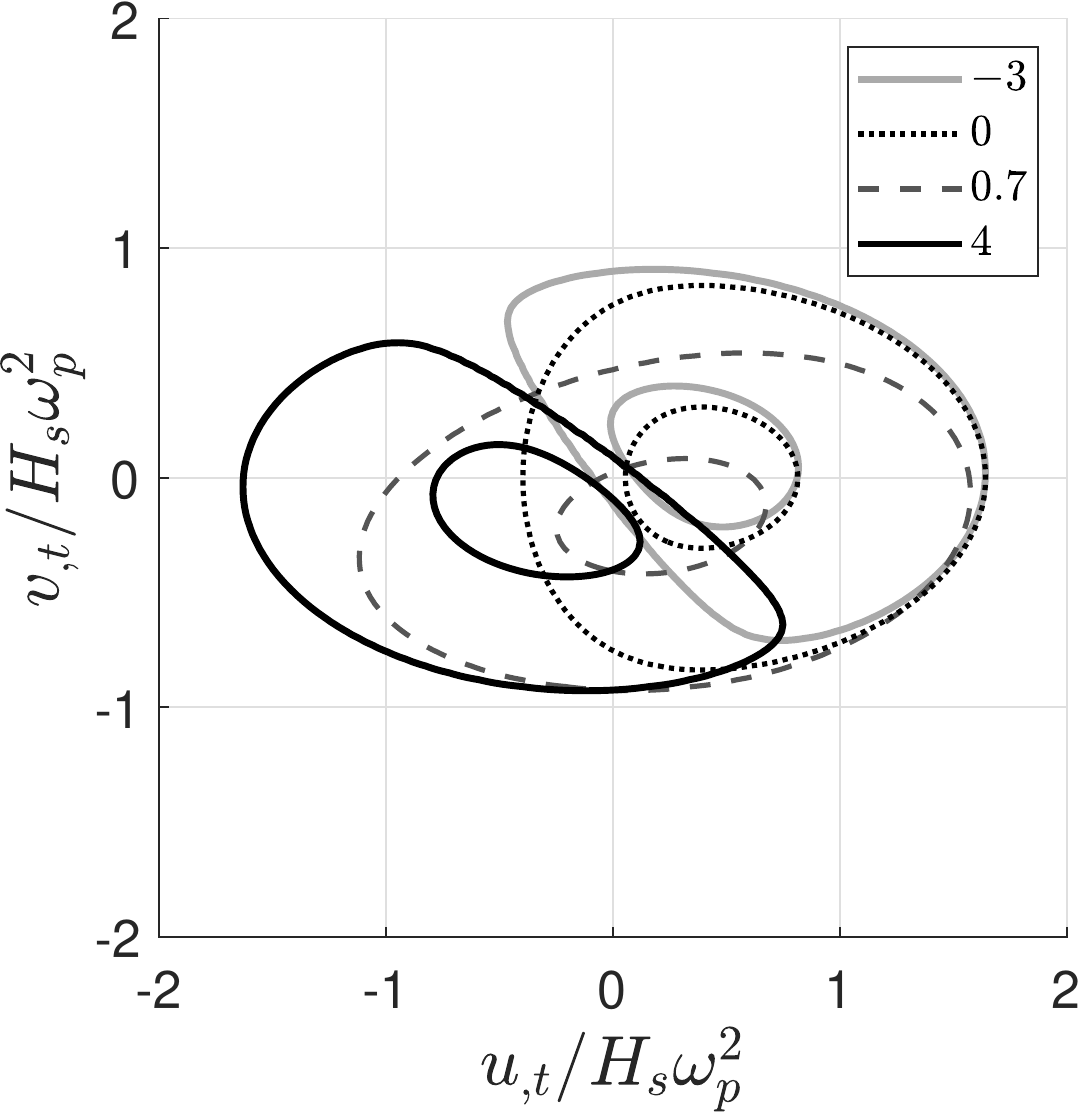} & 
        \includegraphics[width=\scaleF\textwidth]{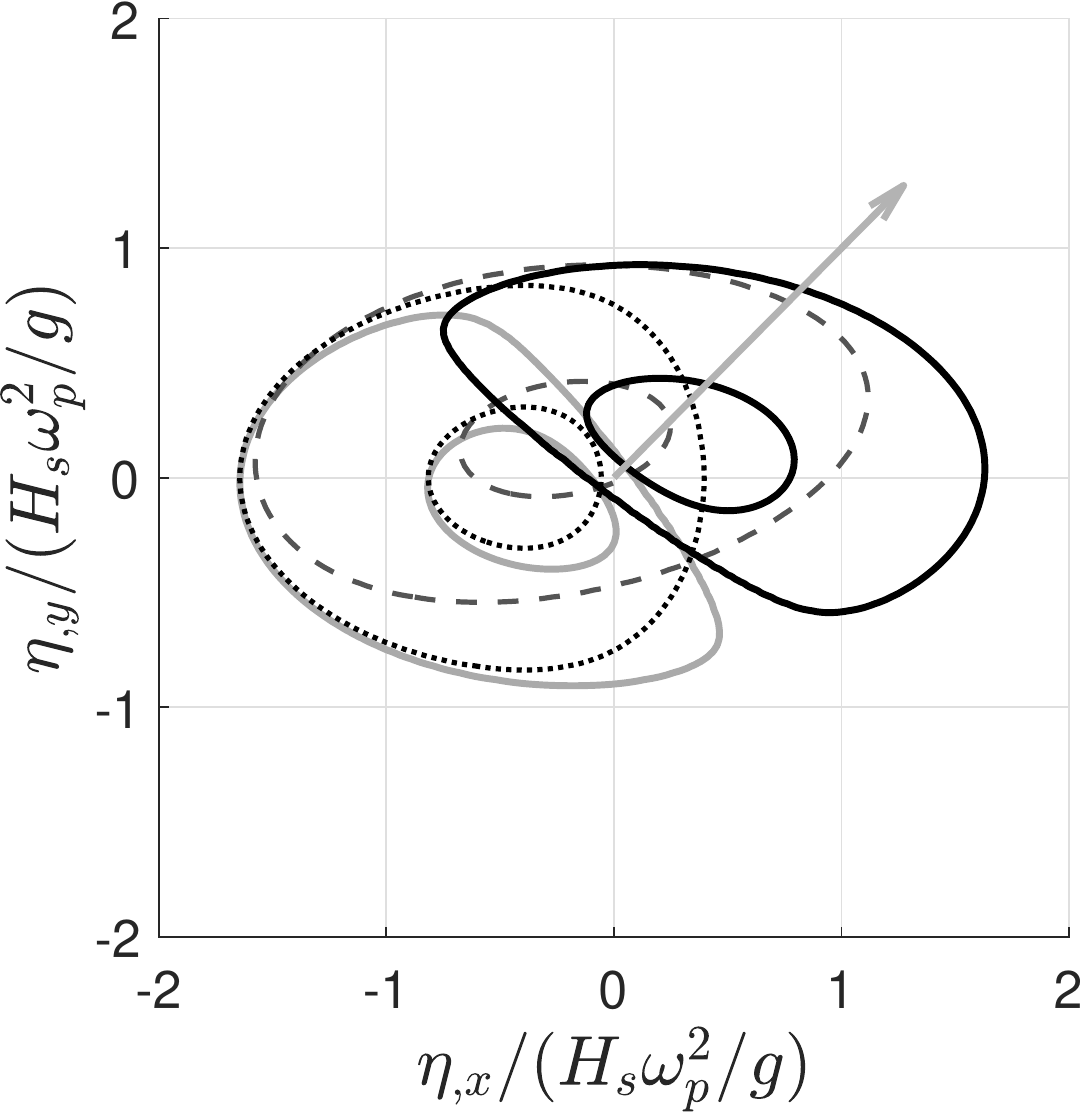} &
        \includegraphics[width=\scaleF\textwidth]{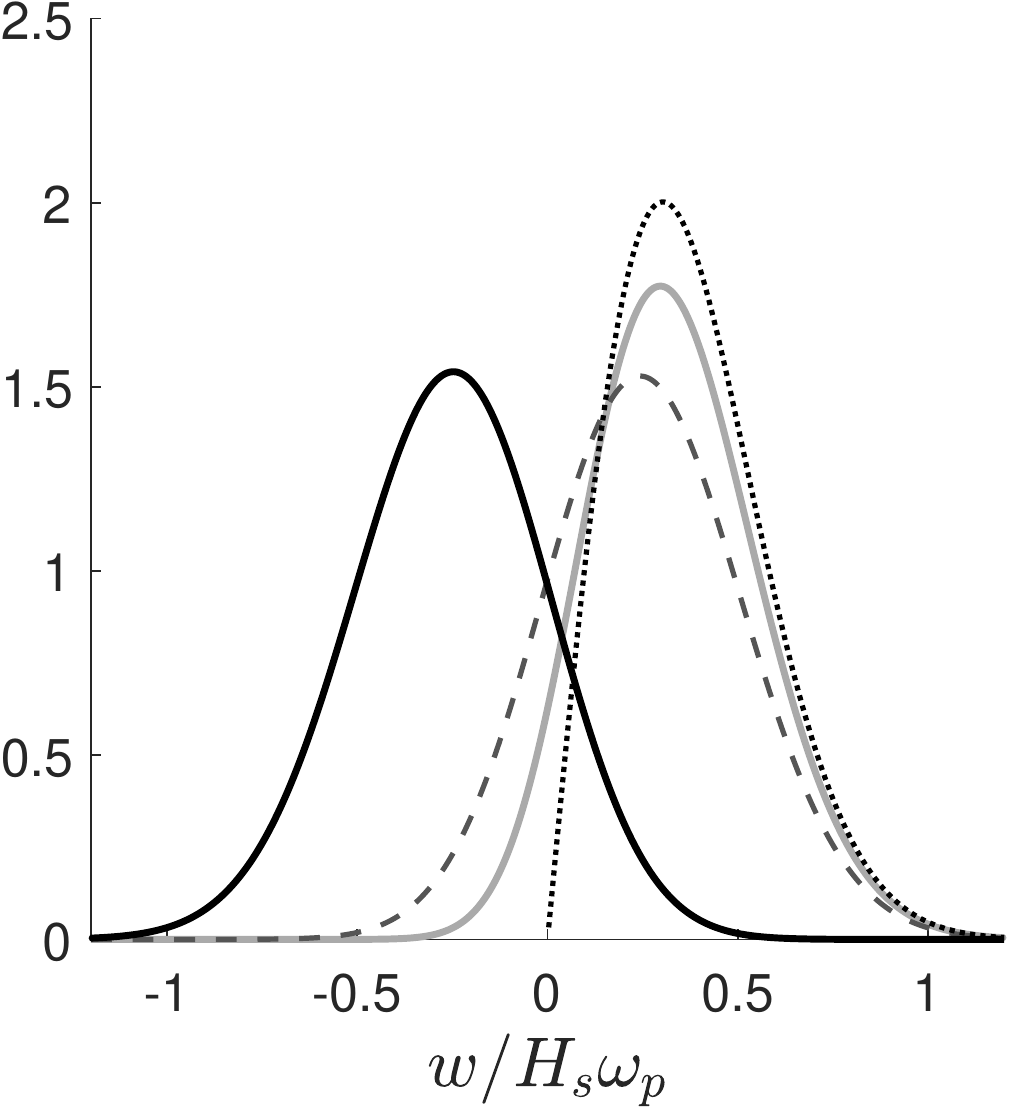}                 
\end{tabular}        
\end{center}
\caption{Continued on next page.}
\label{fig_example_vs_dists}
\end{figure}

% %%%%%%%%%%%%%%%%%%%%%%%%
% %%%%%%%%%%%%%%%%%%%%%%%%
% %%%%%        FIGURE    CONTINUED         %%%%%
% %%%%%%%%%%%%%%%%%%%%%%%%
% %%%%%%%%%%%%%%%%%%%%%%%%

\begin{figure}[h!]
\ContinuedFloat
\captionsetup{list=off,format=cont}
\begin{center}
\begin{tabular}{ccc}
         \multicolumn{3}{c}{\textbf{Configuration $C4$}}  \\    
        \includegraphics[width=\scaleF\textwidth]{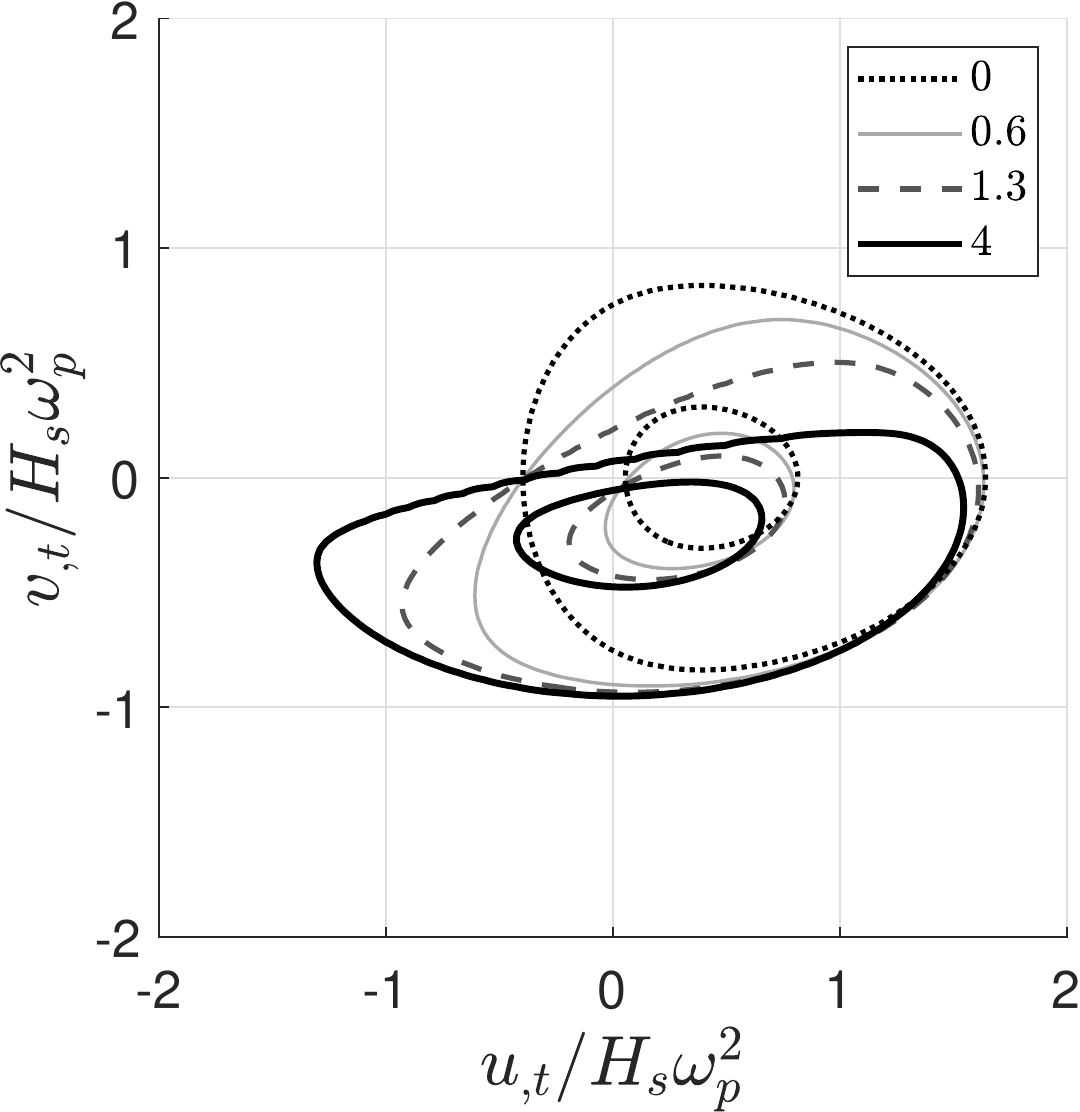} & 
        \includegraphics[width=\scaleF\textwidth]{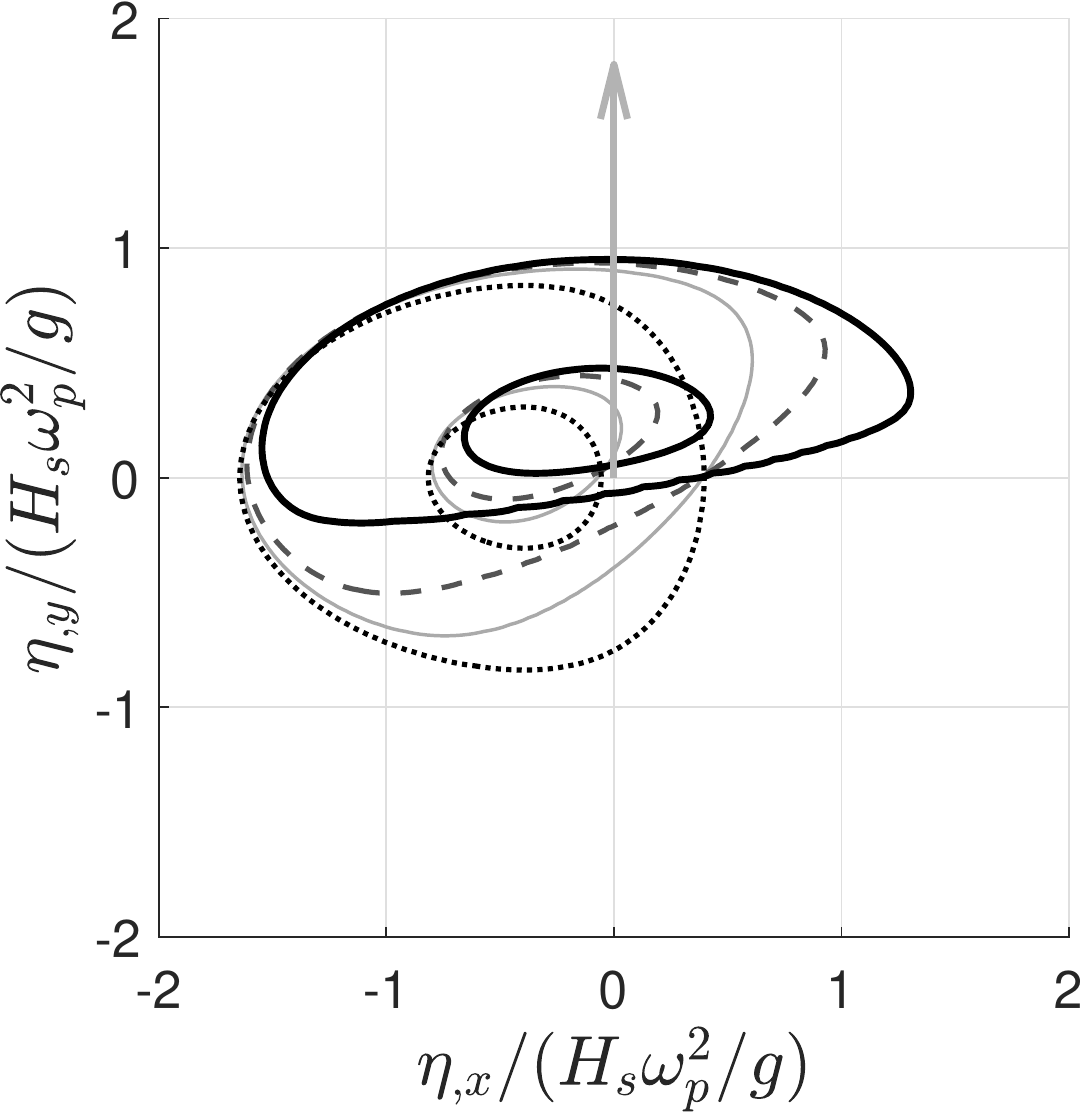} &
        \includegraphics[width=\scaleF\textwidth]{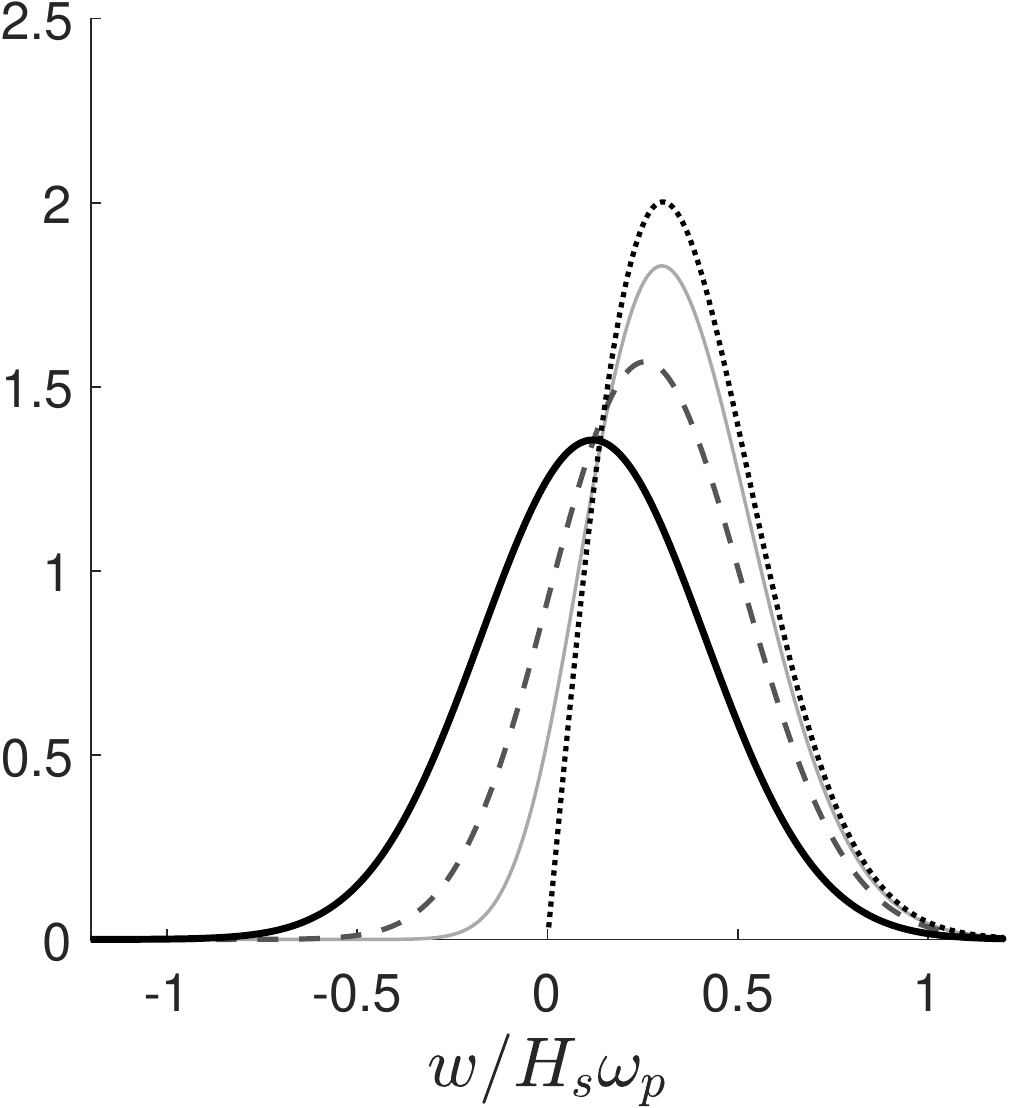} \\        
        	 \multicolumn{3}{c}{\textbf{Configuration $C5$}}  \\
        \includegraphics[width=\scaleF\textwidth]{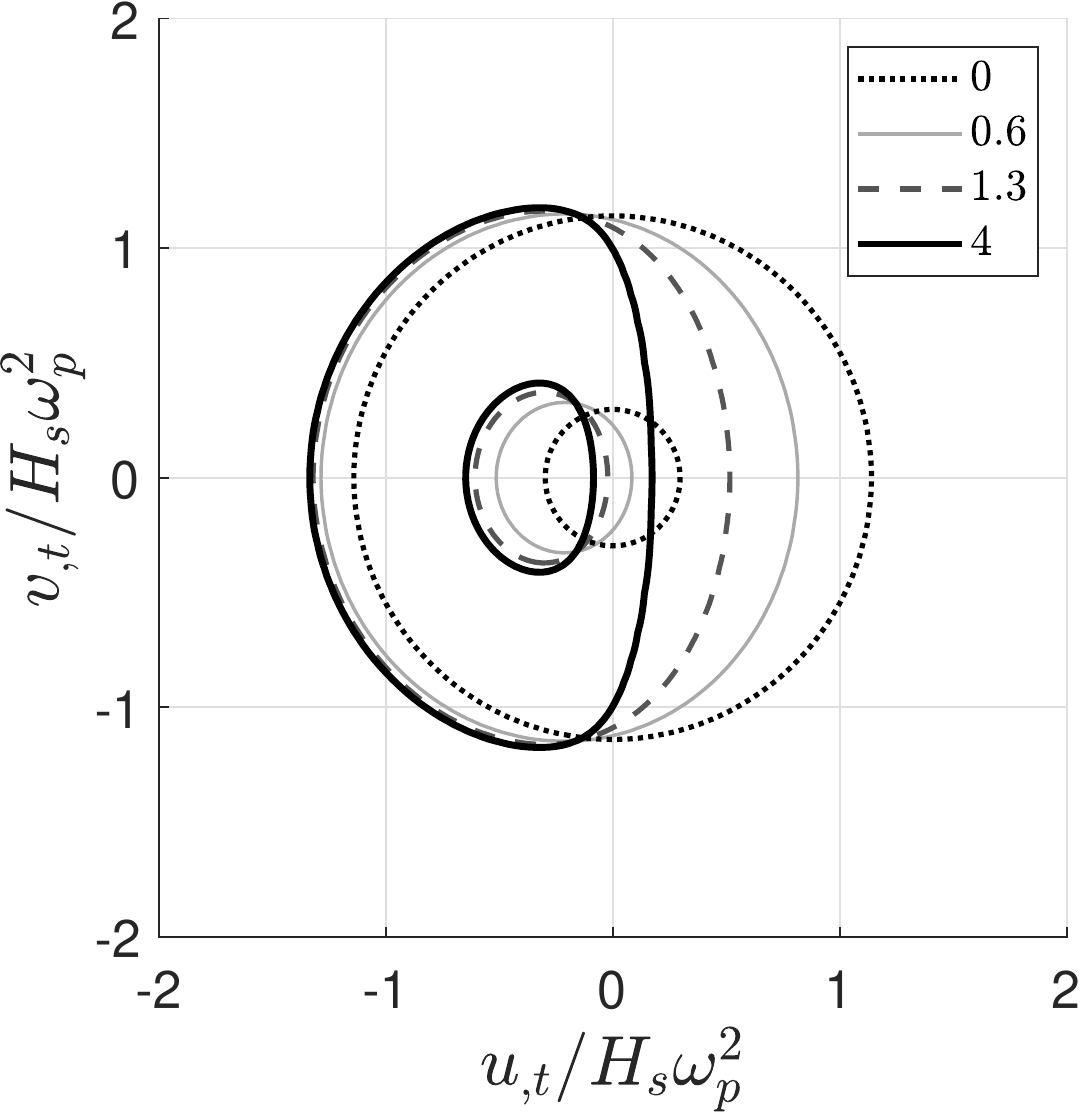} & 
        \includegraphics[width=\scaleF\textwidth]{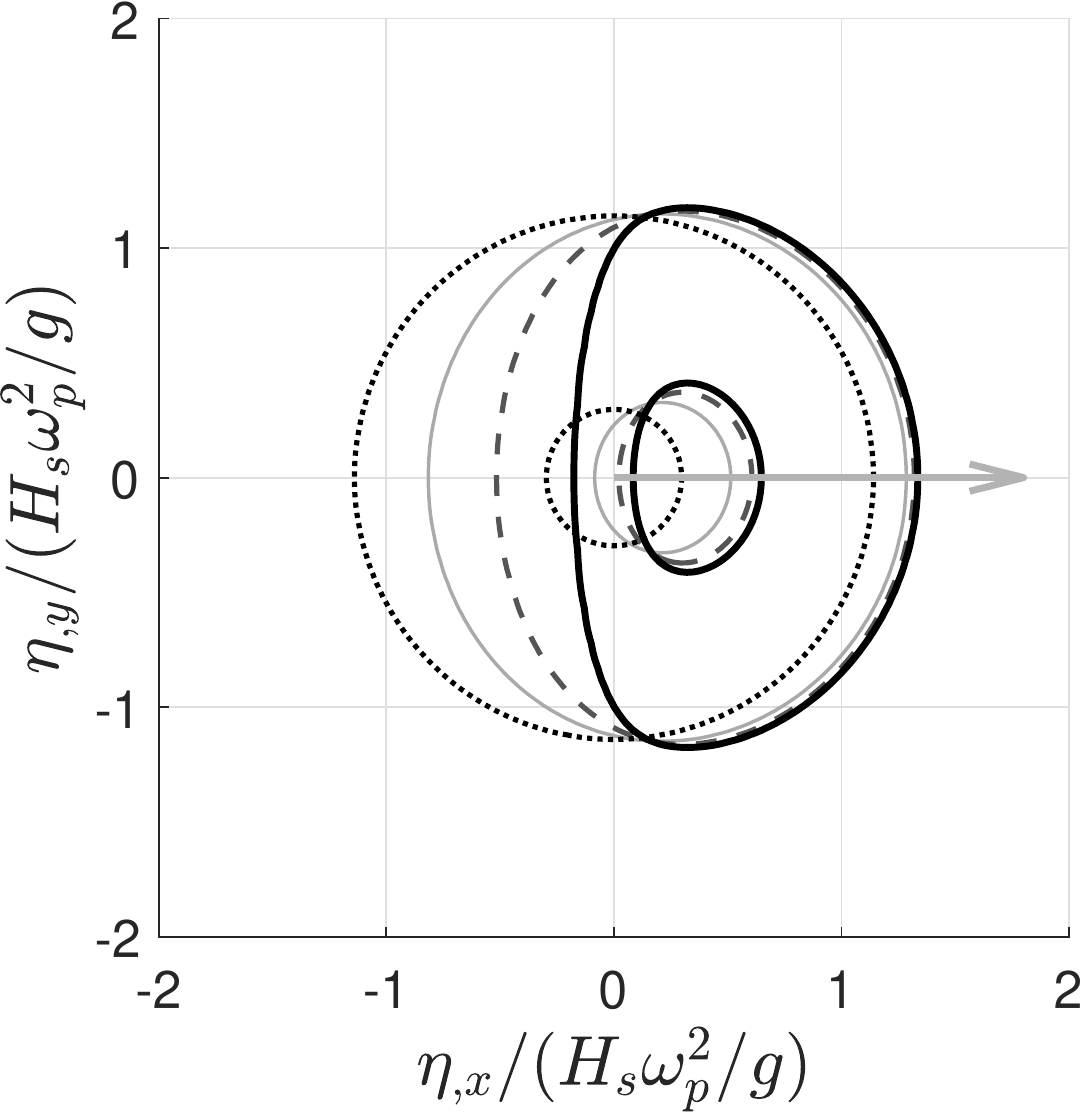} &
        \includegraphics[width=\scaleF\textwidth]{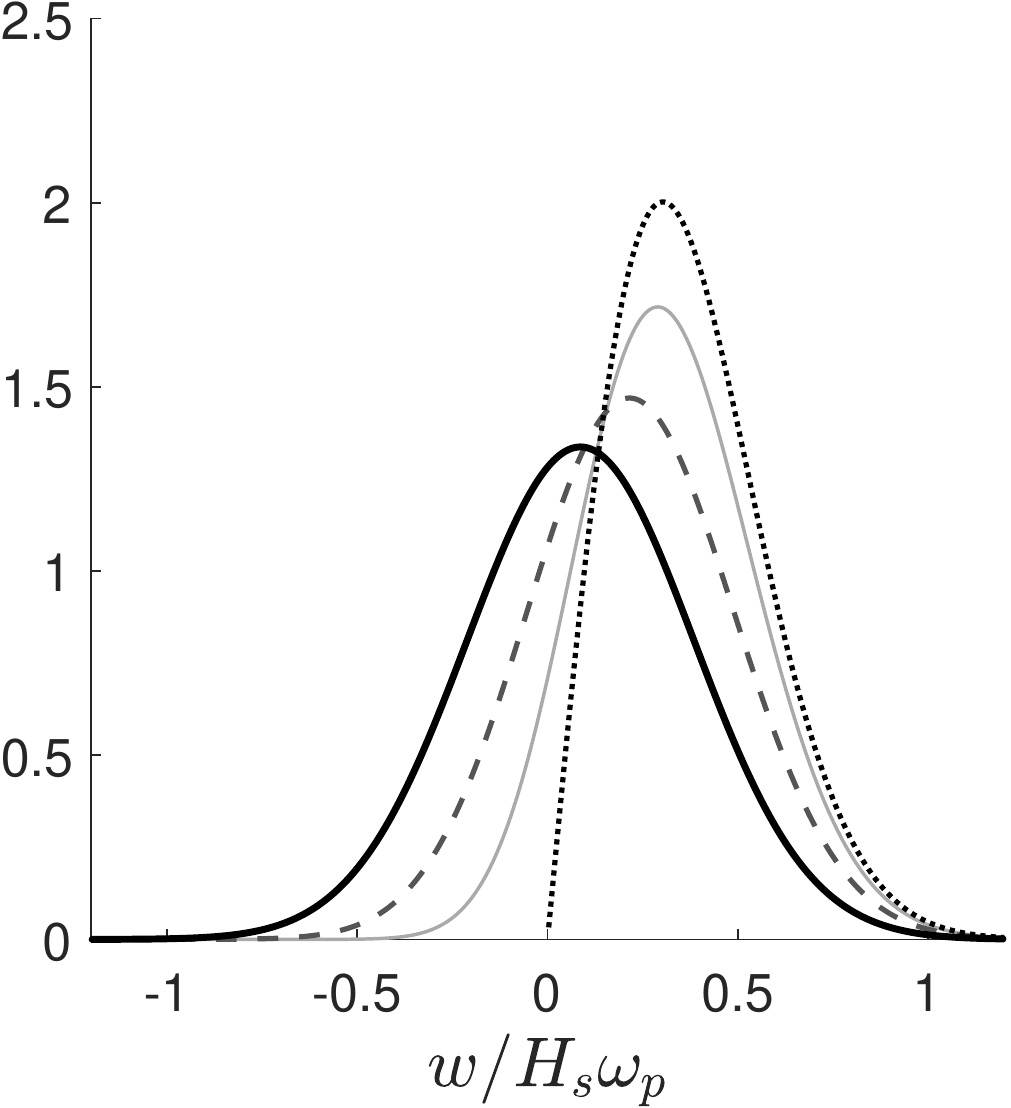}      
\end{tabular}
\end{center}
\caption{
Conditional density functions of kinematic variables, given up-crossing, for a moving material point. 
As in Figs. \ref{fig_example_rest_unidir}-\ref{fig_example_rest_multidir}, 
the density functions are represented in nondimensional form.
Five configurations are considered, as listed in \tab{table_cases}.
For all bivariate distributions, the levels of the two isodensity contours are (0.003;1).
In the case of the unidirectional sea ($C1$), no wave propages along the $y$-axis; therefore, $\ays = 0 $ and $\sys = 0$,
and no bivariate density function is represented.
For each configuration, the legend in the top right corner of the left figure indicates the
nondimensional forward speeds, $V_s / c_p$, for which the density functions have been plotted;
the heading of the moving material point is shown as a grey arrow in the middle figure.
}
\end{figure}

% %%%%%%%%%%%%%%%%%%%%%%%%
% %%%%%%%%%%%%%%%%%%%%%%%%
% %%%%%        FIGURE    END         %%%%%
% %%%%%%%%%%%%%%%%%%%%%%%%
% %%%%%%%%%%%%%%%%%%%%%%%%

\mypar{Configuration $C1$: unidirectional sea.}
As the sea is assumed to be unidirectional along the $x$-axis, 
the variables $\ay$ and $\sy$ are identically zero.
The distributions of $\sxs$, $\axs$ and $\vz$, given up-crossing, 
are all significantly affected by the forward speed of the material point.
In the asymptotic case where $V_s / c_p \rightarrow +\infty$
the conditional distribution of $\sxs$, given up-crossing,
converges towards a Rayleigh distribution of mode $\sqrt{m_4}/g$,
while in the limit $V_s / c_p \rightarrow -\infty$, 
this is $-\sxs$ which converges towards the same Rayleigh distribution.
Mathematically, this asymptotic behaviour can be explained as follows.
The material point heading is $\psi = 0$ and all waves propagate in the direction $\theta = 0$;
then, the transfer function of $\dEtaS$ is asymptotically equivalent to 
\begin{equation}
\mathcal{H}_{\dEtaS} (\omega,\theta=0) 
\ \  \underset{\text{\scriptsize $V_s \rightarrow \pm \infty$}}{\text{\LARGE $\sim$}} \
- i V_s  k(\omega)  
\, ,
\end{equation}
leading to the equivalence relation 
\begin{equation}
\label{eq_slope_equi_detas}
\sx 
\ \  \underset{\text{\scriptsize $V_s \rightarrow \pm \infty$}}{\text{\LARGE $\sim$}} \
\dEtaS / V_s 
\,   .
\end{equation}
As the conditional distribution of $\dEtaS$, given up-crossing, is of Rayleigh type,
\eq{eq_slope_equi_detas} explains the asymptotic behaviour of the conditional distribution of $\sx$.
From a physical viewpoint, \eq{eq_slope_equi_detas} 
reflects the fact
that the free surface can be considered as ``frozen'' at a given time, when $V_s \rightarrow \pm \infty$.
In practice, \fig{fig_example_vs_dists} shows that for $V_s / c_p =+4$ (resp. $V_s / c_p =-3$), 
$\sx$ (resp. $- \sx$) is already very close to be Rayleigh-distributed.

\mypar{Configurations $C2\rightarrow C4$: material point moving in a short-crested anisotropic sea with different headings.}
In the configurations $C2$-$C3$-$C4$, 
the material point is moving in a short-crested anisotropic sea, 
with a heading different for each configuration.
When $\abs{V_s} / c_p \gg 1$, 
the asymptotic behaviour is reminiscent of the one discussed
for configuration $C1$.
In this asymptotic regime, the free surface may be again considered as frozen
(except for waves propagating exactly abeam, 
i.e. with directions satisfying $\cos(\psi-\theta) = 0$).
Let $\eta_{,\ell}$ denote the wave slope 
measured along the direction of motion of the material point.
The transfer function of $\eta_{,\ell}$ is given by
\begin{equation}
\mathcal{H}_{\eta_{,\ell}}(\omega, \theta) = - {\rm sgn} (V_s) i \cos(\theta-\psi) k(\omega) 
\end{equation}
while the transfer function of $\dEtaS$ is asymptotically equivalent to
\begin{equation}
\displaystyle \mathcal{H}_{\dEtaS}(\omega, \theta) 
\ \  \underset{\text{\scriptsize $V_s \rightarrow \pm \infty$}}{\text{\LARGE $\sim$}} \ 
- i V_s  \cos(\theta-\psi) k(\omega)  \, , \ \ \  \text{for $\cos(\theta-\psi) \ne 0$,}
\end{equation}
which induces\footnote{
The only exception would the degenerate situation where all waves propagate exactly abeam, 
with $\cos(\theta-\psi) = 0$, leading to $\eta_{,\ell} = 0$ and $\dEtaS = \dot{\eta}$, 
regardless of the value of $V_s$.
} 
\begin{equation}
\dEtaS 
\ \  \underset{\text{\scriptsize $V_s \rightarrow \pm \infty$}}{\text{\LARGE $\sim$}} \ 
\abs{V_s} \eta_{,\ell} \, .
\end{equation}
As a consequence,
the univariate distribution of $\eta_{,\ell}$, given up-crossing,
converges to a Rayleigh distribution.
This explains the shift of the bivariate density function of $(\sxs,\sys)$ 
toward the direction of motion, as $\abs{V_s}$ increases (\fig{fig_example_vs_dists}, middle column).

For the heading $\psi=0$ (configuration $C2$), $\sys$ 
is independent of $\dEtaS$, which implies that the distribution of $\sys$ 
is not affected by the level-crossing conditioning: 
it remains a centred normal distribution.
Besides, $\sys$ and $\sxs$, non-conditioned,
are independent variables (since $\alpha_{11} = 0$, see \S\ref{subsec_body_rest} and \tab{table_alpha_value}).
Then, the combination of these two features implies 
that $\sxs$ and $\sys$, given up-crossing, are also independent. 
The univariate distribution of $\sxs$, given up-crossing, 
results from the \rev{convolution} of a Rayleigh 
distribution with a normal distribution (see \S\ref{subsubsec_marg_does} in appendix).

For headings $\psi = \pi/4;\pi/2$ (configurations $C3$ and $C4$), $\sxs$ and $\sys$ both depend on $\dEtaS$.
Then, the bivariate density function of $(\sxs,\sys)$, given up-crossing,
takes a more complicated form, 
whose analytical expression is detailed in appendix, \S\ref{subsubsec_biv_does}.
Although $\sxs$ and $\sys$, non-conditioned, are independent,
the level-crossing conditioning introduces a dependence.
For the configuration $C4$, only positive velocities are considered because of the symmetry of this configuration;
a change of sign in $V_s$ would have no effect on the distribution of $\vz$, 
and would change the bivariate density functions  of the pairs $(\sxs,\sys)$ and $(\axs,\ays)$ into their reflexion
through the $x$-axis.

\mypar{Configuration $C5$: isotropic sea.}
In an isotropic wave field, the bivariate density function of $(\sxs,\sys)$, 
given up-crossing, does not depend on the heading of the material point, except for a rotation, 
since the problem has been formulated in a frame which is not aligned with the direction of motion.
In the present example, $\psi = 0$ has been assumed.
Note also that for all density functions shown in \fig{fig_example_vs_dists}, configuration $C5$, 
only positive forward speeds have been assumed.
A change of sign in $V_s$ would have no effect on the distribution of $\vz$, 
and would simply change the bivariate density functions of $(\sxs,\sys)$ and $(\axs,\ays)$ into their reflexion
through the $y$-axis.
The conditional density function of $(\sxs,\sys)$ shifts toward the direction of motion as $V_s$ increases;
this tendency can be understood in the same way as explained above for configurations $C2\rightarrow C4$.
Similarly to configuration $C2$, in configuration $C5$, 
$\sxs$ and $\sys$, given up-crossing, are independent variables.

\section{Discussion about the use of the stochastic model in the context of slamming}
\label{sect_discussion}

\rev{
The conditional distribution of wave kinematic variables, given free-surface up-crossing, 
may be of practical interest for questions related to the resulting water entry events.
One application may be the prediction of the slamming load distribution 
for the design of a marine structure which will be exposed to wave impacts.
This possible application was a motive for the choice of kinematic variables considered in the present study
(see \S\ref{subsec_kin_vars}).
The present section discusses the use of the stochastic framework introduced 
in Sections \ref{sect_body_rest}-\ref{sec_forward},
for applications related to slamming. 
First, the different combinations of kinematic variables which may be considered
are discussed in \S\ref{subsec_ckv},
in the light of existing studies on stochastic slamming in irregular waves.
%(with or without the effect of forward speed).
Then, \S\ref{subsec_vars_we} explains how 
the velocity and acceleration components of the fluid may be expressed 
in a local frame relevant for the water entry problem.
Subsection \ref{subsec_skm} briefly mentions the possibility of taking into account 
seakeeping motions in the stochastic analysis.
Finally, the potential need to take into account the temporal evolution of kinematic variables during water entry events, 
and how it may be implemented,
is discussed in \S\ref{subsec_time_evo}.
}

\subsection{Considered kinematic variables}
\label{subsec_ckv}

The normal component (relative to the free surface) of the fluid velocity is the most decisive variable,
when addressing the question of slamming loads \rev{on a marine structure}.
In many stochastic approaches it is the only considered random variable,
the slamming loads being assumed to be weakly dependent on the other kinematic variables
(see for example \cite{ochi_1973,rassinot_1995,wang_2002,hermundstad_2007,dessi_2013,wang_2016}). 
Helmers et al. (2012) \cite{helmers_2012}
implemented a more comprehensive stochastic approach, 
where the conditional joint distribution of four kinematic variables 
(vertical velocity, vertical acceleration, wave slope, and seakeeping heel angle) was used
to estimate the probability distribution of impact loads on a wedge-shaped body, 
exposed to unidirectional waves, with no forward speed.
As they used an analytical 
Wagner-type \cite{wagner_1932}
water entry model (computationally fast), 
Helmers et al. (2012) could perform the transfer of the distribution of kinematic variables through the impact model
by using a Monte Carlo sampling.
If they are identified as relevant, additional kinematic variables, 
such as the tangential velocity of the fluid (see for example \cite{belik_1982}), 
may be included in the stochastic analysis.
Generally, the more numerous the considered kinematic variables, 
the more elaborate the impact model need to be.
The most advanced analytical models based on Wagner's theory can, 
in principle, take into account all the kinematic variables considered in the present analysis 
(see e.g. \cite{scolan_2015}).
If the water-entry model is computationally demanding (e.g. CFD simulations),
as an alternative to Monte Carlo sampling,
other approaches such as metamodels or reliability methods 
may be used to probe the probability distribution of slamming loads and stresses.

\subsection{Velocity and acceleration components in the local frame of the free surface}
\label{subsec_vars_we}

\begin{figure}[h]
\begin{center}
\begin{tabular}{c}
        \includegraphics[width=0.5\textwidth]{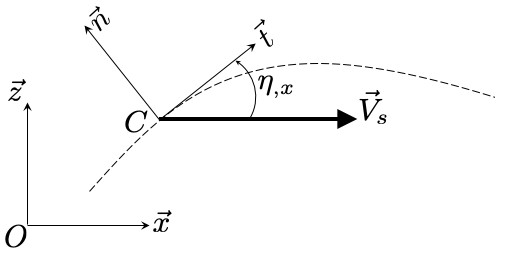}  
\end{tabular}
\end{center}
\caption{
Local frame for the water entry problem. 
The material point moves along the $x$-axis with a constant velocity, $V_s$.
It crosses the water free surface, represented as a dashed line, at a point $C$. 
When addressing a question related to the water entry phenomenon 
(e.g. to estimate slamming loads), 
it is convenient to express kinematic variables 
in a local fixed frame $(C,\vec{t},\vec{n})$, where $\vec{t}$ and $\vec{n}$ 
are locally tangent and normal to the free surface, respectively.
}
\label{fig_frame_slam}
\end{figure}

When the considered solid body moves forward through the wave field, 
some attention is required regarding the fluid motion components 
to be used as an input for the water entry model.
These components should be specified as normal and tangential, 
relative to the local free surface.
For the sake of simplicity, let consider a two-dimensional situation 
(it can be readily generalised to three dimensions), 
where a material point moves through a unidirectional sea,
along the $x$-axis.
Let consider an up-crossing event where the material point crosses the free surface at a point $C$,
and let define a local fixed frame $(C,\vec{t},\vec{n})$, where the vectors $\vec{t}$ and $\vec{n}$ 
are locally tangent and normal to the free surface, at up-crossing.
The situation is sketched in \fig{fig_frame_slam}.
To the leading order 
(consistent with the order of approximation of the linear wave model), the relative fluid velocity,
in the local frame $(C,\vec{t},\vec{n})$, reads
\begin{equation}
\label{eq_vrel_impact}
\vec{V}_{\rm r} \simeq \underbrace{(\vx-V_s)}_{\simeq v_t} \vec{t} + 
\underbrace{(V_s \eta_{,x} +\vz)}_{\simeq v_n} \vec{n} \, .
\end{equation}
Note that the term $V_s \eta_{,x}$ should not be neglected,
since $V_s$ may be significantly larger than the magnitude of $w$,
even for vessels with moderate speeds.
Hence, to the leading order, the relative normal velocity of the fluid
is equal to the relative velocity of the free surface elevation: 
\begin{equation}
\label{eq_vrel_vn}
v_n  \simeq V_s \eta_{,x} + w = \dEtaS \, .
\end{equation}
From a physical standpoint, the relation $v_n \simeq \dot{\eta}_s$ 
can be interpreted as the kinematic free-surface condition, 
expressed to the leading order, in the reference frame of the moving material point.

Regarding the relative fluid acceleration, its components in the local fixed frame $(C,\vec{t},\vec{n})$,
to the leading order, read
\begin{equation}
\label{eq_vrel_impact}
\vec{A}_{\rm r} \simeq \underbrace{\ax}_{\simeq a_t} \vec{t} + 
\underbrace{\az}_{\simeq a_n} \vec{n} \, ,
\end{equation}
where the assumption that the material point moves with a constant velocity 
has been taken into account.
Contrary to the result obtained for the velocity, to the leading order, 
the normal component of the fluid acceleration does not equal 
the acceleration rate of the free surface elevation, 
measured in the frame of the moving material point.

% %%%%%%%%%%%%%%%%%%%%%%%%
% %%%%%%%%%%%%%%%%%%%%%%%%
% %%%%%        FIGURE    BEGIN         %%%%%
% %%%%%%%%%%%%%%%%%%%%%%%%
% %%%%%%%%%%%%%%%%%%%%%%%%

% %%%%%%%%%%%%%%%%%%%%%%%%
% %%%%%%%%%%%%%%%%%%%%%%%%
% %%%%%        FIGURE    END         %%%%%
% %%%%%%%%%%%%%%%%%%%%%%%%
% %%%%%%%%%%%%%%%%%%%%%%%%

\subsection{Accounting for seakeeping motions}
\label{subsec_skm}

The present study focused on the effect of the forward velocity 
on the up-crossing frequency and the related conditional distribution of kinematic variables. 
Accounting for seakeeping motions in the present stochastic analysis
would present no specific difficulties, as long as the seakeeping motions are linearly modelled.
The way forward to include seakeeping motions in the present stochastic approach 
is described in appendix B.

\subsection{Time evolution of kinematic variables during the water entry}
\label{subsec_time_evo}

When considering the level-crossing problem in the context of slamming,
another important question 
is whether the wave kinematics can be considered as fixed during the slamming event.
Indeed, the present study focussed on the joint distribution of kinematic variables at the free-surface crossing,
without considering the subsequent evolution of kinematic variables.
Kinematic variables may be considered as fixed if the characteristic time of the slamming event, $t_s$,
is much smaller than the characteristic encounter period of impacting waves, which may be translated into
\begin{equation}
\label{eq_fixed_vars_condition}
\tilde{\omega} t_{s} \ll 1 \, .
\end{equation}
The characteristic slamming time, $t_s$, 
may be roughly estimated 
as the characteristic height of the exposed structure, $h_e$, 
divided by the characteristic impact velocity, 
\begin{equation}
\label{eq_char_slam_time}
t_s \simeq \frac{h_e}{H\tilde{\omega}/2} \, ,
\end{equation}
where $H$ is the characteristic height of impacting waves.
The combination of Eqs.~(\ref{eq_fixed_vars_condition}-\ref{eq_char_slam_time})
shows that kinematic variables may be considered as fixed during the impact event,
if the vertical extension of the body part exposed to slamming is much smaller than the characteristic height of impacting waves; 
this may be typically the case for appendages, such as stabilising fins on a ship, or diving planes on a surfaced submarine.
Assuming fixed kinematic variables may be also justified for larger bodies,
if the early stage of the water entry can be considered as the most relevant with regard to hydrodynamic loads;
this may be typically the case for blunt bodies such as, for example, 
a bulbous bow or the sections of a flat-bottomed ship.

Conversely, if the late stage of the wave impact and/or the subsequent water exit phase
needs to be investigated, 
the subsequent time evolution of kinematic variables need to be considered for each event.
This time evolution is stochastic by nature.
The realisation of stochastic trajectories, following up-crossing events, 
may be obtained by means of Monte Carlo experiments, 
which may be computationally demanding.
These Monte Carlo experiments may be performed by using two different approaches:
(i) the full realisation of a sea state, from which stochastic trajectories of water entries are extracted, 
after up-crossing detections,
(ii) the realisation of short sea state sequences, starting directly from initial up-crossing conditions 
which are drawn from the conditional distribution of kinematic variables, given up-crossing.
The option (ii) may prove numerically much more efficient, especially when considering rare events; 
for instance, if the body is far from the mean sea level, or if a focus on extreme wave impacts is needed.
This option would also offer more flexibility to implement 
variance reduction techniques, such as importance sampling.

The subsequent evolution of kinematic variables, starting from random initial conditions,  
may be accounted for in a simplified deterministic manner. 
Helmers et al. (2012) \cite{helmers_2012} used such an approach 
to compute the time evolution of the free surface elevation, $\eta_0$, after up-crossing:
starting from random initial up-crossing conditions for $\dot{\eta}_0$ and $\ddot{\eta}_0$, 
and assuming that the third time derivative, $\dddot{\eta}_0$,  is equal to its conditional mean, 
given the instantaneous value of the first time derivative, $\dot{\eta}_0$,
these authors obtained a deterministic expression for $\eta_0$ as the solution of a second order differential equation.
Alternatively, another option is to use regression techniques based on Slepian models (see e.g. \cite{lindgren_1991}).
This latter approach would allow to also insert random intermediate conditions at intermediate timepoints after upcrossing.

\section{Conclusions}
\label{sec_conclusion}

The effect of forward speed on stochastic free-surface crossing has been studied
in the framework of the linear wave model.
The present study has focused on up-crossing events (i.e. the crossing of the free surface by the object, into the water domain);
however, the model may be readily modified to include both up-crossing and down-crossing events,
or focus on down-crossing events only. 
The conditional distribution of kinematic variables, given up-crossing, 
has been shown to be significantly affected by the forward velocity of the considered material point.
The effect on the related up-crossing frequency has been also investigated.
The analysis has been carried out analytically, 
and formulae have been given for the general case of multidirectional waves in water of finite depth.
In the specific case where the water depth can be assumed infinite
and the sea state has a two-dimensional spectrum with separated frequency/direction dependencies,
analytical developments have been furthered to express the conditional distribution of kinematic variables 
and the up-crossing frequency
in terms of wave frequency moments and non-dimensional coefficients accounting for the wave directional spreading.

The prediction of the probability distribution of slamming loads on a ship may 
be a particular application of the present stochastic model.
The considered solid body has to be small compared to water wave wavelengths, 
so the present theory can be used.
This condition may be fulfilled for small crafts, or specific structures on large ships 
(e.g. appendages, bulbous bow, ship sections).
The way forward to include seakeeping motions in the analysis 
has been briefly sketched in appendix.
Different possible approaches to perform the transfer of the probability distribution of kinematic variables,
through a water entry model, have been discussed.

% %%%%%%%%%%%%%%%%%%
% %%%%%%%%%%%%%%%%%%
% %%%%%%%%%%%%%%%%%%
% %%%     APPENDICES
% %%%%%%%%%%%%%%%%%%
% %%%%%%%%%%%%%%%%%%
% %%%%%%%%%%%%%%%%%%

% Defining section format
\setcounter{section}{0}
\renewcommand\thesection{Appendix \Alph{section}}
\renewcommand\thesubsection{\Alph{section}.\arabic{subsection}}

% resetting equation counter and equation numbering format
\setcounter{equation}{0}
\renewcommand{\theequation}{\Alph{section}.\arabic{equation}}

\section{Conditional distributions of kinematic variables, given up-crossing}
\label{sec_ap_cond_dist}

\subsection{Body at rest}
\label{app_marg_dist_rest}

This appendix section gives explicit analytical expressions for the conditional density functions 
of kinematic variables
-- given that $\etaN\rev{(t)}$ up-crosses the level $a$ --
represented in Figs.~\ref{fig_example_rest_unidir}-\ref{fig_example_rest_multidir}. 
The free surface elevation is given, $\eta = a$.
The vertical velocity component, $\vz$, follows a Rayleigh distribution of mode $\sqrt{m_2}$,
whose density function is given by:
\begin{equation}
\label{eq_dist_vz}
f_{\vz|\etaN\rev{(t)} \uparrow a} (\vz) = \frac{\vz}{m_2} \exp\left( - \frac{\vz^2}{2 m_2} \right) 
\, ,  \ \ w \ge 0 \, .
\end{equation}
The other kinematic variables of the Gaussian vector $Z_A$ (see \eqp{eq_Ztot}) 
either depend on $\eta = \etaN$, or on $\vz = \dEtaN$, or on neither;
none of the kinematic variables considered in Section~\ref{sect_body_rest} depends concurrently on both $\eta$ and $\vz$.
Generic expressions for their respective univariate conditional density functions, given up-crossing, are detailed below.

\subsubsection{Variables which depend on $\eta$ (i.e. $\vx$, $\vy$, $\az$)}
\label{subsubsec_doeta}
Variables which depend on $\eta$  
follow conditional distributions of Gaussian type. 
By denoting $\zeta$ the considered variable, its conditional mean is given by 
\begin{equation} 
\label{eq_mean_cn}
E\left\{ \zeta | \etaN\rev{(t)}  \uparrow a \right\} 
=  E\left\{ \zeta | \eta = a \right\} 
= \frac{{\rm Cov}(\zeta,\eta)}{m_0} a  \, , 
\end{equation}
and its conditional variance by 
\begin{equation}
\label{eq_sigma_cn}
{\rm Var} \left\{ \zeta | \etaN\rev{(t)}  \uparrow a \right\} 
= {\rm Var} \left\{ \zeta | \eta = a \right\} 
= {\rm Var}(\zeta) - \frac{{\rm Cov}(\zeta,\eta)^2}{m_0}  \, ,
\end{equation}
where ${\rm Cov}(\zeta,\eta)$ is the non-conditional covariance of $\zeta$ and $\eta$.
The relevant non-conditional covariances may be found 
in the expression of the related covariance matrix, 
$\Sigma_X$ (see \eqp{eq_sigmaX}).

\subsubsection{Variables which depend on $\vz$ (i.e. $\sx$, $\sy$, $\ax$, $\ay$)}
\label{subsubsec_xi0}
The conditional density function of a variable, $\xi$, which depends on $\vz$, is given by 
\begin{equation}
\label{eq_maginal_dotu_0}
f_{\xi|\etaN\rev{(t)}  \uparrow a} (\xi) = \int_0^{+\infty} \sqrt{\frac{2\pi}{m_2}}  \rev{\chi}  f_{\xi, \vz}(\xi, \rev{\chi}) \ {\rm d} \rev{\chi} \, ,
\end{equation}
where $f_{\xi, \vz}$ is the non-conditional bivariate density function of $\xi$ and $\vz$.
\rev{It may be convenient to express $f_{\xi|\etaN\rev{(t)} \uparrow a}$ differently, as follows:
\begin{equation}
\label{eq_maginal_dotu_0_bis}
f_{\xi|\etaN\rev{(t)}  \uparrow a} (\xi) = \int_0^{+\infty} f_{\vz|\etaN\rev{(t)}\uparrow a}(\chi) f_{\xi|\vz=\chi}(\xi) \ {\rm d} \chi \, ,
\end{equation}
where $f_{\vz|\etaN\rev{(t)}\uparrow a}$ is given by \eq{eq_dist_vz} 
and  $f_{\xi|\vz=\chi}$ is the conditional distribution of $\xi$, given $\vz = \chi$. 
%($f_{\xi|\vz=\chi}$ is a normal distribution).
%After some algebra, \eq{eq_maginal_dotu_0} can be expressed as follows
Then, by making the change of variable 
\begin{equation}
z = \frac{\abs{\rho} \sigma_{^\xi}}  {\sqrt{m_2}} \chi  \, ,
\end{equation}
\eq{eq_maginal_dotu_0_bis} may be transformed into
}
\begin{equation}
\label{eq_maginal_dotu_1}
\begin{array}{lll}
f_{\xi | \etaN\rev{(t)}  \uparrow a} (\xi) & = & \displaystyle{
\int_0^{+\infty}   \frac{z}{(\rho\sigma_{^\xi})^2} \exp\left\{-\frac{1}{2}\left( \frac{z}{\rho\sigma_{^\xi}} \right)^2\right\}} \\
& & \displaystyle{\ \ \ \ \ \ \times \frac{1}{\sqrt{2\pi}\sqrt{1-\rho^2}\sigma_{^\xi}} 
\exp\left\{ -\frac{\left[\xi-  {\rm sign}(\rho) z\right]^2}{2(1-\rho^2)\sigma_{^\xi}^2} \right\} 
{\rm d} z} \, , 
\end{array}
\end{equation}
where $\sigma_{^\xi}$ is the non-conditional standard deviation of $\xi$, and $\rho = {\rm Cov}(\xi,\vz) / \sigma_{^\xi}\sqrt{m_2}$
is the non-conditional correlation coefficient between $\xi$ and $\vz$.
In the case where $\rho>0$, 
\eq{eq_maginal_dotu_1} corresponds to the \rev{convolution} of a Rayleigh distribution of mode $\rho\sigma_{^\xi}$
and a centred normal distribution of variance $(1-\rho^2)\sigma_{^\xi}^2$ 
\rev{(which is the conditional variance of $\xi$, given $\vz$)}.
When $\rho<0$, there is a change of sign in \eq{eq_maginal_dotu_1}; 
then,  through the change of variable $z \rightarrow -z$, it can also be expressed as a \rev{convolution}, 
where the Rayleigh distribution is replaced by its reflection about the axis $z=0$.
\eq{eq_maginal_dotu_1} can be further expressed as
\begin{equation}
\label{eq_marg_fin}
\begin{array}{ll}
f_{\xi | \etaN\rev{(t)}  \uparrow a} (\xi) =  & \displaystyle{ \sqrt{\frac{1-\rho^2}{2\pi}}\frac{1}{\sigma_{^\xi}} \exp\left\{ -\frac{\xi^2}{2(1-\rho^2)\sigma_{^\xi}^2} \right\} } \\
& \displaystyle{ \times
\left[
1+\sqrt{\frac{\pi}{2}} \frac{\rho \xi}{\sqrt{1-\rho^2}\sigma_{^\xi}} 
\exp\left\{ \frac{\rho^2\xi^2}{2(1-\rho^2)\sigma_{^\xi}^2} \right\}
\left(1+\erf{\left[\frac{\rho \xi}{ \sqrt{2(1-\rho^2)} \sigma_{^\xi} }\right]}\right)
\right] \,. }
\end{array}
\end{equation}
\rev{
Aberg et al. (2008) \cite{aberg_2008} proposed a more compact form for this type of distribution,
by expressing it in terms of cumulative distribution function, 
and making use of the standard normal distribution (see lemma 5.3 in \cite{aberg_2008}).
}

\rev{
Since the density function $f_{\xi | \etaN\rev{(t)} \uparrow a}$ can be expressed as the convolution 
of two elementary density functions
(see Eq. \ref{eq_maginal_dotu_1}),
the corresponding random variable, $\xi | \etaN\rev{(t)} \uparrow a$, may be 
%more easily 
comprehended 
as the sum (or difference for negative values of $\rho$)
of two independent random variables:
(i) one being Rayleigh-distributed with a mode $\abs{\rho}\sigma_{^\xi}$, 
(ii) and the other being normally distributed, with a mean equal to zero and a variance equal to $(1-\rho^2)\sigma_{^\xi}^2$ 
(see also \cite{aberg_2008}, lemma 5.1).
The ``relative weight'' of these two components is controlled by $\rho$, 
the correlation coefficient  between $\xi$ and $\vz$.
As expected, for $\abs{\rho} = 1$ (resp. $\rho = 0$) 
only the Rayleigh component (resp. the normal component) remains. 
%Therefore, the random variable $\xi | \etaN \uparrow a$ may be comprehended
}

\subsection{Material point with translational motion}
\label{app_marg_dist_vs}

This appendix section gives explicit analytical expressions for the conditional distributions 
-- given that $\etaS\rev{(t)}$ up-crosses the level $a$ --
represented in Fig.~\ref{fig_example_vs_dists}. 
The free surface elevation is given, $\etaS = \eta = a$.
The variable $\dEtaS$, given up-crossing, follows a Rayleigh distribution of mode $\sqrt{\mmt}$
(see Eqs. \ref{eq_m2t}-\ref{eq_m2t_quadratic}).
The conditional distribution of variables which depend on $\etaS$ but not on $\dEtaS$ (namely $\az$, $\vx$, $\vy$), 
is not affected by the horizontal motion of the material point, and results given in \S\ref{subsubsec_doeta} are still valid.
As for the variables which depend on $\dEtaS$ (namely $\vz$, $\ax$, $\ay$, $\sx$, $\sy$), 
the analytical expression of their conditional univariate distributions is given in \S\ref{subsubsec_marg_does}.
An analytical expression for the conditional bivariate distribution of two variables which depend on $\dEtaS$
is given in \S\ref{subsubsec_biv_does}.
This expression has been used to plot 
the isodensity lines of the pairs $(\sx,\sy)$ and $(\ax,\ay)$, 
shown in \fig{fig_example_vs_dists}.

\subsubsection{Conditional distribution of a variable which depends on $\dEtaS$}
\label{subsubsec_marg_does}

The conditional density function, given up-crossing, 
of a variable $\xi$, which depends on $\dEtaS$,
may be expressed as 
\begin{equation}
\label{eq_vz_condi_fs}
f_{\xi|\etaS\rev{(t)}  \uparrow a} (\xi) = \int_0^{+\infty} \sqrt{\frac{2\pi}{m_{\tilde2}}}  \rev{\chi}  f_{\xi, \dEtaS}(\xi, \rev{\chi}) \ {\rm d} \rev{\chi} \, ,
\end{equation}
which is similar to \eq{eq_maginal_dotu_0}. 
Then, in a way similar to the development given in \S\ref{subsubsec_xi0} 
(where the variable $\vz$ should be replaced with $\dEtaS$), 
$f_{\xi|\etaS\rev{(t)} \uparrow a}$ may be expressed as the \rev{convolution} of a Rayleigh distribution 
and a normal distribution;
Eqs.~(\ref{eq_maginal_dotu_1}-\ref{eq_marg_fin}) may be readily reused, 
with $\rho$ now being the non-conditional correlation coefficient between $\xi$ and $\dEtaS$
(see the expression of the related covariance matrix, $\Sigma_{Y_s}$, in \eqp{eq_sigmaY_s}).

\subsubsection{Conditional bivariate distribution of two variables which depend on $\dEtaS$}
\label{subsubsec_biv_does}

The conditional bivariate distributions of $(\axs,\ays)$ and $(\sxs,\sys)$, 
represented in Fig.~\ref{fig_example_vs_dists}, 
require to consider the case of two variables 
which both depend on the level-crossing velocity, $\dEtaS$.
In order to adopt general notations,
let $Q$ be a Gaussian vector  
\begin{equation}
Q = \left[ 
\begin{array}{l}
q_1 \\ 
q_2 \\
q_3
\end{array} 
\right] \, ,
\end{equation}
where $q_1$, $q_2$, $q_3$ denote respectively $\axs$ (or $\sxs$), 
$\ays$ (or $\sys$), and $\dEtaS$.
Then the bivariate density function of $q_1$ and $q_2$, given up-crossing, may be expressed as
\begin{equation}
\label{eq_append_g12}
g(q_1, q_2) = \frac{1}{\sigma_3}\frac{1}{2\pi \
{\rm det}(\Sigma_Q)^{1/2}
} \int_0^{+\infty} {\rm d} q_3 \ q_3 
\displaystyle  \exp\left\{ -\frac{1}{2} Q^\intercal \Sigma_Q^{-1} Q \right\} \, ,
\end{equation}
where $\Sigma_Q$ is the covariance matrix of the Gaussian vector $Q$, 
and $\sigma_3$ is the standard deviation of $q_3$.
\rev{
Similarly to the explanation proposed in \S\ref{subsubsec_xi0},
the conditional bivariate distribution, $g$, 
may be comprehended as resulting from the sum of two independent vectors.
%:one being Rayleigh-distributed and the other being normally distributed.
The first vector may be expressed as 
\begin{equation}
R = \left[ 
\begin{array}{l}
\rho_{13} \sigma_1 \\ 
\rho_{23} \sigma_2
\end{array} 
\right] r \, ,
\end{equation}
where $\rho_{ij}$ is the (unconditioned) correlation coefficient between the $i$-th and $j$-th components
of the vector $Q$,
$\sigma_i$ is the (unconditioned) standard deviation of the $i$-th component of the vector $Q$,
and $r$ is a random variable which is Rayleigh-distributed with a mode equal to $1$.
The second vector is a zero-mean Gaussian vector,
whose covariance matrix is equal to the conditional covariance matrix of $(q_1,q_2)$, given $q_3$.
}

\rev{
After completing the square in the argument of the exponential function in \eq{eq_append_g12},
the following closed-form expression is obtained:
} 
%\eq{eq_append_g12} may be further expressed as:
%
%\begin{align}
%g(q_1, q_2) = & \frac{1}{\sigma_3}\frac{1}{8\pi \
%{\rm det}(\Sigma_Q)^{1/2}
%}  \frac{\exp\left[-C(q_1,q_2)\right]}{{a_{33}}^{3/2}}  \nonumber \\
% & \times \left[ 2 \sqrt{a_{33}} - \sqrt{\pi} B(q_1,q_2)  \exp(\frac{B(q_1,q_2)^2}{4 a_{33}}) {\rm erfc} \left( \frac{B(q_1,q_2)}{2\sqrt{a_{33}}} \right)\right] \, .
%\end{align} 
%
\begin{align}
\displaystyle
g(q_1, q_2) = & \frac{1}{\sigma_3}\frac{1}{8\pi \
{\rm det}(\Sigma_Q)^{1/2}
}  \frac{\exp\left[-C(q_1,q_2)\right]}{{a_{33}}^{3/2}}  \nonumber \\
 & \times \left\{2 \sqrt{a_{33}} - \sqrt{\pi} B(q_1,q_2)  \exp(\frac{B(q_1,q_2)^2}{4 a_{33}}) 
 \rev{\left[ 1 - {\rm erf} \left( \frac{B(q_1,q_2)}{2\sqrt{a_{33}}} \right) \right]}
 \right\} \, .
\end{align} 
The functions $B$ and $C$ are given by 
\begin{equation}
B(q_1,q_2) = 2 a_{13} q_1 + 2 a_{23} q_2 \,
\end{equation}
\begin{equation}
C(q_1,q_2) = a_{11} {q_1}^2 + a_{22} {q_2}^2 + 2 a_{12} q_1 q_2 \, ,
\end{equation}
and $a_{kl}$ are numerical coefficients defined as 
\begin{equation}
a_{kl} = \frac{1}{2} \left[ \Sigma_Q^{-1} \right]_{kl} \, .
\end{equation}
%

% resetting equation counter and equation numbering format
\setcounter{equation}{0}

\section{Stochastic approach including seakeeping motions}
\label{sec_append_seakeeping}

Including seakeeping motions in the present stochastic approach 
would require to consider a new stochastic process, 
\begin{equation}
\label{eq_seakeeping}
\etaB(t) = \eta_s(t) + \zeta_s(t) \, ,
\end{equation}
where $\etaB$ denotes the relative free surface elevation measured in the frame of the moving material point.
It is decomposed as the sum of $\eta_s\rev{(t)}$, which
accounts for the forward motion of the material point (\eqp{eq_etas}), 
and $\zeta_s\rev{(t)}$, which accounts for seakeeping 
motions around the average forward motion.
If the seakeeping motions are linearly modelled, 
$\zeta_s\rev{(t)}$ results from a linear transformation of $\eta_s\rev{(t)}$,
and consequently $\etaB\rev{(t)}$ is also a Gaussian process.
Knowing the response amplitude operators of the floating platform, 
along with the position of the material point on this platform, 
the transfer function of $\zeta_s$ can be readily expressed.
The variable $\zeta_s$ may also account for the diffraction waves as well as the waves generated by
the steady (forward) and unsteady motions of the vessel, if the related transfer functions are known --
see for example Hermundstad and Moan (2005, 2007) \cite{hermundstad_2005, hermundstad_2007}
who include the effect of the waves generated by the forward motion of the ship in their analysis.
Then, in a way similar to \eq{eq_Zs}, 
a new Gaussian vector collecting the relevant kinematic variables should be defined,
\begin{equation}
\label{eq_Zb}
Z_b = \left[ 
\begin{array}{l}
 \etaB \\  
 \dEtaB \\
 Z_{_W} 
\end{array}
\right] \, ,
\end{equation}
where $Z_{_W}$ contains the kinematic variables of interest
(for example the kinematic variables necessary as the input of a water entry model).
The non-conditional probability distribution of $Z_b$ is a multivariate normal distribution,
whose covariance matrix may be expressed in the same way as \eq{eq_covmat_expr}. 
Let $\cZb$ denote the random vector containing the variables of $Z_b$, except for $\etaB$.
The conditional density function of kinematic variables, given up-crossing, can be written as
\begin{equation}
\displaystyle
f_{\cZb | \etaB\rev{(t)}  \uparrow a}(\dEtaB, Z_{_W}) = \frac{\sqrt{2\pi}}{\sigma_{\dEtaB}}\
 \dEtaB \  f_{\cZb | \etaB =  a}(\dEtaB, Z_{_W}) \ , \ {\rm with \ \dEtaB > 0} \, ,
\end{equation}
where $a$ is the altitude of the material point on calm water, 
and $\sigma_{\dEtaB}$ is the non-conditional standard deviation of $\dEtaB$.
The related up-crossing frequency reads
\begin{equation}
\label{eq_fip_sk}
\nub = \frac{1}{2\pi} \frac{\sigma_{\dEtaB}}{\sigma_{\etaB}} \exp \left\{ - \frac{1}{2} \left( \frac{a}{{\sigma_{\etaB}}} \right)^2 \right\} \, ,
\end{equation}
where $\sigma_{\etaB}$ is the standard deviation of $\etaB$ 
(which is not equal to $\sqrt{m_0}$, due to seakeeping motions).

\section*{Acknowledgements}

This work was supported by the French National Agency for Research (ANR) 
and the French Government Defence procurement and technology agency  (DGA); 
ANR-17-ASTR-0026 APPHY.

%\section*{References}

\bibliography{mybibfile}

\end{document}